\RequirePackage{lineno}
\documentclass[aps,prx,10pt,twocolumn,superscriptaddress,longbibliography,nofootinbib]{revtex4-2}

\usepackage{graphicx}
\usepackage{amssymb,amsmath,amsfonts,gensymb}
\usepackage{bm,hyperref}
\usepackage{color}
\usepackage{ulem}
\usepackage{bbm}
\usepackage{subfigure}
\usepackage{dcolumn}
\usepackage{mathtools}
\usepackage{pifont}
\usepackage[symbol]{footmisc}

\newcommand{\ket}[1]{\vert #1 \rangle}
\newcommand{\bra}[1]{\langle #1 \vert}
\newcommand{\bracket}[2]{\langle #1 \vert #2 \rangle}
\newcommand{\eV}{{\text{eV}}}
\newcommand{\eVA}{{\text{eV}\cdot \text{\AA}}}
\newcommand{\bk}{\mathbf{k}}
\newcommand{\bq}{\mathbf{q}}
\newcommand{\btk}{\widetilde{\mathbf{k}}}
\newcommand{\btq}{\widetilde{\mathbf{q}}}
\newcommand{\br}{\mathbf{r}}

\def\I{\uppercase\expandafter{\romannumeral 1}}
\def\II{\uppercase\expandafter{\romannumeral 2}}
\def\III{{\uppercase\expandafter{\romannumeral 3}}}
\def\IV{{\uppercase\expandafter{\romannumeral 4}}}
\def\V{{\uppercase\expandafter{\romannumeral 5}}}
\def\VI{{\uppercase\expandafter{\romannumeral 6}}}
\def\VII{{\uppercase\expandafter{\romannumeral 7}}}

\def\a{\mathbf{a}}
\def\b{\mathbf{b}}
\def\p{\mathbf{p}}
\def\ta{\tilde{\mathbf{a}}}
\def\L{\mathbf{L}}

\def\k{\mathbf{k}}
\def\vr{\mathbf{r}}
\def\G{\mathbf{G}}
\def\Q{\mathbf{Q}}
\def\br{\mathbf{r}}

\def\q{\mathbf{q}}

\def\nn{\nonumber \\}
\def\hc{\hat{c}}
\def\hcd{\hat{c}^{\dagger}}

\def\R{\mathbf{R}}

\makeatletter
\def\@ssect@ltx#1#2#3#4#5#6[#7]#8{%
	\def\H@svsec{\phantomsection}%
	\@tempskipa #5\relax
	\@ifdim{\@tempskipa>\z@}{%
		\begingroup
		\interlinepenalty \@M
		#6{%
			\@ifundefined{@hangfroms@#1}{\@hang@froms}{\csname @hangfroms@#1\endcsname}%
			{\hskip#3\relax\H@svsec}{#8}%
		}%
		\@@par
		\endgroup
		\@ifundefined{#1smark}{\@gobble}{\csname #1smark\endcsname}{#7}%
	}{%
		\def\@svsechd{%
			#6{%
				\@ifundefined{@runin@tos@#1}{\@runin@tos}{\csname @runin@tos@#1\endcsname}%
				{\hskip#3\relax\H@svsec}{#8}%
			}%
			\@ifundefined{#1smark}{\@gobble}{\csname #1smark\endcsname}{#7}%
			\addcontentsline{toc}{#1}{\protect\numberline{}#8}%
		}%
	}%
	\@xsect{#5}%
}%
\makeatother

\begin{document}

	
\title{Theory of  fractional Chern insulator states in pentalayer graphene moir\'e superlattice }

\author{Zhongqing Guo}
\thanks{These  authors contributed equally.}
\affiliation{School of Physical Science and Technology, ShanghaiTech University, Shanghai 201210, China}

\author{Xin Lu}
\thanks{These  authors contributed equally.}
\affiliation{School of Physical Science and Technology, ShanghaiTech University, Shanghai 201210, China}

\author{Bo Xie}
\affiliation{School of Physical Science and Technology, ShanghaiTech University, Shanghai 201210, China}

\author{Jianpeng Liu}
\email{liujp@shanghaitech.edu.cn}
\affiliation{School of Physical Science and Technology, ShanghaiTech University, Shanghai 201210, China}
\affiliation{ShanghaiTech Laboratory for Topological Physics, ShanghaiTech University, Shanghai 201210, China}
\affiliation{Liaoning Academy of Materials, Shenyang 110167, China}
	
\bibliographystyle{apsrev4-2}

\begin{abstract}

The experimental discoveries of fractional quantum anomalous Hall effects  under zero magnetic fields in both  transition metal dichalcogenide and  pentalayer graphene moir\'e superlattices have aroused significant research interest. In this work, we theoretically study the fractional quantum anomalous Hall states (also known as fractional Chern insulator states) in pentalayer graphene moir\'e superlattice. Starting from the highest energy scale ($\sim\!2\,$eV)  of the continuum model, we first construct a renormalized low-energy model that applies to a lower cutoff $\sim\!0.15\,$eV using renormalization group approach. Then, we  study the ground states of the renormalized low-energy model at filling 1 under Hartree-Fock approximation in the presence of  tunable but self-consistently screened displacement field $D$ with  several experimentally relevant background dielectric constant $\epsilon_r$. Two competing Hartree-Fock states are obtained at filling 1, which give rise to two types of topologically distinct isolated flat bands with Chern number 1 and 0, respectively. We continue to explore the interacting ground states of the two types of isolated flat bands at hole dopings of 1/3, 2/5, 3/5, and 2/3 (corresponding electron fillings of 2/3, 3/5, 2/5, and 1/3 with respect to charge neutrality). Setting $\epsilon_r=5$, our exact-diagonalization calculations suggest that the system stays in fractional Chern insulator (FCI) state at 2/3 electron filling when $0.9\,\textrm{V/nm}\leq\!D\!\leq 0.92\,\textrm{V/nm}$; while no robust FCI state is  obtained at 1/3 electron filling.  We have also obtained  composite-fermion type FCI ground states at 3/5 electron filling within $0.9\,\textrm{V/nm}\leq\! D \!\leq\!0.95\,\textrm{V/nm}$  and $\epsilon_r=5$. 
These numerical results are quantitatively consistent with experimental observations.

\end{abstract} 

\maketitle

\section*{Introduction}
Recent experimental discoveries of fractional quantum anomalous Hall effects in both twisted transitional metal dichalcogenides (TMDs) \cite{fqah-nature23,fqah-prx23,fqah-optics-xu-nature23,fqah-mak-nature23} and graphene moir\'e superlattice \cite{fqah-ju-arxiv23} have aroused significant research interest in the condensed matter community. The fractional quantum anomalous Hall state is also known as fractional Chern insulator (FCI) state \cite{fci-prx11,sheng-fci-nc11,murdy-fci-prl11,wen-kagome-prl11,sarma-flatchern-prl11,cooper-fci-prl09}, which is the zero-field analogue of fractional quantum Hall effect \cite{fqhe-prl82,laughlin-prl83,jain-prl89,moore-read,fqhe-rmp99,qhe-book-2012}.  Different types of FCI states have been theoretically studied in  various lattice models \cite{fqh-boson-prl11,fci-zoo-prb12,liuzhao-fci-prl12,vanderbos-fci-prl12,liu-fci-prb13,fiete-ruby-prb11}. However, the experimental realization of such peculiar states in conventional crystalline solids remain elusive so far. 

In order to realize the FCI state, an isolated topological flat band with nonzero Chern number and desirable quantum geometric properties are required \cite{roy-prb14,claassen-prl15,wangjie-prl21,ledwith-vortex-arxiv22}. The two dimensional moir\'e superlattice provides an ideal platform to achieve such topological flat bands with tunable valley Chern numbers. For example, in ``magic-angle" twisted bilayer graphene (TBG), the lowest two bands (per spin per valley) become ultraflat \cite{macdonald-pnas11}, and are found to be topologically nontrivial with Landau-level like wavefunctions \cite{song-tbg-prl19,origin-magic-angle-prl19,jpliu-prb19,yang-tbg-prx19,po-tbg-prb19,zaletel-tbg-2019,wang2021chiral,song-tbg-ii-prb21,song-heavyfermion-prl22,shi-dai-heavy-prb22}. These topological flat bands are also proposed to exist in moir\'e superlattices consisted of nearly aligned hexagonal boron nitride (hBN) and rhombohedral graphene multilayers \cite{senthil-tbg-prr19,jung-hbn-trilayer-prl19,jung-multilayer-arxiv23}, twisted multilayer graphene \cite{jpliu-prx19,koshino-tdbg-prb19,lee-tdbg-nc19,quansheng-tdbg-nanoletter20,eslam-tmg-prl22,wang-tmg-prl22,xie-atmg-npj22,ashvin-atmg-arxiv21,ma-sb2021,zhang-chiral-nl23}, and moir\'e TMDs \cite{wu-tmd-prl19,macdonald-tmd-ll-arxiv23,lin-prr22,liangfu-fqah-tmd-prb23,nicolas-tmd-chiral-arxiv23,xiao-fqah-arxiv23,zhangyang-fqah-tmd-arxiv23,yao-orbitalchern-arxiv23,law-tmd-prl22} systems. By virtue of the interplay between nontrivial topology and strong $e$-$e$ interaction effects in such topological flat bands, integer quantum anomalous Hall effects \cite{young-tbg-science19,Mak-mote2-wse2-nature2021,young-monobi-nature20} as well as field-driven Chern insulators \cite{chen-hbn-trilayer-nature19,andrei-tbg-chern-arxiv20,yazdani-tbg-chern-arxiv20,pablo-tbg-chern-arxiv21,yacoby-ashvin-chern-tbg-natphys21,efetov-tbg-chern-arxiv20}  have been realized in various types of moir\'e superlattice systems. Nevertheless, the conditions to realize  FCIs are more stringent. A flat band with nonzero Chern number is not sufficient; instead, it is proposed that the Berry-curvature distribution and the quantum metric of the flat bands needs to resemble those of the lowest Landau level as much as possible \cite{roy-prb14,claassen-prl15,wangjie-prl21,ashvin-fci-tbg-prr20,ashvin-fci-tbg-arxiv21} and that the flat band needs to be sufficiently separated from other bands in energy. Due to such rigorous conditions, characteristics of FCI states from compressibility measurements and optical measurements are observed only in a few systems \cite{young-science18,xie-tbgfci-nature21}, including heterostructure of bilayer graphene and hBN \cite{young-science18}  and magic-angle TBG \cite{yacoby-tbg-fqah-arxiv21} under magnetic fields, and twisted TMDs \cite{fqah-optics-xu-nature23,fqah-mak-nature23}.  Direct transport evidence of FCIs with fractionally quantized Hall conductivities under zero magnetic field have been observed just a few months ago, in twisted TMDs and hBN-pentalayer graphene moir\'e superlattices \cite{fqah-nature23,fqah-prx23,fqah-ju-arxiv23}. 

In particular, several integer and fractionally quantized plateaus of anomalous Hall resistance $1$, $3/2$, $5/3$, $7/4$, $9/4$, $5/2$, and $7/3$ (in units of $h/e^2$)  have been observed in hBN-pentalayer graphene  moir\'e superlattice at moir\'e band filling factors of $1$, $2/3$, $3/5$, $4/7$, $4/9$, $2/5$, and $3/7$, respectively \cite{fqah-ju-arxiv23}. Transitions between FCIs and other featureless large-resistance states by varying displacement field ($D$) are also observed \cite{fqah-ju-arxiv23}. All of these intriguing phenomena require microscopic understandings. In this work, combining renormalization group (RG), Hartree-Fock (HF), and exact-diagonalization (ED) methods developed based on the continuum model of hBN-pentalayer graphene moir\'e superlattice, we study the interacting ground state of the system at both integer and fractional filling factors. Including effects of  remote-band renormalization and self-consistent Hartree screening of $D$ fields, within the unrestricted Hartree-Fock framework, we obtain a spin-valley polarized integer Chern-insulator state with Chern number $1$ at filling factor of $1$ for $D\lessapprox 0.1\,$V/nm, which competes with another trivial, Chern-number-zero Hartree-Fock state. Either of the two HF states would give rise to a well isolated occupied flat band right below the chemical potential (of filling 1), with the bandwidth $\sim 5\rm{-}30\,$meV, which has Chern number 1 or 0 for the topologically nontrivial or trivial HF state. Then, we consider hole doping the two types of topologically distinct HF flat bands with respect to filling 1, and study the interacting ground states at hole doping levels of 1/3, 2/5, 3/5, and 2/3 (corresponding to electron filling factors of 2/3, 3/5, 2/5, and 1/3 with respect to charge neutrality) in the parameter space of $D$ and background dielectric constant $\epsilon_r$. When $\epsilon_r=5$, our exact-diagonalization calculations suggest that the system stays in FCI state at 2/3 electron doping when $0.9\,\textrm{V/nm}\leq\!D\!\leq 0.92\,\textrm{V/nm}$, which undergoes phase transitions to other trivial gapped states such as charge density wave (CDW) and/or non-degenerate correlated state by varying $D$ field. In contrast, with the same dielectric constant  no robust  FCI state is seen at 1/3 electron doping in the experimentally relevant  regime of $D$ field ($0.7\,\textrm{V/nm}\leq\!D\!\leq 1.1\,\textrm{V/nm}$). Our ED calculations also suggest that there exists robust  composite-fermion type FCI ground state at 3/5 electron filling within $0.9\,\textrm{V/nm}\leq\!D\!\leq\!0.95\,\textrm{V/nm}$ for $\epsilon_r=5$. All of these numerical results are \textit{quantitatively consistent} with the recent experimental observations \cite{fqah-ju-arxiv23}.

To some extent,  our theory can be considered as being  developed \textit{from first principles}. 
We start from the highest energy scale $E_C \sim 2\,$eV  in the problem and progressively integrate out the high-energy degrees of freedom using RG until a low-energy cutoff $E_C^*\sim 0.15\,$eV is reached. We thus obtain a \textit{renormalized low-energy Hamiltonian} including Coulomb interactions from remote-band electrons to low-energy ones, with self consistently screened $D$ field. Within $E_C^*$, we perform fully unrestricted HF calculations at filling 1 based on which isolated HF flat band right below the chemical potential can be obtained.  Then, we consider hole doping the isolated HF flat band, and study the interacting ground states at various partial hole fillings. Moreover, we also compare the energies of ED ground states by hole doping the topologically distinct  flat bands emerging from two competing HF states, and determine the genuine many-body ground state. It is also worthwhile noting that there is only one free parameter $\epsilon_r$ in the entire theoretical framework. Most saliently, even this single free parameter can be unambiguously determined by comparing theoretical results with experimental ones.
 Eventually, we obtain FCI states at both 2/3 and 3/5 fillings for $\epsilon_r=5$, which emerge in a range of $D$ fields that are quantitatively consistent with experiments.


\begin{figure*}[htb]
    \centering
    \includegraphics[width=5in]{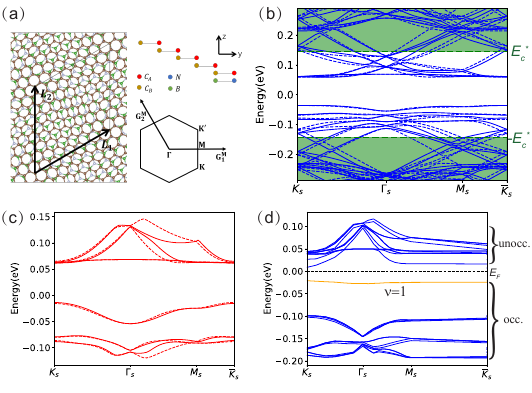}
    \caption{(a) Schematic illustration of the moir\'e superlattice consisted of nearly aligned hBN and pentalayer graphene with rhombohedral stacking. $C_A$, $C_B$ denote carbon atoms in $A$, $B$ sublattices, and $B$, $N$ denote boron and nitrigen atoms. The moir\'e Brillouin zone is plotted. (b)-(d) show continuum-model band structures of hBN-pentalayer graphene moir\'e superlattice with twist angle $0.77\,^{\circ}$ with $D=0.97\,$V/nm and $\epsilon_r=5$. (b) The bare non-interacting band structures, where the green dashed lines mark the low-energy window $E_C^*$ (see text), and the Green area denotes the remote energy bands. The solid and dashed blue lines represent the bands from $K$ and $K'$ valleys, respectively. (c) Low-energy band structures within $E_C^*$  with renormalized continuum model parameters given by Eqs.~\eqref{eq:H-RG}.  The solid and dashed blue lines represent the bands from $K$ and $K'$ valleys, respectively. (d) Hartree-Fock band structures at filling $\nu=1$, where the gray dashed line marks the chemical potential, and the occupied and unoccupied subspaces within $E_C^*$ are marked. The orange line represents the isolated flat band right below the chemical potential of filling 1.}
    \label{fig:1}
\end{figure*}

\section*{Results}

\subsection*{Renormalized continuum model}
\label{sec:continuum}

We consider a moir\'e superlattice consisting of  rhombohedral stacking of pentalayer graphene (PLG) encapsulated by hBN substrates on both sides. On one side of the two interfaces, hBN is nearly aligned with PLG with a twist angle of $0.77\,^{\circ}$, leading to a commensurate moir\'e superlattice with lattice constant $L_s\!=\!109$\,\AA, as schematically shown in Fig.~\ref{fig:1}(a). 
The corresponding moir\'e Brillouin zone is also presented in Fig.~\ref{fig:1}(a). 
The low-energy physics of the system around $K$ or $K'$ valley can be properly described by the following non-interacting continuum model
\begin{equation}
H^{0,\mu}=H_{\textrm{penta}}^{0,\mu}+V_{\rm{hBN}}
\end{equation}
where $H_{\textrm{penta}}^{0,\mu}$ is the non-interacting $\k\cdot\p$ Hamiltonian of rhombohedral PLG of valley $\mu$ ($\mu=\pm$ standing for $K/K'$ valley), and $V_{\rm{hBN}}$ denotes the moir\'e potential exerted on the bottom graphene layer. More specifically, $H_{\textrm{penta}}^{0,\mu}$ is comprised of the intralayer term $h_{\textrm{intra}}^{0,\mu}$  and the interlayer  term $h_{\textrm{inter}}^{0,\mu}$: 
\begin{align}
&H^{0,\mu}_{\rm{penta}}\;\nn
=&\begin{pmatrix}
h_{\rm{intra}}^{0,\mu} &(h_{\textrm{inter}}^{0,\mu})^{\dagger}  & 0 & 0 & 0 \\
h_{\textrm{inter}}^{0,\mu}& h_{\rm{intra}}^{0,\mu} & (h_{\textrm{inter}}^{0,\mu})^{\dagger} & 0 & 0\\
0 & h_{\textrm{inter}}^{0,\mu} & h_{\rm{intra}}^{0,\mu} & (h_{\textrm{inter}}^{0,\mu})^{\dagger}&0\\
 0& 0& h_{\textrm{inter}}^{0,\mu}& h_{\rm{intra}}^{0,\mu} & (h_{\textrm{inter}}^{0,\mu})^{\dagger} \\ 
 0& 0& 0 & h_{\textrm{inter}}^{0,\mu}& h_{\rm{intra}}^{0,\mu} 
\end{pmatrix}\;,
\label{eq:H0}
\end{align}
Here, the intralayer part is just the monolayer $\k\cdot\p$ model 
\begin{equation}
h_{\textrm{intra}}^{0,\mu}=-\hbar v_F^0\,\k\cdot\bm{\sigma}_{\mu}\;,
\end{equation}
 where $\hbar v_F^{0}\approx 5.253\,\rm{eV}\cdot$\AA\ is the non-interacting Fermi velocity of Dirac fermions in monolayer graphene derived from the Slater-Koster tight-binding model \cite{moon-tbg-prb13,supp_info}, $\k$ is the wavevector expanded around the Dirac point from valley $\mu$, and $\bm{\sigma}_{\mu}=(\mu\sigma_x,\sigma_y)$ is the Pauli matrix defined in sublattice space. The interlayer term is expressed as
\begin{equation}
h_{\textrm{inter}}^{0,\mu}=\begin{pmatrix}
\hbar v_{\perp}(\mu k_{x}+ik_y) & t_{\perp}\;\\
\hbar v_{\perp}(\mu k_{x}-ik_y) & \hbar v_{\perp}(\mu k_{x}+ik_y) 
\end{pmatrix}\;,
\end{equation}
where $t_{\perp}=0.34\,$eV, $\hbar v_{\perp}\!=\!0.335$\,eV$\cdot$\AA\ are extracted from the Slater-Koster hopping parameters \cite{moon-tbg-prb13,supp_info}. The moir\'e potential reads \cite{moon-prb14}
\begin{equation}
V_{\rm{hBN}}(\vr)=V_{\rm{eff}}(\vr)+M_{\rm{eff}}(\vr)\sigma_z+ e v_F^0\,\mathbf{A}_{\rm{eff}}(\vr)\cdot\bm{\sigma}_{\mu}
\end{equation}
where $V_{\rm{eff}}$, $M_{\rm{eff}}$, and $\mathbf{A}_{\rm{eff}}$ represent the scalar moir\'e potential, Dirac mass term, and pseudo-vector-potential term, respectively. More details can be found in Ref.~\onlinecite{moon-prb14} and in Supplementary Information \cite{supp_info}. A vertical displacement field would induce an electrostatic potential drop across the pentalayer graphene, which would induce the redistribution of charge density among different layers. This in turn screens the external displacement field. Such screening effects have been calculated self consistently  assuming a background dielectric constant $\epsilon_r$ ($4\leq\epsilon_r\leq 8$), and the details can be found in Ref.~\onlinecite{min-prb23} and Supplementary Information \cite{supp_info}. We note that, for the sake of consistency, the same dielectric constant $\epsilon_r$ has been used in the following RG, HF and ED calculations. 

In Fig.~\ref{fig:1}(b) we show the non-interacting band structures of PLG-hBN moir\'e superlattice with twist angle 0.77\,$^{\circ}$ under (self consistently screened) displacement field $D=0.97\,$V/nm, where the solid and dashed lines represent the bands from the $K$ and $K'$ valleys, respectively. Clearly the valence moir\'e bands are energetically separated from the conduction ones by a sizable gap $\sim\!95\,$meV, while the conduction moir\'e bands are all entangled, which  disfavors FCI state at electron doping at this moment.

The non-interacting continuum model discussed above applies to an energy cutoff $E_C\sim 2\,$eV. Long-range $e$-$e$ interaction effects within this energy window have been  neglected when calculating the band structures of Fig.~\ref{fig:1}(b). However, it is well known that $e$-$e$ Coulomb interactions would significantly renormalize the effective parameters of the non-interacting model, as in the case of monolayer graphene \cite{gonzalez_nuclphysb1993,kotov_rmp2012,castroneto_rmp2009,nair_science2008} and magic-angle TBG \cite{kang-rg-prl20}. For the former, long-range exchange Coulomb interactions would lead to a logarithmic enhancement of Dirac fermion's Fermi velocity; while for the latter, both the Fermi velocity and the intersublattice moir\'e potential are subject to logarithmic corrections under the RG flow such that their ratio remains unchanged.
In order to study the interaction renormalization effects from the high-energy electrons, we need to set up a low-energy  window $\vert E\vert < E_C^*\sim n_{\rm{cut}}\hbar v_F^0/L_s$ ($E_C^*\sim 150\,$meV for $n_{\rm{cut}}=3$)  as marked by the green dashed lines in Fig.~\ref{fig:1}(b).  Within $E_C^{*}$,  the $e$-$e$ Coulomb interactions can no longer be treated by perturbations. Here $2 n_{\rm{cut}}$ may be interpreted as the number of moir\'e bands (per spin per valley) within the low-energy window $E_C^*$. Outside the low-energy window marked by $E_C^*$, the effects of long-range Coulomb interactions are treated by perturbative RG approach \cite{kang-rg-prl20,lu-nc23}, which yields a set of renormalized continuum model parameters as given by Eq.~\eqref{eq:H-RG} in Methods~\ref{methods:rg}.

In Fig.~\ref{fig:1}(c) we present the \textit{renormalized}  band structures within the $n_{\textrm{cut}}=3$ low-energy window, with RG-corrected continuum-model parameters given by Eqs.~\eqref{eq:H-RG}. We see that the lowest conduction moir\'e band is pushed down in energy due to the renormalization effects from the remote bands, and the energy overlap with the higher conduction bands is  reduced compared to the bare bands in Fig.~\ref{fig:1}(b).

\begin{figure}[htb]
    \centering
    \includegraphics[width=3.5in]{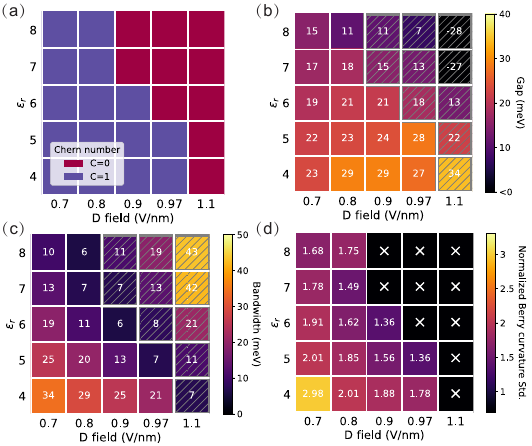}
    \caption{Hartree-Fock phase diagram of hBN-pentalayer graphene moir\'e superlattice at filling factor $\nu=1$. (a) Chern number of the Hartree-Fock ground state, (b) gap of the Hartree-Fock ground state, (c) bandwidth of the isolated flat band right below the chemical potential (for  $\nu=1$), and (d) normalized Berry-curvature standard deviation of the flat Chern band right below chemical potential emerging from the topological HF ground state. The shaded regions in (b) and (c) mark the regions with topologically trivial ground states at $\nu=1$.}
    \label{fig:2}
\end{figure}

\subsection*{Hartree-Fock phase diagram}
\label{sec:hf}
We continue to discuss the $e$-$e$ interaction effects within the $n_{\textrm{cut}}=3$ low-energy window with renormalized continuum model.  Here we consider the dominant intrvalley, long-range Coulomb interaction in 2D moir\'e superlattice systems, where a Thomas-Fermi type of screening form is adopted. Since the thickness of PLG is 1.34\,nm, much thicker than TBG, we also consider the layer dependence of Coulomb interactions. More details can be found in Eqs.~\eqref{eq:coulomb}-\eqref{eq:Vll} in Methods. Then, we project the Coulomb interaction onto the \textit{renormalized wavefunctions} within the low-energy subspace. Specific expressions of the interaction form factors from such low-energy projection can be found in Eq.~\eqref{eq:Hintra-band}-\eqref{eq:form-factor} in Methods~\ref{methods:hf}.

We note that the Coulomb interaction effects between the remote-band electrons and the low-energy electrons have already been taken into account by perturbative RG as discussed above, which precisely lead to the renormalization of the continuum model parameters as given in Eqs.~\eqref{eq:H-RG} in Methods~\ref{methods:rg}. In other words, the RG-rescaled low-energy model describes the behavior of a single low-energy electron within $E_C^*$ which interacts with a background of filled remote-band electrons.
However, the $e$-$e$ interactions \textit{within the low-energy window} have not been included yet. This means that we can just normal order the $e$-$e$ interactions with respect to the \textit{vacuum of the truncated low-energy Hilbert space} within $E_C^*$ ($n_{\textrm{cut}}=3$), which amounts to projecting Eq.~\eqref{eq:coulomb} onto the truncated low-energy Hilbert space.

With the above setup, we study the HF ground states at filling 1. In Fig.~\ref{fig:1}(d) we show  the ground-state Hartree-Fock band structures at filling 1, with $D=0.97\,$V/nm and $\epsilon_r=5$. We see that the highest occupied band marked in orange now is energetically separated from the other moir\'e conduction bands. Moreover, this isolated flat band has a Chern number of $1$, desirable to realize FCI state at partial fillings of the flat bands. We also note that the HF gap at filling $1$ divides the Hilbert space within the $E_C^*$ low-energy window into the occupied and unoccupied subspaces, as marked in Fig.~\ref{fig:1}(d). Such a partition would allow us to consider the electron-doping problem with respect to charge neutrality as the hole-doping one with respect to the gap at filling 1, which will be elaborated in more details in the following section.

We continue to explore the HF phase diagram at filling 1 in the $(D, \epsilon_r)$ parameter space.  As shown in Fig.~\ref{fig:2}(a), the system stays in a Chern-number 1 ground state in a large range of parameter space, which gives rise to a  Chern-number 1  flat band that is well separated from both valence bands (due to large $D$ field) and the higher conduction bands (see Fig.~\ref{fig:1}(d)). In Fig.~\ref{fig:2}(b), we present the gap of the HF ground state at filling 1 in the $(D, \epsilon_r)$ parameter space, which varies from 7 to 29\,meV, and becomes negative in the upper right corner, i.e., metallic state with trivial topological properties. In Fig.~\ref{fig:2}(c) we show the Hartree-Fock bandwidth of the lowest conduction moir\'e band, which varies from $6\,$meV to $34\,$meV (when the HF ground state has positive gap at filling 1). We have also calculated the normalized Berry-curvature standard deviation of the isolated flat Chern band emerging from the topologically nontrivial HF ground state, as shown in Fig.~\ref{fig:2}(d).  It has been argued that smaller Berry-curvature standard deviation would imply more resemblance between the flat Chern band and the lowest Landau level, which may favor FCI ground state at partial fillings. Combining the Chern number, energy gap, bandwidth, and Berry-curvature standard deviation obtained from HF calculations, we conclude that 0.8\,V/nm$< D < 1\,$V/nm and $5 \leq \epsilon_r \leq 7$ may be the best  regime to realize FCI at partial fillings.

\begin{figure*}[htb]
    \centering
    \includegraphics[width=5in]{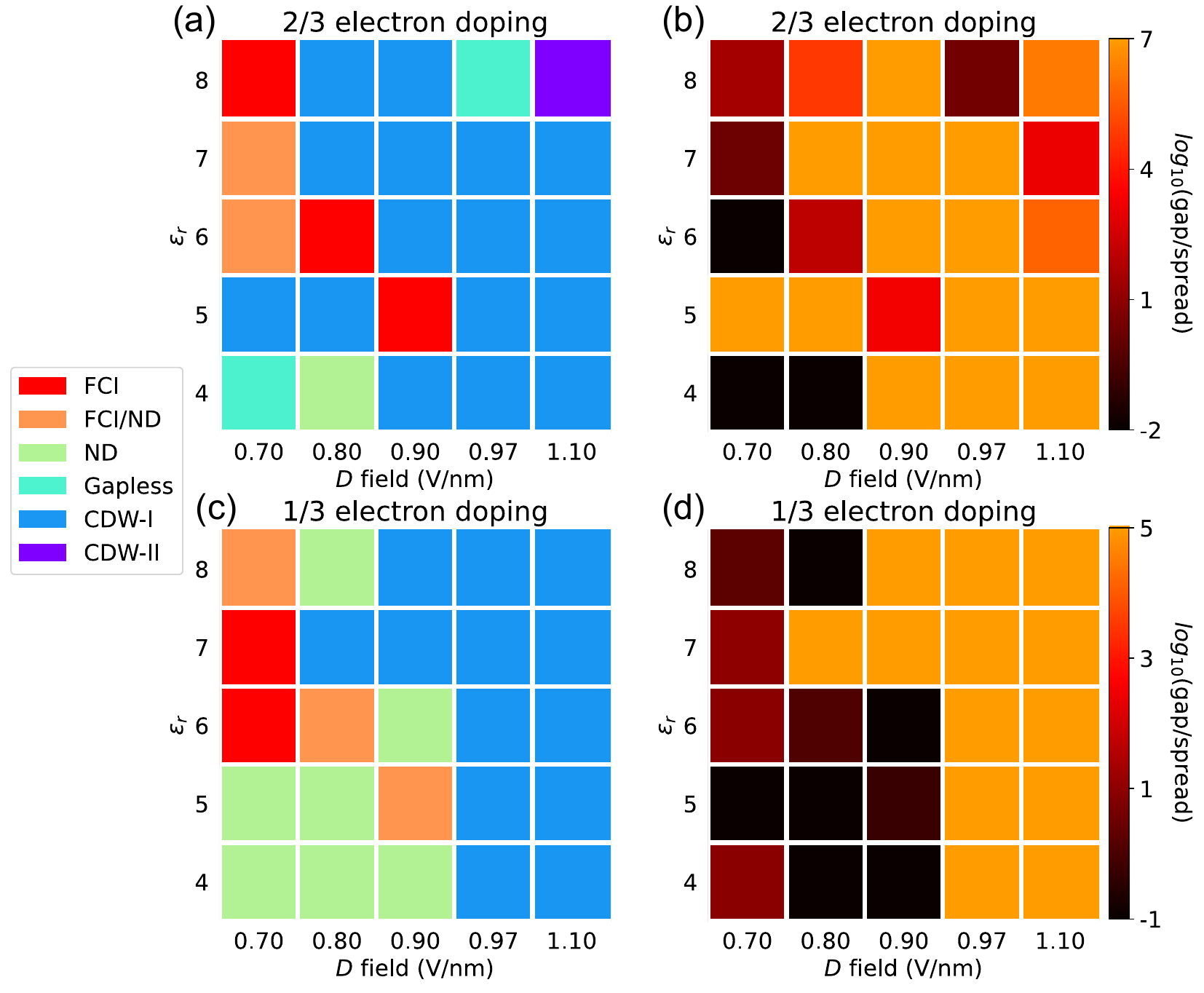}
    \caption{Phase diagram at 2/3 and 1/3 electron fillings obtained from exact-diagonalization calculations. (a) Phase diagram at 2/3 electron filling, (b) the ``gap/spread" (cf. the subsection on FCI) at 2/3 electron filling plotted on a logarithmic scale, (c) phase diagram at 1/3 electron filling, and (d) the ``gap/spread" (cf. the subsection on FCI) at 1/3 electron filling plotted on a logarithmic scale. In (a) and (c), ``FCI" stands for fractional Chern insulator, ``ND" stands for non-degenerate state, ``CDW" stands for charge density wave, while ``FCI/ND" denotes cross-over state between FCI and ND.}
    \label{fig:3}
\end{figure*}

\subsection*{Fractional Chern insulators}
\label{sec:ed}

We further consider hole doping the isolated Hartree-Fock flat  band (see Fig.~\ref{fig:1}(d)), and explore the interacting ground states of this flat and at hole doping levels of 1/3, 2/5, 3/5, and 2/3 (corresponding to electron dopings of 2/3, 3/5, 2/5, 1/3) in the parameter space of $D$ and  $\epsilon_r$. As discussed above, there are two types of topologically distinct  isolated flat bands which emerge from two competing HF ground states: the Chern-number 1 isolated flat band from the topological HF ground state, and the Chern-number 0  flat band from the topologically trivial HF ground state.  We consider hole doping the two types of topologically distinct flat bands emerging from the two types of HF states, and determine the genuine many-body ground state.  Such an RG+HF+ED approach is explained in greater details in Methods~\ref{methods:workfolow}.

\begin{figure*}[htb]
    \centering
    \includegraphics[width=7in]{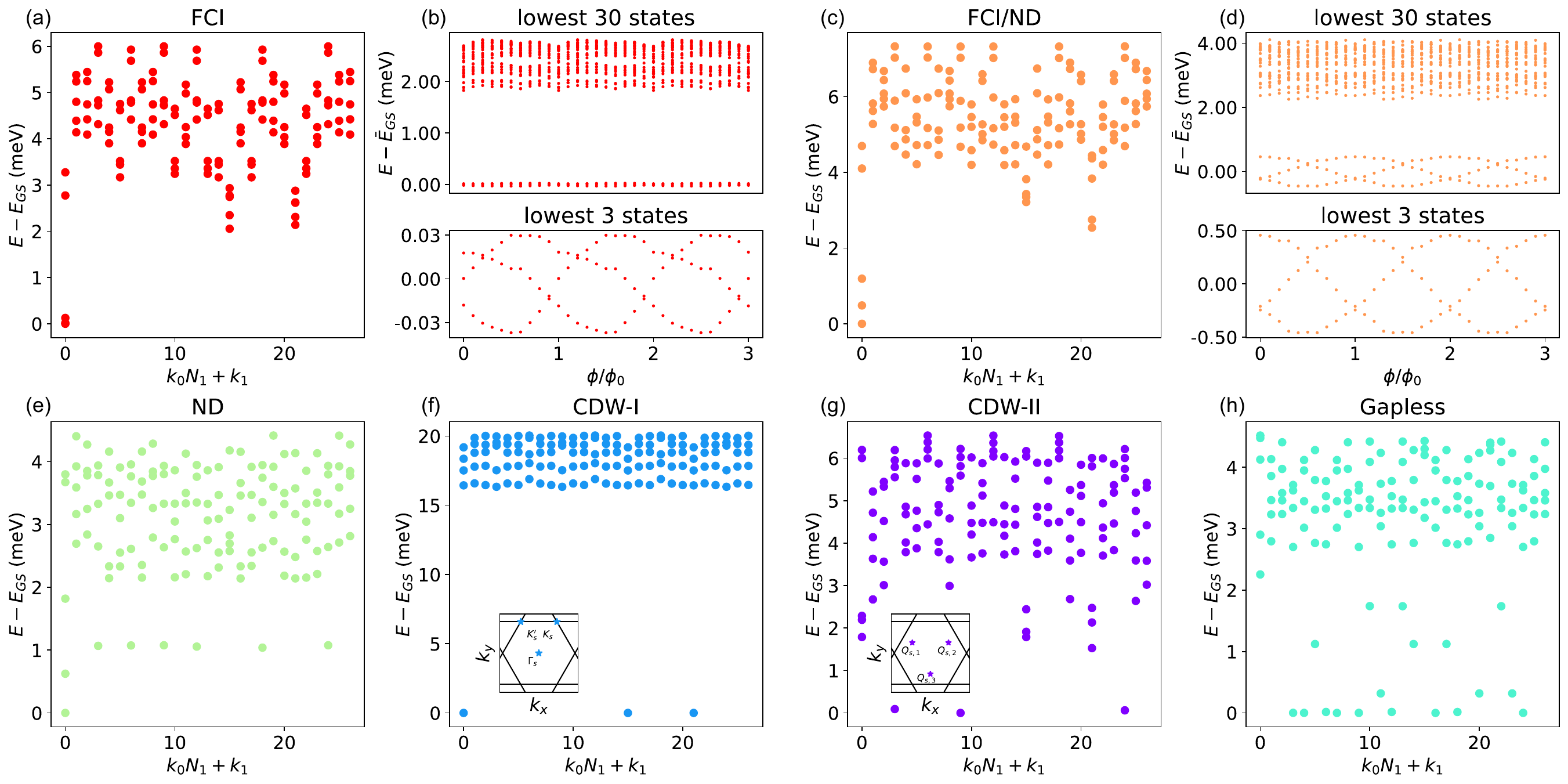}
    \caption{(a) Many-body band structures of Laughlin-type fractional Chern insulator state at 2/3 electron doping. (b) The spectral-flow behavior of  Laughlin-type FCI state, where the upper panel shows the spectral flow of the lowest 30 states, and the lower panel zooms in the spectral flow within the (nearly) threefold degenerate ground-state manifold. (c) Many-body band structures of a cross-over state between  fractional Chern insulator and non-degenerate state (FCI/ND), which shows three low-lying states separated by a gap from other excited states. But the gap is comparable to the energy  spread of the three low-lying states. (d) Spectral-flow behavior of the cross-over FCI/ND state. (e) Many-body band structures of non-degenerate state. Many-body band structures of (f) CDW1 state, and (g) CDW2 state, where their insets mark the ground-state crystalline momenta within moir\'e Brillouin zone. (h) Many-body bandstructures of the gapless phase.}
    \label{fig:4}
\end{figure*}

Practically, we project the long-range $e$-$e$ interactions (Eq.~\eqref{eq:coulomb}) onto the wavefunctions of the isolated flat band emerging from the corresponding HF ground state at filling 1 ($\nu=1$), and calculate the interacting ground states using ED under different hole doping levels. Then, we compare their total energy to determine the genuine many-body ground state at given sets of parameters and doping levels.

\subsubsection*{1/3 and 2/3 fillings}
\label{sec:ed-laughlin}

With the workflow sketched in Methods~\ref{methods:workfolow}, in the end we only need to perform ED calculations within an isolated moir\'e band from one spin one valley which is right below the chemical potential of  HF state at filling 1. However, we need to perform ED calculations at partial hole fillings with respect to two types of topologically distinct HF states at $\nu=1$ (see Methods~\ref{methods:workfolow}). For the topological nontrivial HF state, an isolated flat band with Chern number 1 is generated; while for the topological trivial case, a flat band with zero Chern number is obtained. These two HF states are  competing with each other with energy difference $\sim 0.01\rm{-}1\,$meV. 
Therefore, one needs to hole dope both types of isolated flat bands, and compare the energies of the many-body ground states in the two cases.  One needs to be careful that, when performing the particle-hole transformation: $\hat{c}_{\k}\to \hat{h}_{-\k}^{\dagger}$ (where $\hat{c}_{\k}$  refers to electron annhiliation operator and $\hat{h}_{-\k}^{\dagger}$ refers to hole creation operator), the two-particle interaction Hamiltonian Eq.~(\ref{eq:coulomb}) is unchanged. However, the single-particle Hamiltonian is changed by a constant due to normal ordering of the hole operators, i.e. 
\begin{equation}
\hat{c}_{\k}^{\dagger}\,\hat{c}_{\k}\to\hat{h}_{-\k}\,\hat{h}_{-\k}^{\dagger}=1-\hat{h}_{-\k}^{\dagger}\,\hat{h}_{-\k}\;.
\end{equation}
Thus, the effective single-particle Hamiltonian of the isolated flat band generated from $\nu=1$ HF state is also subjected to the above transformation:
\begin{equation}
\hat{H}^{\textrm{HF}}_{\nu=1}=\sum_{\k} E^{\textrm{HF}}_{\k}\,(1-\hat{h}_{-\k}^{\dagger}\,\hat{h}_{-\k})
\label{eq:hf-ham}
\end{equation}
where $E^{\textrm{HF}}_{\k}$ is the valley-spin polarized HF energy dispersion of the isolated flat band right below the chemical potential of the $\nu=1$ HF state, $\k$ is the wavevector  within the moir\'e Brillouin zone.  We note that in order to compare the energies of many-body states at partial hole fillings of two different HF flat bands, one has to include the constant HF band energy sum in Eq.~(\ref{eq:hf-ham}), which may be interpreted as ``effective vacuum energy" of the two different HF vacua in our calculations.

Keeping this subtlety  in mind,  at 2/3 electron doping (1/3 hole doping), we perform ED calculations on a 27-site cluster with periodic boundary condition (see Methods~\ref{methods:ed}), and dope 9 holes into the two types of isolated flat band. The phase diagram at 2/3 electron filling is shown in Fig.~\ref{fig:3}(a).
We see that Laughlin-type FCI state (marked in red) is fiercely competing with two types of CDW states (marked by ``CDW1" in blue and ``CDW2" in purple) and non-degenerate state  (``ND" marked in light green) . Both the FCI state and the CDW states are characterized by a (nearly) three-fold degenerate ground-state manifold, which is separated from other excited states by a finite energy gap on the order of $1\textrm{-}10\,$meV. The ND, however, is characterized by a non-degenerate ground state separated from excited states by a gap on the order of 1\,meV. We note that there are three blocks in the $(D,\epsilon_r)$ space exhibiting FCI ground state at 2/3 filling, i.e., ($\epsilon_r=5$, $D= 0.9\,$V/nm), ($\epsilon_r=6$, $D= 0.8\,$V/nm), and ($\epsilon_r=8$, $D= 0.7\,$V/nm), as indicated by red color in Fig.~\ref{fig:3}(a). Moreover, it is interesting to note that there is also a finite region in the phase diagram of Fig.~\ref{fig:3}(a) in which the system seems to undergo a cross-over from a non-degenerate state to FCI. Such a cross-over region is marked as ``FCI/ND" by orange color in Fig.~\ref{fig:3}(a). The presence of such a cross-over regime may be attributed to finite size effects. In the thermodynamics limit, the cross-over regime is expected to shrink to a line of phase boundary. 

In order to better capture the characters of the different types of many-body states shown in the phase diagram at 2/3 electron filling, we extract the energy gap between the fourth and third eigenstates, and  divide it by the energy spread of the first three lowest-energy eigenstates. Such ``gap/spread" function is plotted in ($D$,$\epsilon_r$) space on a logarithmic scale, as shown in Fig.~\ref{fig:3}(b). If the log of this quantity is large, it means that there are three low-energy states which are separated from the fourth state by a notable gap, which implies either FCI or CDW state. Otherwise, it could either be a gapless state or a ND state. Practically, in our calculations a  threefold quasi-degenerate many-body state is identified as a FCI if it is topologically nontrivial (to be discussed in the following) and meanwhile the log of its gap/spread is greater than 0.5. We find that the log of this gap/spread quantity is largest ($\sim 4$) at $D=0.9\,$V/nm and $\epsilon_r=5$, which means that the FCI state around this point is most stable.
More detailed characterizations of the different types of correlated states are to be discussed later.

In Fig.~\ref{fig:3}(c) we present the phase diagram at 1/3 electron filling. FCI ground state emerges at ($D=0.7\,$V/nm, $\epsilon_r=6$) and ($D=0.7\,$V/nm, $\epsilon_r=7$). However, experimentally no signature of fractional quantum anomalous Hall effect has been observed at 1/3 electron filling, while it is observed for $0.95\,\rm{V/nm}\lessapprox D\lessapprox 0.965\,$V/nm at 2/3 filling \cite{fqah-ju-arxiv23}. Comparing our theoretical results with experimental observations at both 2/3 and 1/3 fillings, we find that the background dielectric constant can be unambiguously determined as $\epsilon_r=5$, which lead to results that are consistent with experiments at both fillings.  In Sec.~\ref{sec:ed-composite} we will show that using the same dielectric constant $\epsilon_r=5$, we are also able to obtain numerical results that are quantitatively consistent with experimental observations at 3/5 and 2/5 fillings.  In Fig.~\ref{fig:3}(d) we present the ``gap/spread" (as defined above) at 1/3 filling on a logarithmic scale. Clearly, the CDW region is marked by large and positive ``log$_{10}$(gap/spread)" value $\sim 5$; while the FCI region  has a small  value $\sim 0.5\rm{-}1$, much smaller than that of the FCI state at 2/3 filling ($\sim 2 \rm{-}4$).  This indicates that the FCI state obtained at 1/3 filling for $D=0.7\,$V/nm and $\epsilon_r=6, 7$ is not as robust as that at 2/3 filling. Again, here we identify a threefold quasi-degenerate state as FCI if it is topologically nontrivial (to be discussed later) and if the log of its gap/spread quantity is greater than 0.5.

\begin{figure*}[htb]
    \centering
    \includegraphics[width=5in]{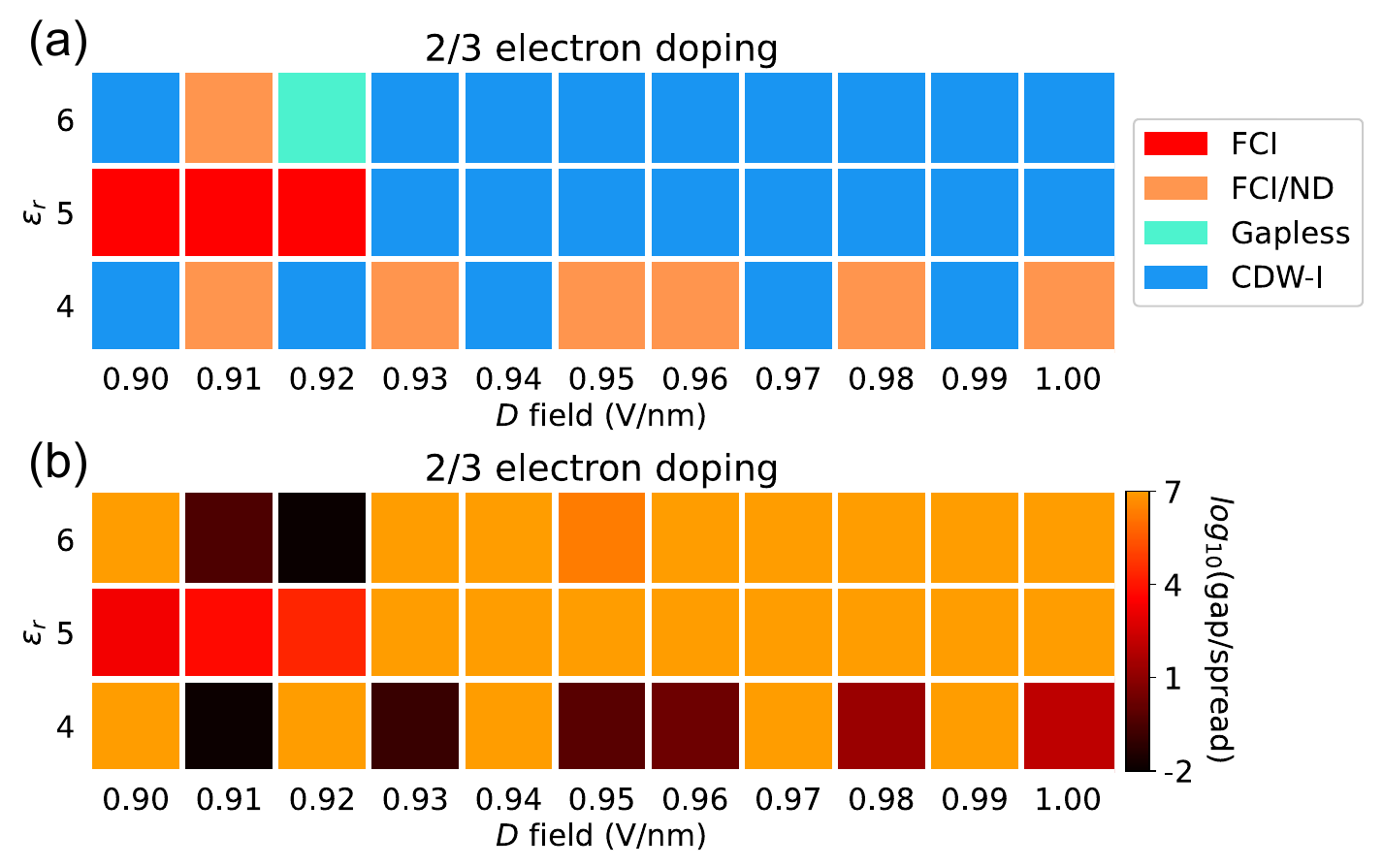}
    \caption{(a) Finer phase diagrams in a smaller range of parameters $0.9\textrm{V/nm}\leq D\leq 1.0\,$V/nm and $4\leq\epsilon_r\leq 6$ at 2/3 electron filling. (b) The corresponding ``gap/spread" (cf. the subsection on FCI) at 2/3 electron filling plotted on a logarithmic scale.}
    \label{fig:5}
\end{figure*} 

As mentioned above, both FCI and CDW states are characterized by (nearly) threefold degenerate ground states and finite energy gaps. One has to look at more detailed properties  to distinguish them. To this end, in Fig.~\ref{fig:4} we show the total energy \textit{vs.} total crystalline momenta dispersions (dubbed as ``many-body band structures" hereafter) of various types of correlated states. In particular, as shown in Fig.~\ref{fig:4}(a), the many-body band structures of FCI exhibits three (nearly) degenerate ground states  in the zero momentum sector, which  satisfies the ``generalized Pauli principle" \cite{fci-prx11} for 27 sites at 1/3 hole doping. Upon the adiabatic insertion of magnetic flux $\phi$, the three degenerate ground states of FCI would flow into each other, and come back to the original configuration when $\phi=3\phi_0$ ($\phi_0=h/e$ is the flux quantum). With Landau gauge, this amounts to change the crystalline momentum along one direction by  $\delta k_1=(\phi/\phi_0)/N_1$, where $N_i$ ($i=1, 2$) is the number of lattice sites along the two primitive reciprocal vector directions, $k_1$ is the corresponding reduced crystalline momentum along the first reciprocal vector direction. 
 Such  spectral-flow behavior of the FCI state at 2/3 electron filling is clearly shown in the lower panel of Fig.~\ref{fig:4}(b). For the purpose of spectral-flow calculation at 2/3 electron filling, we take a $4\times 6$ site with 8 holes (see Methods~\ref{methods:ed}), which also yields the  FCI state same as  the case of 27 sites and 9 holes.

In Fig.~\ref{fig:4}(c) we present the typical many-body band structures of the FCI/ND cross-over states at 2/3 electron filling. We see that, although there are three low-lying states which are separated from the fourth state by a gap $\sim 1\,$meV, the energy spread of the three low-lying state is also about $1\,$meV.  In Fig.~\ref{fig:4}(d) we show the energy of the FCI/ND cross-over state as a function of inserted fluxes (for 24-site system with 8 holes). There is similar spectral-flow behavior as FCI. Nevertheless, the large energy spread of the three low-lying states and small excitation gap hinder us to call it FCI.  In Fig.~\ref{fig:4}(e) we present the many-body band structures of ND state, which clearly shows a non-degenerate ground state. 

In Fig.~\ref{fig:4}(f) and (g), we  show  typical many-body band structures of the two types of CDW states. Both types of  CDW states are three-fold degenerate and are gapped from the excited states. However, for the first type of CDW (CDW1), the total crystalline momenta of the three ground states occur at $\Gamma_s$, $K_s$ and $K_s'$ as marked in the inset of Fig.~\ref{fig:4}(f), which are the characteristic wavevectors of CDW states  at 1/3 and 2/3 fillings preserving threefold rotational symmetry.   While for the second type of CDW (CDW2), the total reduced crystalline momenta of the three degenerate ground states are (0,1/3), (1/3,0) and (2/3,2/3), as marked in the inset of Fig.~\ref{fig:4}(g). The CDW2 state is a stripe state that breaks the threefold rotational symmetry.  In Fig.~\ref{fig:4}(h),  typical many-body band structures for the gapless phase are also presented.


\begin{figure*}[htb]
    \centering
    \includegraphics[width=5in]{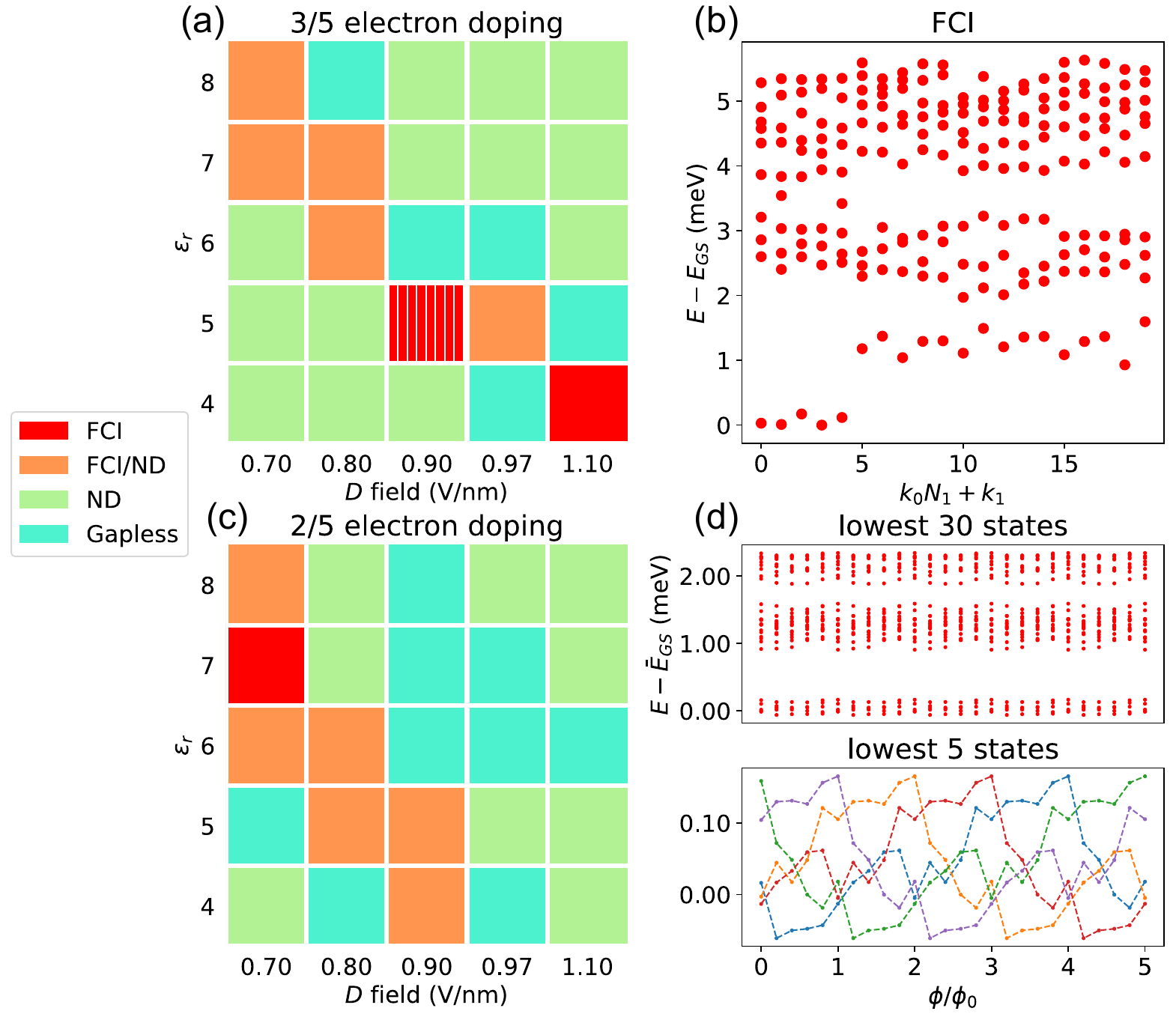}
    \caption{(a) Phase diagram and many-body energy spectra at 3/5 and 2/5 electron fillings obtained from exact-diagonalization calculations. (a) Phase diagram at 3/5 electron filling. (b) Many-body band structures of the composite-fermion type FCI state at 3/5 electron filling at $D=0.94\,$V/nm and $\epsilon_r=5$. (c) Phase diagram at 2/5 electron filling. (d) Spectral-flow behavior of the composite-fermion type FCI state at 3/5 electron filling, where the upper panel shows the spectral flow of lowest 30 states, while the lower panel zooms in the spectral flow within the (nearly) fivefold degenerate ground-state manifold. The fine white mesh within the red FCI block in (a) means that we have performed ED calculations in a finer range of $D$ field at 3/5 electron filling for $0.9\,\rm{V/nm}\!\leq\!D\!\leq\!0.96\,$V/nm (with interval of 0.1\,V/nm) at $\epsilon_r=5$.}
    \label{fig:6}
\end{figure*}

We continue to search for the phase diagram at 2/3 electron filling (1/3 hole filling) within a smaller parameter space $0.9\,\textrm{V/nm}\leq D\leq 1.0\,\textrm{V/nm}$ and $4\leq\epsilon_r\leq 6$, and we compare the ground-state energies by 1/3 hole doping two topologically distinct flat bands derived from two competing HF states at $\nu=1$. 
This finer  phase diagram is presented in Fig.~\ref{fig:5}(a). 
When $0.9\,\rm{V/nm}\leq D\leq 0.92\,$V/nm and $
epsilon_r=5$, FCI is the ground state at 2/3 electron filling. In Fig.~\ref{fig:5}(b) we present the ``gap/spread" on a logarithmic scale, which show peak values in the FCI and CDW states.  It is worthwhile noting that experimentally fractional quantum anomalous Hall effect shows up at 2/3 filling for $0.95\,\rm{V/nm}\lessapprox D\lessapprox 0.965\,$V/nm \cite{fqah-ju-arxiv23}, which is quantitatively consistent with the fine phase diagram in Fig.~\ref{fig:5}(a).

\subsubsection*{2/5 and 3/5 fillings}
\label{sec:ed-composite}

We further consider the situations of 2/5 and 3/5 hole dopings of the isolated flat band with respect to $\nu=1$, corresponding to 3/5 and 2/5 electron dopings with respect to charge neutrality point. We perform ED calculations on a  20-site ($4\times 5$, see Methods~\ref{methods:ed}) cluster under periodic boundary conditions  with 8 (2/5 hole doping) and 12 (3/5 hole doping) doped holes. Fractional quantum anomalous Hall effect  has been observed at both of these fillings \cite{fqah-ju-arxiv23}, implying the presence of composite-fermion  type FCI states \cite{liu-fci-prb13} at these fillings. In Fig.~\ref{fig:6} we present the phase diagram at 3/5 electron filling, which is obtained by 2/5 hole doping the two types of isolated flat bands emerging from the two competing  HF states  at $\nu=1$. We see that when $D=0.9\,$V/nm and $\epsilon_r = 5$, and $D = 1.1\,$V/nm and $\epsilon_r= 4$, the system may stay in composite-fermion type FCI state marked by red blocks in Fig.~\ref{fig:6}(a). The fine white mesh within the red FCI block at ($D = 0.9\,$V/nm, $\epsilon_r = 5$) in Fig.~\ref{fig:6}(a) means that within this red block, we have performed ED calculations in a finer range of $D$ field at 2/5 hole filling (3/5 electron filling) of the two distinct flat bands emerging from the two competing HF states at filling 1. We find that the system stays in composite-fermion FCI state for $D=0.9\,\rm{V/nm}$ and for $D=0.93\,\rm{V/nm}\!\leq\!D\!\leq\!0.95\,$V/nm (with interval of 0.1\,V/nm) with $\epsilon_r=5$. 
Such a state is characterized by five-fold (nearly) degenerate ground state and a sizable gap $\sim\!1\,$meV from the excited states, as shown in Fig.~\ref{fig:6}(b). Moreover, upon the adiabatic flux insertion, these five nearly degenerate ground states would interchange with each other and return to the original configurations when $\phi=5\phi_0$, as clearly shown in Fig.~\ref{fig:6}(d). 
This further confirms the topological nature of such a degenerate many-body state. There are also some other correlated ground states in the $(D, \epsilon_r)$ phase diagram at 3/5 electron filling, such as ND, gapless, and  FCI/ND cross-over states, as shown in Fig.~\ref{fig:6}(a). 

In contrast, at 2/5 electron filling, according to our ED calculation on a relatively rough mesh in the $(D,\epsilon_r)$ space as shown in Fig.~\ref{fig:6}(c), FCI state only shows up at $D=0.7\,$V/nm and $\epsilon_r=7$. And this FCI state is not very robust, with the log of the gap/spread $\sim 1.18$.  When $\epsilon_r=5$, FCI/ND cross-over state emerge for $D=0.8, 0.9\,$V/nm at 2/5 electron filling. 
We note that, in experiments fractional quantum anomalous Hall effect has been observed over a relatively large range of $D$ field  at 3/5 filling ($0.91\,\rm{V/nm}\lessapprox D\lessapprox 0.95\,$V/nm), but only observed within a small range of $D$ field at 2/5 filling ($0.92\,\rm{V/nm}\lessapprox D\lessapprox 0.93\,$V/nm) \cite{fqah-ju-arxiv23}.
 Therefore, our numerical results with $\epsilon_r=5$ at 2/5 and 3/5 fillings are also fully consistent with experimental observations. 

\section*{Summary}
\label{sec:summary}

\begin{table*}
	\caption{The range of displacement field ($D$, in units of V/nm) within which the fractional quantum anomalous Hall effect shows up. The second row shows results from our theoretical calculations with dielectric cosntant $\epsilon_r=5$, and the third row shows the results from experimental measurements \cite{fqah-ju-arxiv23}.}
	\label{table:range}
	\centering
	\begin{tabular}{c|c|c|c|c}
		\hline
Electron filling factor  & 2/3  & 1/3  & 3/5  & 2/5    \\		
		\hline
Theory    &   $0.9\leq D\leq 0.92$  & None & $0.9\leq D\leq 0.95$ &  None \\
        \hline
Experiment &  $0.95\lessapprox D \lessapprox 0.965$	 &  None & $0.91\lessapprox D \lessapprox 0.95$   &  $0.92\lessapprox D \lessapprox 0.93$  \\ 	        
\hline
	\end{tabular}
\end{table*}

To summarize, in this work, combining renormalization group, Hartree-Fock, and exact-diagonalization methods developed based on continuum model, we have theoretically studied the correlated and topological ground states at both integer and fractional electron fillings of nearly aligned hBN-pentalayer graphene moir\'e superlattice. Including renormalization effects due to Coulomb interactions with  remote-band electrons,  we manage to construct an effective continuum model with \textit{renormalized parameters} that applies to a lower energy cutoff $E_C^{*}\sim 150\,$meV. Then, we project $e$-$e$ Coulomb interactions to the renormalized non-interacting wavefunctions  and perform unrestricted Hartree-Fock calculations within $E_C^*$. We have obtained a spin-valley polarized integer Chern-insulator state with Chern number $1$ at filling factor of $1$ for $D\lessapprox 0.1\,$V/nm, which competes with another trivial, Chern-number-zero Hartree-Fock state. Both of the two HF states would give rise to  well isolated occupied flat bands right below the chemical potential, with the bandwidth $\sim 5\rm{-}30\,$meV. Such emerging HF flat bands have Chern number  1 and 0 for the topologically nontrivial or trivial HF states, respectively. Then, we consider hole doping the two types of topologically distinct HF flat bands, and study the interacting ground states at hole doping levels of 1/3, 2/5, 3/5, and 2/3 (corresponding to electron dopings of 2/3, 3/5, 2/5, 1/3 with respect to charge neutrality) in the parameter space of $D$ and background dielectric constant $\epsilon_r$. Comparing our theoretical phase diagrams with experimental results at various fillings, it is concluded that the dielectric constant can be determined as $\epsilon_r=5$, which give rise to theoretical results that are quantitatively consistent with experiments at most of the filling factors. Specifically, our exact-diagonalization calculations suggest that the system stays in FCI state at 2/3 electron doping when $0.9\,\textrm{V/nm}\leq\!D\!\leq 0.92\,\textrm{V/nm}$ and $\epsilon_r= 5$, which undergoes phase transitions to other trivial gapped states such as charge density wave (CDW) and/or non-degenerate correlated state by varying $D$ field. In contrast,  no robust  FCI ground state is seen at 1/3 electron doping in the experimentally relevant  regime of $D$ field ($0.7\,\textrm{V/nm}\leq\!D\!\leq 1.1\,\textrm{V/nm}$) with $\epsilon_r=5$. Our ED calculations also suggest that there exists robust  composite-fermion type FCI ground state at 3/5 electron filling within $0.9\,\textrm{V/nm}\!\leq\!D\!\leq\!0.95\,\textrm{V/nm}$ and $\epsilon_r= 5$. These numerical results are fully consistent with the recent experimental measurements \cite{fqah-ju-arxiv23}.  In Table.~\ref{table:range} we compare the theoretically calculated range of $D$ fields (within which the FCI states show up) with the corresponding values extracted from transport measurements \cite{fqah-ju-arxiv23}, which exhibit remarkble consistency with each other at most of the filling factors.

Combining molecular dynamics simulations  and atomistic tight-binding modelling, we have also studied effects of lattice relaxations on hBN-pentalayer moir\'e superlattice \cite{supp_info}. However, we find that the effects of lattice relaxations are so weak such that the  non-interacting band structures can be barely changed, especially under large displacement fields \cite{supp_info}. Thus, lattice relaxations may not play an important role in determining the correlated and topological states in this system.

Our work provides a microscopic  theory for the fractional quantum anomalous Hall effect and other correlated states observed in hBN-pentalayer graphene moir\'e superlattice. Our theory may be considered as being developed ``from first principles" in the sense that we have derived the low-energy theory by progressively integrating out the high-energy degrees of freedom, and that there is only one free parameter ($\epsilon_r$) in the entire theoretical framework. Even this free parameter can be unambiguously determined after comparing theoretical results with experimental ones.
 Most of the  results obtained in this work shows quantitative consistency with the recent experimental measurements \cite{fqah-ju-arxiv23}. We have also predicted the presence of various types of ``featureless" correlated states including the non-degenerate correlated state and two different types of gapped CDW states, which await experimental verifications.  The methodology developed in this work can be readily applied to various other moir\'e superlattices to explore potentially richer correlated and topological physics.

\section*{Methods}

\subsection*{Remote-band renormalization effects}
\label{methods:rg}
In order to study the interaction renormalization effects from the high-energy electrons, we set up a low-energy  window $\vert E\vert < E_C^*\sim n_{\rm{cut}}\hbar v_F^0/L_s$ ($E_C^*\sim 150\,$meV for $n_{\rm{cut}}=3$)  as marked by the green dashed lines in Fig.~\ref{fig:1}(b).  Within $E_C^{*}$,  the $e$-$e$ Coulomb interactions can no longer be treated by perturbations. 
Outside the low-energy window marked by $E_C^*$, the effects of long-range Coulomb interactions are treated by perturbative RG approach \cite{kang-rg-prl20,lu-nc23}, which yields the following renormalized continuum-model parameters:
\begin{subequations}
\begin{align}
 v_F(E_C^*)&=v_F^0 \left(1+\frac{\alpha_0}{4\epsilon_r}\log{\frac{E_C}{E_C^*}} \right)\; \label{eq:H-RG-a}\\
 e v_F\mathbf{A}_{\textrm{eff}}(E_C^*)&=e v_F\mathbf{A}_{\textrm{eff}}\,\left(1+\frac{\alpha_0}{4\epsilon_r}\log{\frac{E_C}{E_C^*}}\right)\; \label{eq:H-RG-b}\\
 t_{\perp}(E_C^*)&=t_{\perp}\,\left(1+\frac{\alpha_0}{4\epsilon_r}\log{\frac{E_C}{E_C^*}}\right)\; \label{eq:H-RG-c}\\
 M_{\textrm{eff}}(E_C^*)&=M_{\textrm{eff}}\,\left(1+\frac{\alpha_0}{4\epsilon_r}\log{\frac{E_C}{E_C^*}}\right)^2\; \label{eq:H-RG-d}\\
 V_{\textrm{eff}}(E_C^*)&=V_{\textrm{eff}}\; \label{eq:H-RG-e}\\
 v_{\perp}(E_C^*)&=v_{\perp}\;
 \label{eq:H-RG-f}
\end{align}
\label{eq:H-RG}
\end{subequations}
where $\alpha_0=e^2/4\pi \epsilon_0 \hbar v_F^0$ is the effective fine-structure constant. We see that the scalar potential $V_{\textrm{eff}}$ and interlayer Fermi velocity $v_{\perp}$ remain unchanged under the RG flow, while the intralayer Fermi velocity $v_F$, the nearest-neighbor interlayer hopping $t_{\perp}$, the Dirac mass term $M_{\textrm{eff}}$, and the pseudo vector potential $\mathbf{A}_{\textrm{eff}}$ are subject to  logarithmic enhancement. Among them, $M_{\textrm{eff}}$ undergoes the largest correction due to interactions with remote-bands as compared to other parameters. Derivations of Eqs.~\eqref{eq:H-RG} are given in Supplementary Information \cite{supp_info}. Similar derivations can also be found in the Supplementary Information of Ref.~\onlinecite{kang-rg-prl20} and that of Ref.~\onlinecite{lu-nc23}.
In the remaining part of the paper, we set $n_{\rm{cut}}=3$ such that our low-energy windows includes 3 valence and 3 conduction moir\'e bands per spin per valley.


\subsection*{Hartree-Fock method}
\label{methods:hf}
Here we consider the dominant intra-valley long-range Coulomb interactions, 
\begin{align}
&H^{\textrm{int}}\;\nn
\!=&\!\frac{1}{2N_s}\sum _{\lambda \lambda^{\prime},\alpha\alpha',ll'}\sum _{\mathbf{k} \mathbf{k^{\prime}}\mathbf{q}}\,V_{l l'}(\mathbf{q})\,\hat{c}^{\dagger}_{\mathbf{k+q},\lambda l \alpha }\,\hat{c}^{\dagger}_{\mathbf{k^{\prime}-q}, \lambda^{\prime} l^{\prime} \alpha^{\prime}}
\,\hat{c}_{\mathbf{k^{\prime} l^{\prime} \alpha^{\prime}}}\,\hat{c}_{\mathbf{k},\lambda l \alpha}
\label{eq:coulomb}
\end{align}
where $N_s$ denotes the total number of moir\'e primitive cells in the system, $\mathbf{k}$, $\k'$ and $\mathbf{q}$ represent  wavevectors relative to the Dirac points, $\lambda\equiv(\mu,\sigma)$ is a composite index denoting valley $\mu$ and spin $\sigma$, while $l$ and $\alpha$ are layer and sublattice indices. Since the PLG has a thickness of $1.34\,$nm, four times thicker than TBG, the difference between intralayer and interlayer Coulomb interactions may no longer be neglected, thus we keep the layer dependence of Coulomb interactions, as manifested by the layer-index dependent Coulomb interaction $V_{ll'}(\q)$ \cite{supp_info}
\begin{align}
    &V_{ll} (\mathbf{q})=\frac{e^2}{2 \Omega_0 \epsilon_r \epsilon_0 \sqrt{q^2+\kappa^2}}\;\nn
    &V_{ll'} (\mathbf{q})=\frac{e^2}{2 \Omega_0 \epsilon_r \epsilon_0 q } e^{-q|l-l'|d_0} \;,\hspace{3pt} l\neq l'\;
\label{eq:Vll}    
\end{align}
A Thomas-Fermi-type screened Coulomb interaction is adopted for the intralayer Coulomb interaction $V_{ll} (\mathbf{q})$, with a fixed inverse screening length $\kappa=0.0025\,$\AA$^{-1}$ \cite{supp_info}. The interlayer Coulomb interaction $V_{ll'} (\mathbf{q})$ ($l\neq l'$) decays exponentially in momentum space with $d_0=3.35\,$\AA. We then project the long-range Coulomb interactions onto the wavefunctions of the 3 conduction and 3 valence moir\'e bands of the \textit{renormalized continuum model} (Eqs.~\eqref{eq:H-RG}), and perform unrestricted HF calculations within this 24-band (including valley and spin) low-energy window \cite{supp_info}. We have considered 32 trial initial states characterized by the order parameters $\{s_{0,z}\tau_{a}\sigma_b$ ($a, b=x, y, z$)$\}$, where $\mathbf{s}$, $\bm{\tau}$, and $\bm{\sigma}$ denote Pauli matrices defined in the spin, valley, and sublattice subspaces, respectively. 
The background dielectric constant $\epsilon_r$ and displacement field $D$ are treated as  parameters, where $4\leq\epsilon_r\leq 8$ and $0.7\,\textrm{V/nm}\leq D\leq 1.1\,\textrm{V/nm}$. Here the only free parameter is $\epsilon_r$, since $D$ can be fixed by experiments. The screening of displacement field has been treated self consistently \cite{min-prb23,supp_info}.  

After projecting the Coulomb interaction to the renormalzied low-energy wavefunctions, it can be written in the band basis
\begin{align}
&\hat{V}^{\rm{intra}}\nn
=&\frac{1}{2N_s}\sum _{\btk \btk'\btq}\sum_{\substack{\mu\mu' \\ \sigma\sigma'\\l l'}}\sum_{\substack{nm\\ n'm'}} \sum _{\mathbf{Q}}\,V_{ll'} (\mathbf{Q}+\btq)\,\Omega^{\mu l ,\mu' l'}_{nm,n'm'}(\btk,\btk',\btq,\mathbf{Q}) \nonumber \\
&\times \hat{c}^{\dagger}_{\sigma\mu,n}(\btk+\btq) \hat{c}^{\dagger}_{\sigma'\mu',n'}(\btk'-\btq) \hat{c}_{\sigma'\mu',m'}(\btk') \hat{c}_{\sigma\mu,m}(\btk)
\label{eq:Hintra-band}
\end{align}
where the form factor $\Omega ^{\mu  l,\mu' l'}_{nm,n'm'}$ is expressed as
\begin{align}
&\Omega ^{\mu l ,\mu' l'}_{nm,n'm'}(\btk,\btk',\btq,\mathbf{Q})\nn
=&\sum _{\alpha\alpha'\mathbf{G}\mathbf{G}'}\,C^*_{\mu l \alpha\mathbf{G}+\mathbf{Q},n}(\btk+\btq) C^*_{\mu'l'\alpha'\mathbf{G}'-\mathbf{Q},n'}(\btk'-\btq)\;\nn
&\times C_{\mu'l'\alpha'\mathbf{G}',m'}(\btk')C_{\mu l \alpha\mathbf{G},m}(\btk).
\label{eq:form-factor}
\end{align}
The band indices $n,m,n', m'$ run over the low-energy Hilbert space within $E_C^{*}$ including 3 valences and 3 conduction bands (per spin per valley).  		$\{C_{\mu l \alpha\mathbf{G},m}(\btk)\}$ denote the renormalized low-energy wavefunctions of band $m$ at moir\'e wavevector $\btk$, where $\mu$, $l$, $\alpha$ represent the valley, layer, and sublattice indices, respectively. $\mathbf{G}$, $\G'$ and $\Q$ denote the moir\'e reciprocal vectors, while $\btk,\btk'$ and $\widetilde{\mathbf{q}}$ denote wavevectors within moir\'e Brillouin zone.
Then we make Hartree-Fock approximation to Eq.~\eqref{eq:Hintra-band}, and more details can be found in Supplementary Information \cite{supp_info}.

\subsection*{Exact diagonalization}
\label{methods:ed}
\begin{figure}[bth!]
    \includegraphics[width=3.5in]{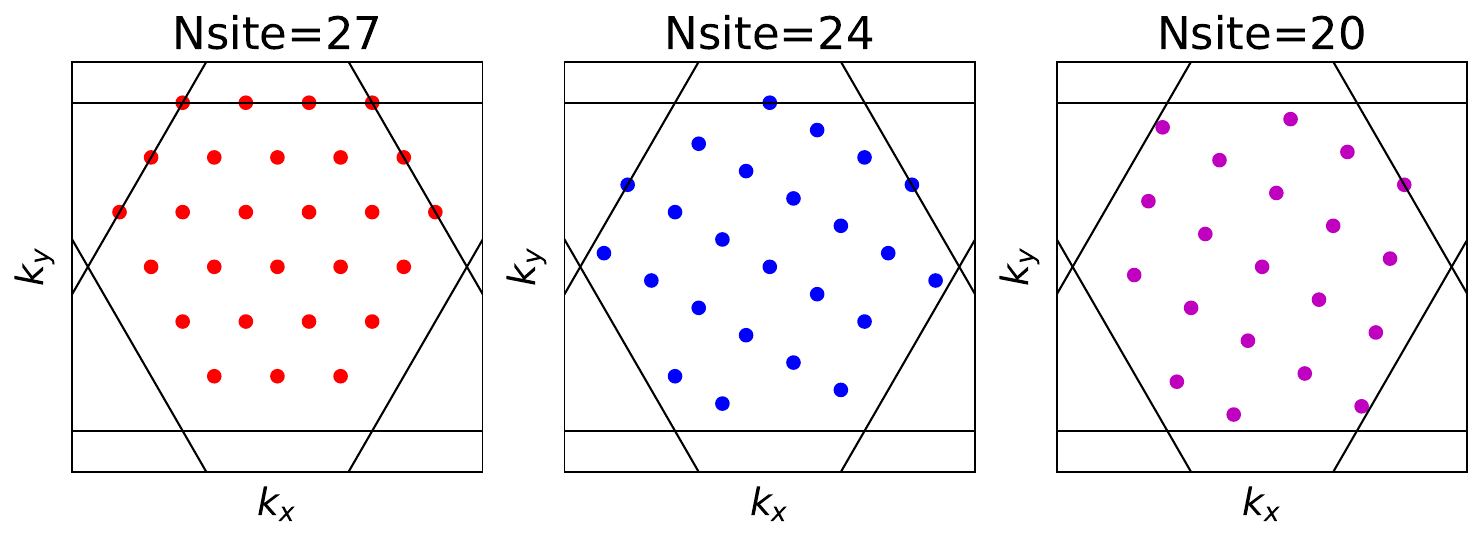}
\caption{
Brillouin zone samplings for the ED calculations.}
\label{fig:EDkmesh}
\end{figure}
In our numerical implementations, the ED calculations at 2/3 and 1/3 electron doping are performed on a 27-site cluster generated from three sets of 9-site clusters, and those at 3/5 and 2/5 electron doping are done on a 20-site (4$\times$5) cluster. The spectral-flow calculations of (2/3, 1/3) and (3/5, 2/5) electron fillings are done on 24-site (4$\times$6) and 20-site (4$\times$5) cluster, respectively. The Brillouin zone samplings for the ED calculations are shown in Fig.~\ref{fig:EDkmesh}. The Coulomb interactions in the ED calculations are identical to those adopted in the HF calculations in the previous section, which are then projected to the isolated flat band right below the chemical potential for HF states at filling 1. As discussed in the main text, there are two competing HF states, which give rise to two types of topologically distinct flat bands, and we have performed ED calculations by hole doping both types of flat bands. We have found the global many-body ground state by comparing their many-body total energy.

\subsection*{Workflow to solve the many-body ground state}
\label{methods:workfolow}

Practically, we  project the long-range $e$-$e$ interactions (Eq.~\eqref{eq:coulomb}) onto the wavefunctions of the isolated flat  band emegring from the corresponding Hartree-Fock ground state at filling 1, and calculate the interacting ground states using ED under different hole doping levels. Such an approach is justified as follows:
\begin{itemize}
\item
First, we have set up a low-energy window $E_C^*$ above which the effects of $e$-$e$ interactions have been treated by perturbative RG, which leads to a \textit{renormalized} low-energy effective model with its parameters given by Eqs.~\eqref{eq:H-RG}. Such a renormalized continuum model applies to the low-energy cutoff $E_C^*\sim n_{\textrm{cut}}\hbar v_F^0/L_s\sim 150\,$meV for $n_{\textrm{cut}}=3$, as shown in  Fig.~\ref{fig:1}(b).

\item
Second, within $E_C^*$, $e$-$e$ Coulomb interactions can no longer be treated by perturbations. Instead, we first perform HF calculations at some integer filling $\nu$, which divides the low-energy Hilbert space within $E_C^*$ into the occupied and unoccupied subspaces, as  shown in  Fig.~\ref{fig:1}(d).  The integer filling factor $\nu$ is made in such a way that the occupied and unoccupied subspaces are well separated by a gap. In the present study, we make the choice $\nu=1$ (see Fig.~\ref{fig:1}(d)). The presence of such a gap at $\nu=1$ is not only theoretically confirmed by HF calculations, but also  justified by the experimental observation of  robust integer quantum Hall effect at filling 1 over a large range of $D$ field (0.75\,V/nm$\lessapprox D\lessapprox 1.1\,$V/nm) in the system \cite{fqah-ju-arxiv23}.

\item
Third, we further consider \textit{either hole doping the occupied subspace with respect to the gap at $\nu=1$ or electron-doping the unoccupied subspace with respect to the gap at $\nu=0$}. This requires the projection of the Coulomb interactions onto either the occupied or unoccupied HF wavefunctions within $E_C^*$.  Then, in principle, one needs to perform \textit{multi-band} ED calculations within either subspace. For the case of $\nu=1$, there are 13 occupied bands and 11 unoccupied bands within $E_C^*$ as shown in Fig.~\ref{fig:1}(d). Thus, in principle one needs to do a 13-band ED calculation if considering hole-doping the $\nu=1$ HF ground state.

\item
Performing a 13-band ED calculation is certainly unrealistic. Nevertheless, we note that within the occupied subspace at $\nu=1$, the highest occupied HF band originates from the spin-valley polarized first moir\'e conduction band, which is well separated from the valence moir\'e bands by a single-particle gap $\sim 80\rm{-}100\,$meV at large $D$ fields, as shown in Fig.~\ref{fig:1}(d), which is greater than the $e$-$e$ Coulomb interaction energy scale in the system ($\sim 20\,$meV). Thus, when hole doping the highest occupied HF band with respect to $\nu=1$, it is legitimate to frozen the 12 occupied valence moir\'e bands and project Coulomb interactions only to the isolated, spin-valley polarized conduction flat band, which is highlighted in orange in Fig.~\ref{fig:1}(d).

\item
Lastly, it is worthwhile noting that the choice of integer filling $\nu$ is not unique. Another natural choice is $\nu=0$, which divides the low-energy Hiblert space $E_C^*$ into 12 occupied bands and 12 unoccupied bands. With such a choice, one needs to electron dope the unoccupied subspace. However, since the 12 unoccupied bands are entangled together (see Fig.~\ref{fig:1}(c)), one has to project Coulomb interactions onto all of the 12 unoccupied bands and perform 12-band ED calculation, which is barely possible. So the best choice is to hole dope the $\nu=1$ HF ground state.
\end{itemize}

\acknowledgements


We thank Y. Zhang, Z. Lu, and J. Yu for helpful discussions. This work is supported by the National Key R \& D program of China (grant No. 2020YFA0309601), the National Natural Science Foundation of China (grant No. 12174257), the Science and Technology Commission of the Shanghai Municipality (Grant No. 21JC1405100), and the start-up grant of ShanghaiTech University. \textit{Note added:} in the process of preparing this manuscript, we are aware of several closely related manuscripts posted on arXiv \cite{senthil-fqah-arxiv23,zhang-fqah-arxiv23,ashvin-fqah-arxiv23,bernevig-fci-plg-arxiv23}




\widetext
\clearpage

\makeatletter
\def\@fnsymbol#1{\ensuremath{\ifcase#1\or \dagger\or \ddagger\or
		\mathsection\or \mathparagraph\or \|\or **\or \dagger\dagger
		\or \ddagger\ddagger \else\@ctrerr\fi}}
\makeatother

\begin{center}
\textbf{\large Supplementary Information for “Theory of fractional Chern insulator states in pentalayer graphene moir\'e superlattice"} \\
\vspace{0.5cm}
Xin Lu, Zhongqing Guo, Bo Xie, and Jianpeng Liu \footnote{liujp@shanghaitech.edu.cn}
\end{center}


%
%
%
%
%
%


\setcounter{equation}{0}
\setcounter{figure}{0}
\setcounter{table}{0}
\setcounter{section}{0}
\makeatletter
\renewcommand{\theequation}{S\arabic{equation}}
\renewcommand{\thesection}{S\arabic{section}}
\renewcommand{\figurename}{Supplementary Figure}
\renewcommand{\tablename}{Supplementary Table}

\def\bibsection{\section*{References}} 
\tableofcontents

\section{Continuum model for twisted pentalayer graphene-hBN moir\'e superlattice}
\label{sec:continuum_model}

In our theoretical study, we adopt the continuum model derived by Moon and Koshino \cite{moon-hBNgr-prb-2014} to twisted pentalayer graphene-hBN moir\'e superlattice, the Hamiltonian of which can be divided in the most general way into three parts:  hBN intralayer Hamiltonian $H_{\rm{hBN}}$, pentalayer graphene Hamiltonian $H_{\rm{pGr}}$ and the moir\'e coupling between them. 

For $H_{\rm{pGr}}$, we start from the tight-binding model for rhombohedral pentalayer graphene. We define $\a_1=a_0 (1,0)$ and $\a_2 = a_0 (1/2,\sqrt{3}/2)$ as the lattice vectors of graphene with lattice constant $a_0=2.46$\,\AA. The corresponding reciprocal lattices are $\b_1= 4\pi/\sqrt{3}a_0 (\sqrt{3}/2,-1/2)$ and $\b_2 = 4\pi/\sqrt{3} a_0 (0,1)$. Two sublattices $A/B$ form a honeycomb lattice. The two sublattices in Layer 1, namely the graphene layer that is the closest to the underlying hBN layer, are supposed to be seated at $\bm{\tau}_{1,A} =  (0,a_0/\sqrt{3})$ and $\bm{\tau}_{1,B} = (0,0)$, respectively. The rhombohedral stacking of multilayer graphene is set by shifting each layer in the in-plane direction $(0,-a_0/\sqrt{3})$ with respect to its nearest underlying layer. Then, every two nearest graphene layers are Bernal-stacked with top $A$ site sitting vertically over bottom $B$ site. In other words, for Layer $l$, we have $\bm{\tau}_{l,A} =  (0,(2-l) \times a_0/\sqrt{3})$ and $\bm{\tau}_{l,B} =  (0,(1-l) \times a_0/\sqrt{3})$ with $l=1,2,3,4,5$. The Slater-Koster tight-binding Hamiltonian constructed from carbon's $p_z$ atomic orbitals is readily written as
\begin{align}
    H_{\rm{pGr}} = \sum_{\substack{i,j,\\ l_1,l_2,\\ \alpha,\beta}} -t(\R_i + \bm{\tau}_{l_1,\alpha} + l_1 d_0  \mathbf{e}_z - \R_j - \bm{\tau}_{l_2,\beta} - l_2 d_0 \mathbf{e}_z ) \, \hcd_{l_1,\alpha} (\R_i) \hc_{l_2,\beta} (\R_j)
    \label{eq:generalTB}
\end{align}
where $(i,j)$ stands for unit-cell's indices, $(l_1,l_2)$ for layer and $(\alpha,\beta)$ for sublattice indices, respectively. The interlayer distance is set to be $d_0 = 3.35$\,\AA \  and $\mathbf{e}_z$ is the unit vector perpendicular to the graphene plane. The Slater-Kolster-type hopping parameters are employed here 
\begin{subequations}
    \begin{align}
        -t(\R) &= V_{pp\pi} \left[ 1- \left(\frac{\R \cdot \mathbf{e}_z }{|\R|}\right)^2 \right] + V_{pp\sigma} \left(\frac{\R \cdot \mathbf{e}_z }{|\R|}\right)^2 , \\
        V_{pp\pi} &= V_{pp\pi}^0 \exp \left(-\frac{|\R|-a_0/\sqrt{3}}{r_0}\right) , \\
        V_{pp\sigma} &= V_{pp\sigma}^0 \exp \left(-\frac{|\R|-d_0}{r_0}\right) .
    \end{align}
\end{subequations} 
where we take $V_{pp\pi}^0 = -2.7$\,eV, $V_{pp\sigma}^0=0.48$\,eV and $r_0 = 0.184 a_0$ \cite{moon-hBNgr-prb-2014}. 

In our modelling, we consider only interlayer hopping between two nearest graphene sheets. After the following in-plane lattice Fourier transformation
\begin{subequations}
    \begin{align}
        \hc_{l,\alpha} (\R) &= \frac{1}{\sqrt{N}} \sum_{\k} e^{i \k \cdot (\R + \bm{\tau}_{l,\alpha})} \hc_{l,\alpha} (\k) \\
        \hc_{l,\alpha} (\k) &= \frac{1}{\sqrt{N}} \sum_{\k} e^{-i \k \cdot (\R + \bm{\tau}_{l,\alpha})} \hc_{l,\alpha} (\R)
    \end{align}
\end{subequations}
with the total number of unit-cell $N$, we expand the tight-binding Hamiltonian at the vicinity of $\mathbf{K}/\mathbf{K}'$ points to obtain the $\k \cdot \p$ Hamiltonian for pentalayer graphene:
\begin{align}
    H^{0,\mu}_{\rm{penta}} = 
    \begin{pmatrix}
        h^{0,\mu}_{\rm{intra}} & (h^{0,\mu}_{\rm{inter}})^\dagger & 0 & 0 & 0 \\
        h^{0,\mu}_{\rm{inter}} & h^{0,\mu}_{\rm{intra}} & (h^{0,\mu}_{\rm{inter}})^\dagger & 0 & 0 \\
        0 & h^{0,\mu}_{\rm{inter}} & h^{0,\mu}_{\rm{intra}} & (h^{0,\mu}_{\rm{inter}})^\dagger & 0 \\
        0 & 0 & h^{0,\mu}_{\rm{inter}} & h^{0,\mu}_{\rm{intra}} & (h^{0,\mu}_{\rm{inter}})^\dagger \\
        0 & 0 & 0 & h^{0,\mu}_{\rm{inter}} & h^{0,\mu}_{\rm{intra}}
    \end{pmatrix}
\label{eq:Htot_penta}
\end{align}
where $\mu=\pm 1$ is the valley index respectively for $\mathbf{K}_\mu = -\mu (4\pi/3 a_0,0)$ ($\mathbf{K}\equiv \mathbf{K}_{+}$ and $\mathbf{K}' \equiv \mathbf{K}_{-}$). The intra- and inter-layer blocks are 
\begin{align}
    h^{0,\mu}_{\rm{intra}} &= - \hbar v_F^0 \k \cdot \bm{\sigma}_\mu \\
    h^{0,\mu}_{\rm{inter}} &=
    \begin{pmatrix}
       \hbar v_\perp (\mu k_x + i k_y) & t_\perp \\
        \hbar v_\perp (\mu k_x - i k_y) & \hbar v_\perp (\mu k_x + i k_y)
    \end{pmatrix}
\end{align}
where $\bm{\sigma}_\mu = (\mu \sigma_x, \sigma_y)$ are the Pauli matrices and the value of the parameters are $\hbar v_F^0 = 5.253\,\eVA$, $\hbar v_\perp=0.335\,\eVA$ and $t_\perp = 0.34\,\eV$. 
The lattice structure for hBN is also honeycomb-like but with a different lattice constant $a_{\rm{hBN}} = 1.018 a_0$. The two sublattices $A$ and $B$ can be then occupied by the two different atoms N and B, or vice versa. In our study, we define the stacking geometry of the pentalayer graphene-hBN system by starting from a non-rotated arrangement, where a $B/A$ site of graphene and a boron/nitrogen site of hBN share the same in-plane position $(0,0)\,/\,(0,a_0/\sqrt{3})$, so that the in-plane $A$-$B$ bonds are parallel to each other. Note that the presence of hBN breaks the spatial inversion symmetry. If we rotate hBN by $180\degree$ by exchanging the role of N and B in hBN, then we obtain a different system. Nevertheless, it turns out that the two different configurations result in similar energy spectra. Especially when we apply a strong electric field to simulate what has been done in the experiment \cite{lu-fqahe_pGr-arxiv-2023}, the typical energy difference in the non-interacting conduction bands, in which we are interested, is below 0.1\,meV, legitimately negligible compared to the strength of Coulomb potential ($\sim$ 20\,meV taking the dielectric constant $\epsilon_r=5$, more details in the following sections on Hartree-Fock calculations). So we focus on the first configuration in our study. 

After a rotation, the lattice vectors of hBN becomes
\begin{align}
    \ta_{1,2} = M R(\theta) \a_{1,2}
\end{align}
where $M = a_{\rm{hBN}}/a_0$ and $R(\theta)$ is the rotation matrix. Therefore, the mismatch between the two lattices generates moir\'e pattern with the moir\'e lattice vectors $\L_i$ and reciprocal lattice vectors $\G_i$ ($i=1,2$)
\begin{align}
    \L_{i} &= (1-R(-\theta) M^{-1})^{-1} \a_i \ , \\     
    \G_{i} &= (1- M^{-1}R(\theta)) \b_i \  . 
\end{align}
For $\theta=0.77\,\degree$, the moir\'e length scale is $L_s\!\equiv\!|\L_{i}| = 109.16$\,\AA \  and $|\G_{i}|=4 \pi / \sqrt{3} L_s$. Following the derivation in \cite{moon-hBNgr-prb-2014} and considering only the hopping between bottom-layer graphene and hBN, the moir\'e interlayer hopping terms from the bottom-layer graphene to hBN are
\begin{align}
    U = u_0 \left[ \begin{pmatrix}
            1 & 1 \\
            1 & 1
        \end{pmatrix}+ \begin{pmatrix}
            1 & e^{-i\mu\frac{2 \pi}{3}} \\
            e^{i\mu\frac{2 \pi}{3}} & 1
    \end{pmatrix} e^{i \mu \G_1 \cdot \br}+ \begin{pmatrix}
        1 & e^{i\mu\frac{2 \pi}{3}} \\
        e^{-i\mu\frac{2 \pi}{3}} & 1
\end{pmatrix} e^{i \mu (\G_1+\G_2) \cdot \br}
    \right]
    \label{eq:u-moire}
\end{align}
where $u_0=0.152\,\eV$. Since the band edges of insulating hBN and its gap are much larger than the long-wavelength moir\'e coupling at small twist angle, we can safely neglect completely the dispersion in hBN by dropping the $\k$ dependence in $H_{\rm{hBN}}$, namely
\begin{align}
    H_{\rm{hBN}} = 
    \begin{pmatrix}
        V_{\rm{N}} & 0 \\
        0 & V_{\rm{B}}
    \end{pmatrix}
\end{align}
with $V_{\rm{N}} = -1.40\,\eV$ and $V_{\rm{B}}=3.34\,\eV$. For the same reason, the low-energy physics of pentalayer graphene-hBN heterostructure is dominated by the graphene's side so that the interlayer moir\'e coupling (Eq.~\eqref{eq:u-moire}) can be integrated out into an effective moir\'e superlattice potential $V_{\rm{hBN}}$ acting directly on Layer 1 of pentalayer graphene, namely
\begin{align}
    V_{\rm{hBN}} = - U^\dagger H_{\rm{hBN}} U = V^{\rm{eff}}(\br) + M^{\rm{eff}}(\br) \sigma_z + e v_F \mathbf{A}^{\rm{eff}}(\br) \cdot \bm{\sigma}_\mu.
\end{align}
where we classify different terms in the effective potential by their sublattice structure. Simple algebra calculations give
\begin{subequations}
    \begin{align}
        V^{\rm{eff}}(\br) &= V_0 - V_1 \sum_{j=1}^{3} \cos \alpha_j(\br) \\
        M^{\rm{eff}}(\br) &= \sqrt{3} V_1 \sum_{j=1}^{3} \sin \alpha_j(\br) \\
        e v_F \mathbf{A}^{\rm{eff}}(\br) &= 2 \mu V_1 \sum_{j=1}^{3} \begin{pmatrix}
            \cos [2\pi (j+1)/3] \\
            \sin [2\pi (j+1)/3]
        \end{pmatrix}\cos \alpha_j(\br) \\
        \alpha_j (\br) &=  \G_j \cdot \br + \psi + \frac{2 \pi}{3} \quad \rm{with} \quad \G_3 = -\G_1 - \G_2
    \end{align}
\end{subequations} 
where $V_0=0.0289\,\eV$, $V_1 = 0.0210\,\eV$ and $\psi = -0.29$\,rad. Finally, the continuum model for twisted pentalayer graphene-hBN heterostructure is readily written as
\begin{align}
    H^{0,\mu} = 
    \begin{pmatrix}
        h^{0,\mu}_{\rm{intra}} + V_{\rm{hBN}} & (h^{0,\mu}_{\rm{inter}})^\dagger & 0 & 0 & 0 \\
        h^{0,\mu}_{\rm{inter}} & h^{0,\mu}_{\rm{intra}} & (h^{0,\mu}_{\rm{inter}})^\dagger & 0 & 0 \\
        0 & h^{0,\mu}_{\rm{inter}} & h^{0,\mu}_{\rm{intra}} & (h^{0,\mu}_{\rm{inter}})^\dagger & 0 \\
        0 & 0 & h^{0,\mu}_{\rm{inter}} & h^{0,\mu}_{\rm{intra}} & (h^{0,\mu}_{\rm{inter}})^\dagger \\
        0 & 0 & 0 & h^{0,\mu}_{\rm{inter}} & h^{0,\mu}_{\rm{intra}}
    \end{pmatrix}
\label{eq:Htot_all}
\end{align}
In the following calculations,  we always consider $e v_F \mathbf{A}^{\rm{eff}}(\br)$ as a whole, instead of treating $\mathbf{A}^{\rm{eff}}(\br)$ separately. 
Also, the constant energy shift $V_0$ can be absorbed in chemical potential so we will omit it in the following.

Conventionally, the effect of applied electric field is incorporated in Hamiltonian Eq.~\eqref{eq:Htot_all} by setting a layer-dependent on-site energy, linearly increasing with layer index. In the experiment, the typical $D$ field is around 1\,V/nm. Given that $\epsilon_r=5$ and the vertical distance between hBN and Layer 1 graphene is $\sim d_0$, the energy difference induced by electric field is at most $\sim 70$\,meV (the actual electric field would be screened as shown in the next section), which would shift the relative band position between graphene and hBN through changing $V_{\rm{B}}$ and $V_{\rm{N}}$. However, this energy shift is negligible compared to $|V_{\rm{B}}|$ and $|V_{\rm{N}}|$ so that $V_{\rm{hBN}}$ is scarcely affected. So, we will omit the influence of applied electric field on the moir\'e potential.

\section{Self-consistent Hartree screening of applied electric field}
\label{sec:hartree_screening}
In multilayer graphene, it is well-known that an externally applied out-of-plane electric field is significantly screened due to the redistribution of electrons within different layers \cite{jang-multiGr-prblett-2023,mccann-bGrEfield-prb-2006,avetisyan-multiGrEfield1-prb-2009,avetisyan-multiGrEfield2-prb-2009,koshino-triGrEfield-prb-2009,zhang-triGrEfield-prb-2010}. To simulate $D\sim 1$\,V/nm in the experiments, we need to take into account such effect to find the actual on-site energy differences between layers in pentalayer graphene at various background dielectric constant $\epsilon_r$. 

To encounter the screening effect, we start enforcing a fixed external electric field $E_{\rm{ext}} = D/\epsilon_{\rm{hBN}}$ to pentalayer graphene with the dielectric constant of hBN $\epsilon_{\rm{hBN}}=4$. This would induce a constant electrostatic potential difference between the two adjacent layers $eE_{\rm{ext}} d_0$ ($e>0$), which is added into the diagonal part in the Hamiltonian of pentalayer graphene Eq.~\eqref{eq:Htot_penta}. Then, we solve the Hamiltonian and retrieve the distribution of excess electron in each layer. The unevenly distributed electrons would induce an additional electric field to counteract $E_{\rm{ext}}$ so as to cancel out  part of the electrostatic potential difference between layers. 
This screening process is treated by solving the classical Poisson equation in electrostatics, while the charge density is calculated quantum mechnically using the continuum model described in Sec.~\I. This is equivalent to making Hartree approximation to $e$-$e$ interactions assuming homogeneous in-plane charge density within each layer.
Once the total electric field is readjusted, we compute the new electrostatic potential difference and the new layer-dependent charge density distribution. We continue to proceed in the same way until a self-consistent layer-resolved on-site energy is obtained.

\begin{figure}
    \includegraphics[width=0.6 \textwidth]{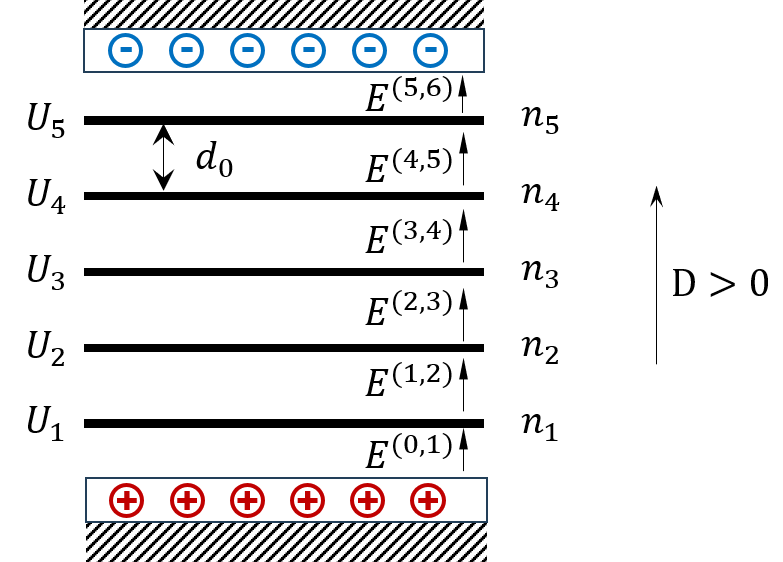}
    \caption{Schematic for a device of pentalayer graphene sandwiched by the top and bottom hBN gates.}
    \label{fig:device}
\end{figure}

In our screening theory, we consider a device of pentalayer graphene encapsulated by top and bottom hBN substrates, as shown in Supplementary Figure~\ref{fig:device}. We consider the situation that the device is charge neutral, and  that the top and bottom metallic gates above and below the top and bottom hBN substrates  have equal density but opposite sign of charges, which generates vertical displacement field $D$ and induces equal polarized charge densities of opposite sign in the top and bottom hBN substrates as shown in Supplementary Figure~\ref{fig:device}.  This enforces pentalayer graphene to be also charge neutral 
\begin{equation}
\sum_{i=1}^{5}\,n_i=0\;.
\end{equation}
From Gauss' law, the electric field $E^{(i,i+1)}$ between the $i$-th and $(i+1)$-th graphene layers is 
\begin{equation}
    E^{(i,i+1)} - E^{(i-1,i)} = \frac{-e n_i}{\epsilon_0 \epsilon_r}
\end{equation} 
for $i=1, ... ,5$,
where $\epsilon_r$ is the dielectric constant inside pentalayer graphene and $n_i$ is defined with respect to intralayer charge neutral point, i.e., positive if there is excess electron and vice versa. 
Then the layer-resolved on-site energies $U_i$ are set given that
\begin{subequations}
    \begin{align}
        0 &= \sum_{i=1}^{5} U_i , \\
        U_i & = U_{i-1} + e E^{(i-1,i)} d_0 \quad \rm{for} \quad i \geq 2 .
    \end{align}
\end{subequations}
where the first equation is to set the average electrostatic potential energy of the pentalayer to be zero.
In practice, we use a plane-wave cut-off covering $ n_1 \G_1 + n_2 \G_2$ with $n_{1,2} \in [-4,5]$ and a $18 \times 18$ $\k$-mesh for each valley. We numerically checked that the non-interacting band structure and the obtained $U_i$ converge ($\sim 10^{-3}$\,meV) already for these parameters. To match the experiments, we select the ranges $\epsilon_r=4-8$ and $D = 0.7 - 1.1$\,V/nm. Note that $\epsilon_r$ has to be at least larger than $\epsilon_{\rm{hBN}}$ because of additional screening in pentalayer graphene besides the environmental screening. Furthermore, it turns out that changing the filling of pentalayer graphene from charge neutral to $\nu=1$ or adding $V_{\rm{hBN}}$ would only alter $U_i$ by around 1\,meV, which is far smaller than $|U_i| \sim 40$\,meV. So, it is legitimate to do the self-consistent calculations simply using Eq.~\eqref{eq:Htot_penta}.

\section{Perturbative renormalization group calculations of Coulomb potential from remote bands}
\label{sec:RG}
Although we focus on the low-energy physics, the presence of filled high energy valence bands indeed play an important role considering $e$-$e$ interactions. The electrons in the filled remote bands will act through long-ranged Coulomb potential upon the dynamics of low-energy electrons such that an effective low-energy Hamiltonian Eq.~\eqref{eq:Htot_all} would have parameters in general larger in amplitudes than the non-interacting ones. For example, it is well known that the Fermi velocity around the Dirac point in graphene would be amplified by the filled Dirac Fermi sea \cite{gonzalez_nuclphysb1993,elias_natphys2011}. We take into account this effect using perturbative renormalization group (RG) approach. The derivation shown in this section is deeply inspired from Ref.~\cite{vafek_prl2020} and it is parallel to Ref.~\cite{lu-GrCrOCl-natcom-2023}. 

The $e$-$e$ Coulomb interaction operator in our derivations is written as
\begin{equation}
    \hat{V}_{\text{int}} = \frac{1}{2} \int d^2 \br d^2 \br ' V_c(\br - \br ') \hat{\rho} (\br) \hat{\rho} (\br ') 
\end{equation}
where $V_c(\br) = e^2/4 \pi \epsilon_0 \epsilon_r r$ and $\hat{\rho}(\br)$ is the density operator of electrons at $\br$. The Hamiltonian Eq.~\eqref{eq:Htot_all} is defined at some high energy cutoff $\pm E_c$. Remember that the parameters should be thought of as being fixed by a measurement at $\pm E_c$  with fully empty or occupied Hilbert space and without $e$-$e$ interactions (in the framework of continuum model). 

This also amounts to write $\hat{\rho}(\br)= \hat{\psi}^\dagger (\br) \hat{\psi} (\br)$ with the non-interacting field operator $\hat{\psi} (\br)$
\begin{equation}
    \hat{\psi} (\br) = \sum_{\substack{\sigma, n, 
    \k;\\ |\epsilon_{n,\k}| \leq E_c }}  \phi_{\sigma n \k} (\br) \hc_{\sigma n}(\k) 
\end{equation}
where $\phi_{\sigma n \k} (\br)$ is the wavefunction of an eigenstate of the non-interacting Hamiltonian Eq.~\eqref{eq:Htot_all} with energy $\epsilon_{n,\k}$ and its associated annihilation operator is $\hc_{\sigma n}(\k)$. The band index $n$ includes the bands from the two valleys $\mu=\pm 1$.

\subsection{Electron-electron interaction in a lower energy window}
Now we change the cutoff $E_c$ to a smaller one $E_c '$ and see how these parameters are modified by $\hat{V}_{\text{int}}$, which can be treated perturbatively when $E_c '$ is much larger than any other energy scale in the system. To do so, we split the field operator $\hat{\psi} (\br) = \hat{\psi}^{<} (\br) + \hat{\psi}^{>} (\br)$ where 
\begin{align}
    \hat{\psi}^{<} (\br) &= \sum_{\substack{\sigma, n, 
    \k;\\ |\epsilon_{n,\k}| \leq E_c ' }} \phi_{\sigma n \k} (\br) \hc_{\sigma n}(\k) \\
    \hat{\psi}^{>} (\br) &= \sum_{\substack{\sigma, n, 
    \k;\\ E_c ' < |\epsilon_{n,\k}| \leq E_c }} \phi_{\sigma n \k} (\br) \hc_{\sigma n}(\k). 
\end{align}
where $\sigma$, $n$, $\k$ refer to spin, band index, and wavevector respectively. Note that for simplicity, here the valley index has been included into the band index $n$.
Then, we integrate out the fast modes $\hat{\psi}^{>} (\br)$ in the expansion of $\hat{\rho}(\br) \hat{\rho}(\br')$. Note that $\hat{\psi}^{>} (\br)$ and $\hat{\psi}^{> \dagger} (\br)$ must appear equal times in each terms of the expansion otherwise it would vanish by taking the non-interacting mean value $\langle \dots \rangle_0$. 

Since we need to shrink the energy window for the valence and conduction band at the same time, this requires particle-hole symmetry \cite{vafek_prl2020} such that we can always get the Coulomb interaction in the same form. In our system described by Eq.~\eqref{eq:Htot_all}, several terms break particle-hole symmetry, which actually matters knowing the fact that the nontrivial correlated states are experimentally observed only among conduction bands but absent among valence bands. However, the energy scales of the particle-hole breaking terms turn out to be negligible outside the energy window $[-E_c^*, E_c^*]$ ($E_c^*\sim\,0.15\,$eV) at which the RG flow is stopped, and we start to explicitly take into account these particle-hole breaking terms in unrestricted Hartree-Fock calculations within $E_C^*$. More details will be given at the end of this section. Therefore, it is legitimate at least for now in our RG derivation to neglect such particle-hole asymmetry. Finally, it is useful to write the Coulomb interaction in the RG process using the approximate particle-hole symmetry as
\begin{align}
    \hat{V}_{\text{int}} (E_c) &= \frac{1}{2} \int d^2 \br d^2 \br ' V_c(\br - \br ') (\hat{\rho} (\br) - \bar{\rho}_{E_c} (\br)) ( \hat{\rho} (\br ') -\bar{\rho}_{E_c} (\br ')) \\
    \bar{\rho}_{E_c} (\br) &= \frac{1}{2}\sum_{\substack{\sigma, n, 
    \k;\\ |\epsilon_{n,\k}| \leq E_c}} \phi_{\sigma n \k}^* (\br) \phi_{\sigma n \k} (\br).
    \label{eq:drho-drho}
\end{align}
After integrating out the fast modes, this formulation assures that
\begin{align*}
    (\hat{\rho} (\br) - \bar{\rho}_{E_c} (\br)) ( \hat{\rho} (\br ') -\bar{\rho}_{E_c} (\br ')) &\, \to\, (\hat{\rho}^{<} (\br) - \bar{\rho}_{E_c'} (\br))  (\hat{\rho}^{<} (\br') - \bar{\rho}_{E_c'} (\br))\\ 
    & \quad +  \underbrace{\hat{\psi}^{< \dagger} (\br) \  \langle \hat{\psi}^{>} (\br) \hat{\psi}^{> \dagger} (\br ') \rangle_0 \ \hat{\psi}^{<} (\br ') + \hat{\psi}^{<} (\br) \  \langle \hat{\psi}^{> \dagger} (\br) \hat{\psi}^{>} (\br ') \rangle_0 \ \hat{\psi}^{< \dagger} (\br ')}_{(*)}
\end{align*}
with constants omitted and 
\begin{align}
    \hat{\rho}^{<} (\br) &= \hat{\psi}^{< \dagger} (\br) \hat{\psi}^{<} (\br) \\
    \bar{\rho}^{>} (\br) &= \sum_{\substack{\sigma, n, 
    \k;\\ E_c ' < |\epsilon_{n,\k}| \leq E_c}} \phi_{\sigma n \k}^* (\br) \phi_{\sigma n \k} (\br).
\end{align}
The first term gives the $e$-$e$ Coulomb interactions between electrons of the slow modes $\hat{\psi}^{<} (\br)$ below the new cutoff $E_c '$. 

Then, we evaluate the rest of the terms in the expansion, which represents precisely the correction to Eq.~\eqref{eq:Htot_all} from the fast modes $\hat{\psi}^{>} (\br)$ via $e$-$e$ Coulomb interactions. Let us write
\begin{align*}
    (*) &= \hat{\psi}^{< \dagger} (\br) \left(  \sum_{\substack{\sigma, n, 
    \k;\\ E_c ' < \epsilon_{n,\k} \leq E_c}} \phi_{\sigma n \k} (\br) \phi_{\sigma n \k}^* (\br ')  \right) \hat{\psi}^{<} (\br ') + \hat{\psi}^{<} (\br) \left(  \sum_{\substack{\sigma, n, 
    \k;\\ -E_c ' > \epsilon_{n,\k} \geq -E_c}} \phi_{\sigma n \k}^* (\br) \phi_{\sigma n \k} (\br ') \right) \hat{\psi}^{< \dagger} (\br ') \\
    &= \hat{\psi}^{< \dagger} (\br) \left(  \sum_{\substack{\sigma, n, 
    \k;\\ E_c ' < \epsilon_{n,\k} \leq E_c}} \phi_{\sigma n \k} (\br) \phi_{\sigma n \k}^* (\br ')  \right) \hat{\psi}^{<} (\br ') + \hat{\psi}^{< \dagger} (\br ') \left(  \sum_{\substack{\sigma, n, 
    \k;\\ -E_c ' > \epsilon_{n,\k} \geq -E_c}} -\phi_{\sigma n \k}^* (\br) \phi_{\sigma n \k} (\br ') \right) \hat{\psi}^{<} (\br) 
\end{align*}
where the minus sign in the second line comes from the permutation of the two fermionic operators and the constant arising from the permutation is omitted. Then, the $e$-$e$ interaction $\hat{V}_{\text{int}}$ in the lower energy window delimited by $E_c '$ is
\begin{equation}
    \hat{V}_{\text{int}} (E_c') = \frac{1}{2} \int d^2 \br d^2 \br ' V_c(\br - \br ')  (\hat{\rho}^{<} (\br) - \bar{\rho}_{E_c'} (\br))  (\hat{\rho}^{<} (\br') - \bar{\rho}_{E_c'} (\br))+ \frac{1}{2} \int d^2 \br d^2 \br ' V_c(\br - \br ') \hat{\psi}^{< \dagger} (\br) \mathcal{F}(\br, \br ') \hat{\psi}^{<} (\br ') 
    \label{eq:vint_RG}
\end{equation}
with 
\begin{equation}
   \mathcal{F}(\br, \br ') = \sum_{\substack{\sigma, n, 
    \k;\\ E_c ' < |\epsilon_{n,\k}| \leq E_c}} \text{sign} (\epsilon_{n,\k}) \phi_{\sigma n \k} (\br) \phi_{\sigma n \k}^* (\br ').
    \label{eq:Frrp}
\end{equation}

\subsection{Evaluation of the correction to the non-interacting Hamiltonian from the fast modes}
In the following, we set $\hbar=1$ for the simplicity in mathematical expressions. Note that $\mathcal{F}(\br, \br ')$ has the structure of the residue of the Green's function $\hat{G}(z) = (z - H^{0,\mu})^{-1}$, namely
\begin{equation}
    \mathcal{F}(\br, \br ') = \oint_\mathcal{C} \frac{dz}{2 \pi i} \bra{\br} \hat{G}(z) \ket{\br'}
\end{equation}
where the contour $\mathcal{C}$ encloses the $z$-plane real line segment $[-E_c, -E_c']$ in the clockwise, and segment $[E_c ', E_c ]$ in the counterclockwise, sense. As long as $E_c '$ dominates over all other energy scales such as $v_F^0 |\G_i|$ or $t_\perp$ ($\sim 300$\,meV, the two largest parameters), the dominant contribution to the contour integral can be evaluated perturbatively using $\hat{G}(z) \approx \hat{G}_0(z) + \hat{G}_0(z) H_{\rm{pert}} \hat{G}_0(z) + \mathcal{O}\left( t_\perp^2/E_c'^2, (v_F^0 |\G_i|)^2/E_c'^2 \right)$. Here,  $H_{\rm{pert}}$ represents moir\'e potential terms such as $V^{\rm{eff}}(\br)$, $M^{\rm{eff}}(\br) \sigma_z$ and $e v_F \mathbf{A}^{\rm{eff}}(\br) \cdot \bm{\sigma}_\mu$ and interlayer hopping terms:
\begin{equation}
    H^{0,\mu}_{\rm{inter}} = \begin{pmatrix}
        0 & (h^{0,\mu}_{\rm{inter}})^\dagger \\
        h^{0,\mu}_{\rm{inter}} & 0
    \end{pmatrix}.
\end{equation}
The non-interacting Green's function of decoupled pentalayer graphene $G_0(z)$ is expressed as
\begin{equation}
    \hat{G}_0 (z) = \begin{pmatrix}
        \hat{g}_0 (z-2\Delta_D) & 0 & 0 & 0 & 0\\
        0 & \hat{g}_0 (z-\Delta_D) & 0 & 0 & 0\\
        0 & 0 & \hat{g}_0 (z) & 0 & 0 \\
        0 & 0 & 0  & \hat{g}_0 (z+\Delta_D) & 0 \\
        0 & 0 & 0 &  0& \hat{g}_0 (z+2\Delta_D)
    \end{pmatrix}
\end{equation}
where $\hat{g}_0 (z) = (z - h^{0,\mu}_{\rm{intra}})^{-1}$ is the non-interacting Green's function of monolayer graphene. We note that the on-site energy difference between the two adjacent layers $\Delta_D \sim 40$\,meV (for self consistently screened $D\sim 1\,$V/nm) is negligible compared to $E_c'$ so we can omit it. 

In the following, we focus in the valley $\mu=+1$ since the derivation for the valley $\mu=-1$ is immediate and the results are identical. It is easier to calculate the Green's function in the plane wave basis $\ket{\k}$ 
\begin{equation}
   \mathcal{F}(\br, \br ') = \int \frac{d^2 k d^2 k'}{(2 \pi)^4} e^{i (\k \cdot \br -  \k ' \cdot \br ' )} \oint_\mathcal{C} \frac{dz}{2 \pi i} \bra{\k} \hat{G}(z) \ket{\k'} 
\end{equation}
with
\begin{align}
    \ket{\br} &= \int \frac{d^2 k}{(2 \pi)^2} e^{-i \k \cdot \br} \ket{\k} \\
    \bracket{\br}{\br '} &= \delta^{(2)} (\br - \br') \\
    \bracket{\k}{\k '} &= (2 \pi)^2 \delta^{(2)} (\k - \k')
\end{align}
where $\delta^{(2)}(\mathbf{x})$ is the 2D Dirac distribution. In the plane wave basis, the evaluation of Green's functions is straightforward. Let us first focus on the intralayer terms:
\begin{align}
    \bra{\k} \hat{G}_0(z) \ket{\k'} = (2 \pi)^2 \delta^{(2)} (\k - \k') \frac{1}{2} \sum_{\lambda=\pm} \frac{1+\lambda \frac{\k}{k} \cdot \bm{\sigma}}{z + \lambda v_F k} 
\end{align}
\begin{align}
    & \quad\bra{\k} \hat{G}_0(z) V^{\rm{eff}}(\br) \hat{G}_0(z) \ket{\k'} \\
    &=  -\frac{V_1}{2} \sum_{j=1}^{3}  \left( e^{i \psi} (2 \pi)^2 \delta^{(2)} (\k - \k' - \G_j) \frac{1}{4} \sum_{\lambda,\lambda' = \pm} \frac{\left(1+\lambda \frac{\k}{k} \cdot \bm{\sigma} \right) \left(1+\lambda' \frac{\k-\G_j}{|\k - \G_j|} \cdot \bm{\sigma} \right)}{ \left( z + \lambda v_F k \right) \left( z + \lambda ' v_F |\k-\G_j| \right)} \right. \nonumber \\
    & \quad + \left.  e^{-i \psi} (2 \pi)^2 \delta^{(2)} (\k - \k' + \G_j) \frac{1}{4} \sum_{\lambda,\lambda' = \pm} \frac{\left(1+\lambda \frac{\k}{k} \cdot \bm{\sigma} \right) \left(1+\lambda' \frac{\k+\G_j}{|\k + \G_j|} \cdot \bm{\sigma} \right)}{ \left( z + \lambda v_F k \right) \left( z + \lambda ' v_F |\k+\G_j| \right)}   \right) \nonumber\\
    & \quad\bra{\k} \hat{G}_0(z) M^{\rm{eff}}(\br) \sigma_z \hat{G}_0(z) \ket{\k'} \\
    &=  \frac{\sqrt{3} V_1}{2i} \sum_{j=1}^{3}  \left( e^{i \psi} (2 \pi)^2 \delta^{(2)} (\k - \k' - \G_j) \frac{1}{4} \sum_{\lambda,\lambda' = \pm} \frac{\left(1+\lambda \frac{\k}{k} \cdot \bm{\sigma} \right) \sigma_z  \left(1+\lambda' \frac{\k-\G_j}{|\k - \G_j|} \cdot \bm{\sigma} \right)}{ \left( z + \lambda v_F k \right) \left( z + \lambda ' v_F |\k-\G_j| \right)} \right. \nonumber \\
    & \quad - \left.  e^{-i \psi} (2 \pi)^2 \delta^{(2)} (\k - \k' + \G_j) \frac{1}{4} \sum_{\lambda,\lambda' = \pm} \frac{\left(1+\lambda \frac{\k}{k} \cdot \bm{\sigma} \right) \sigma_z  \left(1+\lambda' \frac{\k+\G_j}{|\k + \G_j|} \cdot \bm{\sigma} \right)}{ \left( z + \lambda v_F k \right) \left( z + \lambda ' v_F |\k+\G_j| \right)}   \right) \nonumber \\
    & \quad\bra{\k} \hat{G}_0(z) e v_F \mathbf{A}^{\rm{eff}}(\br) \cdot \bm{\sigma} \hat{G}_0(z) \ket{\k'} \\
    &=  V_1 \sum_{j=1}^{3}  \cos \frac{2(j+1)\pi}{3} \left( e^{i \psi} (2 \pi)^2 \delta^{(2)} (\k - \k' - \G_j) \frac{1}{4} \sum_{\lambda,\lambda' = \pm} \frac{\left(1+\lambda \frac{\k}{k} \cdot \bm{\sigma} \right) \sigma_x  \left(1+\lambda' \frac{\k-\G_j}{|\k - \G_j|} \cdot \bm{\sigma} \right)}{ \left( z + \lambda v_F k \right) \left( z + \lambda ' v_F |\k-\G_j| \right)} \right. \nonumber \\
    & \quad + \left.  e^{-i \psi} (2 \pi)^2 \delta^{(2)} (\k - \k' + \G_j) \frac{1}{4} \sum_{\lambda,\lambda' = \pm} \frac{\left(1+\lambda \frac{\k}{k} \cdot \bm{\sigma} \right) \sigma_x  \left(1+\lambda' \frac{\k+\G_j}{|\k + \G_j|} \cdot \bm{\sigma} \right)}{ \left( z + \lambda v_F k \right) \left( z + \lambda ' v_F |\k+\G_j| \right)}   \right) \nonumber \\
    &\quad + V_1 \sum_{j=1}^{3}  -i \sin \frac{2(j+1)\pi}{3} \left( e^{i \psi} (2 \pi)^2 \delta^{(2)} (\k - \k' - \G_j) \frac{1}{4} \sum_{\lambda,\lambda' = \pm} \frac{\left(1+\lambda \frac{\k}{k} \cdot \bm{\sigma} \right) \sigma_y  \left(1+\lambda' \frac{\k-\G_j}{|\k - \G_j|} \cdot \bm{\sigma} \right)}{ \left( z + \lambda v_F k \right) \left( z + \lambda ' v_F |\k-\G_j| \right)} \right. \nonumber \\
    & \quad - \left.  e^{-i \psi} (2 \pi)^2 \delta^{(2)} (\k - \k' + \G_j) \frac{1}{4} \sum_{\lambda,\lambda' = \pm} \frac{\left(1+\lambda \frac{\k}{k} \cdot \bm{\sigma} \right) \sigma_y \left(1+\lambda' \frac{\k+\G_j}{|\k + \G_j|} \cdot \bm{\sigma} \right)}{ \left( z + \lambda v_F k \right) \left( z + \lambda ' v_F |\k+\G_j| \right)}   \right) \nonumber
\end{align}
Then, the contour integral can be done:
\begin{align}
    \oint_\mathcal{C} \frac{dz}{2 \pi i} \bra{\br} \hat{G}_0(z) \ket{\br'} = \int_{ E_c ' < v_F k \leq E_c} \frac{d^2 k}{(2 \pi)^2} e^{i \k \cdot (\br - \br ')} \frac{\k}{k} \cdot \bm{\sigma}
\end{align}
\begin{align}
   & \quad \oint_\mathcal{C} \frac{dz}{2 \pi i} \bra{\br} \hat{G}_0(z) V^{\rm{eff}}(\br) \hat{G}_0(z) \ket{\br'} \\
   &= -\frac{V_1}{8}\int_{ E_c ' < v_F k \leq E_c} \frac{d^2 k}{(2 \pi)^2} e^{i \k \cdot (\br - \br ')} \sum_{j=1}^{3} \left(e^{i \G_j \cdot \br ' + i \psi}\, \mathcal{I}_0(\k,-\G_j) +  e^{- i \G_j \cdot \br ' - i \psi}\, \mathcal{I}_0(\k,\G_j) \right)  \nonumber \\
   &\quad \oint_\mathcal{C} \frac{dz}{2 \pi i} \bra{\br} \hat{G}_0(z) M^{\rm{eff}}(\br) \sigma_z \hat{G}_0(z) \ket{\br'}  \\
   &= \frac{\sqrt{3}V_1}{8i}\int_{ E_c ' < v_F k \leq E_c} \frac{d^2 k}{(2 \pi)^2} e^{i \k \cdot (\br - \br ')} \sum_{j=1}^{3} \left(e^{i \G_j \cdot \br ' + i \psi}\, \mathcal{I}_z(\k,-\G_j) -  e^{- i \G_j \cdot \br ' - i \psi}\, \mathcal{I}_z(\k,\G_j) \right)  \nonumber \\
   & \oint_\mathcal{C} \frac{dz}{2 \pi i} \bra{\br} \hat{G}_0(z) e v_F \mathbf{A}^{\rm{eff}}(\br) \cdot \bm{\sigma} \hat{G}_0(z) \ket{\br'} \\
   &= \frac{V_1}{4}\int_{ E_c ' < v_F k \leq E_c} \frac{d^2 k}{(2 \pi)^2} e^{i \k \cdot (\br - \br ')} \sum_{j=1}^{3} \cos \frac{2(j+1)\pi}{3} \left(e^{i \G_j \cdot \br ' + i \psi}\, \mathcal{I}_x(\k,-\G_j) +  e^{- i \G_j \cdot \br ' - i \psi}\, \mathcal{I}_x(\k,\G_j) \right)  \nonumber \\
   &\quad + \frac{V_1}{4i}\int_{ E_c ' < v_F k \leq E_c} \frac{d^2 k}{(2 \pi)^2} e^{i \k \cdot (\br - \br ')} \sum_{j=1}^{3} \cos \frac{2(j+1)\pi}{3} \left(e^{i \G_j \cdot \br ' + i \psi}\, \mathcal{I}_y(\k,-\G_j) -  e^{- i \G_j \cdot \br ' - i \psi}\, \mathcal{I}_y(\k,\G_j) \right)  \nonumber
\end{align}
with
\begin{align}
    \mathcal{I}_0(\k,\G) &= \frac{4}{v_F k + v_F |\k + \G|} \left( 1 - \frac{\k \cdot (\k + \G)}{k|\k + \G|} - \frac{i \sigma_z (\k \times \G) \cdot \bm{\hat{z}}}{k|\k + \G|} \right) \\
    \mathcal{I}_z(\k,\G) &= \frac{4 \sigma_z}{v_F k + v_F |\k + \G|} \left( 1 + \frac{\k \cdot (\k + \G)}{k|\k + \G|} + \frac{i \sigma_z (\k \times \G) \cdot \bm{\hat{z}}}{k|\k + \G|} \right) \\
    \mathcal{I}_{x,y}(\k,\G) &= \frac{4}{v_F k + v_F |\k + \G|} \left( \sigma_{x,y} + \frac{\pm(k_x^2 -k_y^2) \sigma_{x,y} + 2 k_x k_y \sigma_{y,x}}{k^2}\right)
\end{align}

For interlayer terms, we have
\begin{align}
    \hat{G}_0 (z) H^{0,\mu}_{\rm{inter}} \hat{G}_0 (z) = \begin{pmatrix}
        0 & \hat{g}_0 (z+\Delta_D) (h^{0,\mu}_{\rm{inter}})^\dagger \hat{g}_0 (z) \\
        \hat{g}_0 (z+\Delta_D) h^{0,\mu}_{\rm{inter}} \hat{g}_0 (z) & 0
    \end{pmatrix}.
\end{align}
It is practical to write 
\begin{align}
    h^{0,\mu=+}_{\rm{inter}} = v_\perp (k_x + i k_y) + t_\perp \frac{\sigma_x + i \sigma_y}{2} + v_F( k_x - i k_y) \frac{\sigma_x - i \sigma_y}{2}.
\end{align}
We proceed in the same way to get
\begin{align}
    \bra{\k} \hat{g}_0(z+\Delta_D) h^{0,\mu}_{\rm{inter}} \hat{g}_0(z) \ket{\k'} &=  (2 \pi)^2 \delta^{(2)} (\k - \k') v_\perp ( k_x + i k_y) \frac{1}{4} \sum_{\lambda,\lambda' = \pm} \frac{\left(1+\lambda \frac{\k}{k} \cdot \bm{\sigma} \right) \left(1+\lambda' \frac{\k}{k} \cdot \bm{\sigma} \right)}{ \left( z + \lambda v_F k + \Delta_D \right)  \left( z + \lambda ' v_F k \right)} \\
    &\quad + (2 \pi)^2 \delta^{(2)} (\k - \k') \frac{1}{4} \sum_{\lambda,\lambda' = \pm} \frac{1+\lambda \frac{\k}{k} \cdot \bm{\sigma}}{ z + \lambda v_F k + \Delta_D} \begin{pmatrix}
        0 & t_\perp \\
        0 & 0
    \end{pmatrix} \frac{1+\lambda' \frac{\k}{k} \cdot \bm{\sigma}}{z + \lambda ' v_F k} \\
    &\quad + (2 \pi)^2 \delta^{(2)} (\k - \k') \frac{1}{4} \sum_{\lambda,\lambda' = \pm} \frac{1+\lambda \frac{\k}{k} \cdot \bm{\sigma}}{ z + \lambda v_F k + \Delta_D} \begin{pmatrix}
        0 & 0 \\
        v_F( k_x - i k_y) & 0
    \end{pmatrix} \frac{1+\lambda' \frac{\k}{k} \cdot \bm{\sigma}}{z + \lambda ' v_F k}
\end{align}
Then, we compute the contour integral:
\begin{align}
    \quad &\oint_\mathcal{C} \frac{dz}{2 \pi i} \bra{\br} \hat{g}_0(z+\Delta_D) h^{0,\mu}_{\rm{inter}} \hat{g}_0(z) \ket{\br'} \\
    &= \int_{ E_c ' < v_F k \leq E_c} \frac{d^2 k}{(2 \pi)^2} e^{i \k \cdot (\br - \br ')} \frac{1}{2 v_F k} \left( \begin{pmatrix}
        0 & t_\perp \\
        v_F( k_x - i k_y) & 0
    \end{pmatrix}  - \frac{\k \cdot \bm{\sigma}} {k} \begin{pmatrix}
        0 & t_\perp \\
        v_F( k_x - i k_y) & 0
    \end{pmatrix} \frac{\k \cdot \bm{\sigma}} {k} \right) \nonumber
 \end{align}
where we keep the leading order term and neglect the influence of $\Delta_D$, which is a second order effect $\mathcal{O}(\Delta_D^2/E_c'^2)$.

\subsection{Renormalization group flow equations}

Now we only have to insert the previous results into the second term in Eq.~\eqref{eq:vint_RG} to derive the RG equations for all the parameters. Let us compute first the integral for $\bra{\br} \hat{G}_0(z) \ket{\br'}$. After writing the 2D Coulomb potential in Fourier space $\widetilde{V}_{\text{2D}}(\q) = e^2 / 2 \epsilon_0 \epsilon_r q$, we have
\begin{align*}
      &\quad \frac{1}{2} \int d^2 \br d^2 \br ' V_c(\br - \br ') \oint_\mathcal{C} \hat{\psi}^{< \dagger} (\br) \frac{dz}{2 \pi i} \bra{\br} \hat{G}_0(z) \ket{\br'} \hat{\psi}^{<} (\br ') \\
      &= \int \frac{d^2 q}{(2 \pi)^2} \hat{\widetilde{\psi}}^{< \dagger} (\q) \underbrace{\left( \int_{ E_c ' < v_F k \leq E_c} \frac{d^2 k}{(2 \pi)^2} \frac{e^2}{4 \epsilon_0 \epsilon_r |\q-\bk|} \frac{\bk}{k} \cdot \bm{\sigma} \right)}_{\text{(A)}} \hat{\widetilde{\psi}}^{<} (\q) 
\end{align*}
with $\hat{\widetilde{\psi}}^{<} (\q)$ is the Fourier transform of $\hat{\psi}^{<} (\br)$. Since $v_F q \ll E_c'$, we can Taylor expand (A) in terms of $q/k$. The leading order reads
\begin{equation}
    \text{(A)} =  \frac{e^2}{16 \pi \epsilon_0 \epsilon_r} \log \left( \frac{E_c}{E_c'} \right) \,\q \cdot \bm{\sigma}.
\end{equation}
Therefore, the RG equation reads
\begin{equation}
    \frac{d v_F}{ d \log E_c} = - \frac{e^2}{16 \pi \epsilon_0 \epsilon_r}.
\end{equation}
Actually, we find the famous result of the Fermi velocity renormalization in graphene due to the e-e interactions. 

In the same way, we calculate the integral for all the other parameters. It turns out that the results are deeply connected to the matrix structure and the $\k$-dependence. This can be seen in the form of $\mathcal{I}_{\alpha}$ with $\alpha=0,x,y,z$. The integral over $\k$ of $\mathcal{I}_{\{0,x,y,z\}}$ gives, to the first order, $\{0,\sigma_{x}/ v_F,\sigma_{y}/ v_F, 2\sigma_{z}/ v_F\}$. If there is additional $(k_x,k_y)$-dependence, the angle integration will give zero. Therefore, retaining only the first order term, the rest of RG flow equations are  
\begin{subequations}
    \begin{align}
        \frac{d V^{\rm{eff}}}{d \log E_c} &= 0 \\
        \frac{d M^{\rm{eff}}}{d \log E_c} &= - 2 \times \frac{e^2}{16 \pi \epsilon_0 \epsilon_r v_F} \\
        \frac{d (e v_F \mathbf{A}^{\rm{eff}})}{d \log E_c} &= - \frac{e^2}{16 \pi \epsilon_0 \epsilon_r v_F} \\
        \frac{d t_\perp}{d \log E_c} &= - \frac{e^2}{16 \pi \epsilon_0 \epsilon_r v_F} \\
        \frac{d v_\perp}{d \log E_c} &= 0
    \end{align}
    \label{eq:RG_flow}
\end{subequations}

In our numerical study of e-e interactions, we use the renormalized parameters in the Hartree-Fock calculations, where we have to take a cut-off $n_{\text{cut}}$ to the number of bands, to include the contributions from the higher energy bands outside the cut-off. Technically, we use
\begin{subequations}
    \begin{align}
        v_F^* &= v_F^0 \left(1 + \frac{e^2}{16 \pi \epsilon_0 \epsilon_r v_F^0} \log \left( \frac{L_s}{n_{\text{cut}} a_0} \right) \right) \\
        M^{\rm{eff},*} &= M^{\rm{eff}} \left(1 + \frac{e^2}{16 \pi \epsilon_0 \epsilon_r v_F^0} \log \left( \frac{L_s}{n_{\text{cut}} a_0} \right) \right)^2 \\
        t_\perp^* &= t_\perp \left(1 + \frac{e^2}{16 \pi \epsilon_0 \epsilon_r v_F^0} \log \left( \frac{L_s}{n_{\text{cut}} a_0} \right) \right) \\
        e v_F^* \mathbf{A}^{\rm{eff},*} &= e v_F^0 \mathbf{A}^{\rm{eff}} \left(1 + \frac{e^2}{16 \pi \epsilon_0 \epsilon_r v_F^0} \log \left( \frac{L_s}{n_{\text{cut}} a_0} \right) \right).
    \end{align}
    \label{eq:RG_eqs}
\end{subequations}
Here, the ratio $L_s/n_{\text{cut}} a_0$ plays the role of $E_c/E_c'$. In the Hartree-Fock calculations, we use $n_{\text{cut}}=3$ taking into account 3 valence and 3 conduction band per valley per spin. In other words, we stop our RG process at $n_{\text{cut}} v_F^0 |\G_i| \approx 1\,\eV $ much greater than all the energy scales in Eq.~\eqref{eq:Htot_all} so that the perturbative calculations are safe. However, we should admit that in the Hartree-Fock calculations, we extrapolate Eqs.~\eqref{eq:RG_eqs} to a smaller energy window $E_c^* \sim 150\,$meV $\sim \mathcal{O}(t_\perp)$. Nevertheless, the small prefactor $e^2/16 \pi \epsilon_0 \epsilon_r v_F^0 \sim 0.1$ and the $\log$-dependence guarantee that our perturbative RG equations Eqs.~\eqref{eq:RG_eqs} are still good approximations. 

\subsection{Comparison of particle-hole symmetry breaking term with $E_c^*$}
We have mentioned previously that the RG process requires particle-hole (p-h) symmetry of the Hamiltonian. However, several terms in Eq.~\eqref{eq:Htot_all} break p-h symmetry, which is represented by $l_x \sigma_z$ in the layer-sublattice space:
\begin{align}
    l_x = \begin{pmatrix}
        0 & 0 & 0 & 0 & 1 \\
        0 & 0 & 0 & 1 & 0 \\
        0 & 0 & 1 & 0 & 0 \\
        0 & 1 & 0 & 0 & 0 \\
        1 & 0 & 0 & 0 & 0
    \end{pmatrix}
\end{align} 
where $l_x$ inverts the stacking order of pentalayer graphene and $\sigma_z$ flips the sign of $\sigma_{x,y}$. Let us list all the terms in the Hamiltonian Eq.~\eqref{eq:Htot_all} that breaks p-h symmetry and the energy scale of p-h asymmetry they can bring:
\begin{itemize}
    \item $v_\perp (\mu k_x + i k_y)$ in the diagonal part of $h^{0,\mu}_{\rm{inter}}$: $v_\perp |\G_i| \sim 20$\,meV
    \item $V^{\rm{eff}}$: $ V_1^2 /v_F^0 |\G_i| \sim 0.005$\,meV
    \item $M^{\rm{eff}}$: $ 3 V_1^2 /v_F^0 |\G_i| \sim 0.015$\,meV
    \item $A^{\rm{eff}}$: $ 4 V_1^2 /v_F^0 |\G_i| \sim 0.02$\,meV
    \item Non-linearity in layer-resolved on-site energy: $\sim 1$\,meV
\end{itemize} 
We see that the most important term of the diagonal part of $h^{0,\mu}_{\rm{inter}}$, which is far smaller than $n_{\text{cut}} v_F^0 |\G_i| \approx 1\,\eV$. We have also numerically checked that for $\Delta_D \sim 40$ meV, the p-h asymmetry is indeed around 20\,meV at the energy scale $n_{\text{cut}} v_F^0 |\G_i|$. This concludes our RG derivations.

\section{Hartree-Fock approximations to electron-electron interactions}
\label{sec:HF}
We consider the Coulomb interactions
\begin{equation}
\hat{V}_\text{ee}=\frac{1}{2}\int d^2 r  d^2 r' \sum _{\sigma, \sigma '} \hat{\psi}_\sigma ^{\dagger}(\br)\hat{\psi}_{\sigma '}^{\dagger}(\br ') V_c (|\br -\br '|) \hat{\psi}_{\sigma '}(\br ') \hat{\psi}_{\sigma}(\br)
\label{eq:coulomb}
\end{equation}
where $\hat{\psi}_{\sigma}(\br)$ is real-space electron annihilation operator at $\br$ with spin $\sigma$. This interaction can be written as 
\begin{equation}
\hat{V}_\text{ee}=\frac{1}{2}\sum _{i i' j j'}\sum_{l l' m m'}\sum _{\alpha \alpha '\beta \beta ' }\sum _{\sigma \sigma '} \hat{c}^{\dagger}_{i, \sigma l \alpha}\hat{c}^{\dagger}_{i', \sigma ' l' \alpha '} V^{\alpha \beta l m \sigma , \alpha ' \beta ' l' m' \sigma '} _{ij,i'j'}\hat{c}_{j', \sigma ' m' \beta '} \hat{c}_{j, \sigma m \beta}\;,
\end{equation}
where
\begin{align}
&V^{\alpha \beta l m \sigma , \alpha ' \beta ' l' m' \sigma '} _{ij,i'j'} \nonumber \\
&=\int d^2 r d^2 r'  V_c (|\br -\br '|) \,\phi ^*_{l,\alpha} (\mathbf{r}-\mathbf{R}_i-\bm{\tau}_{l,\alpha})\,\phi_{m,\beta} (\mathbf{r}-\mathbf{R}_j- \bm{\tau}_{m,\beta}) \phi^*_{l', \alpha  '}(\br-\mathbf{R}_i'-\bm{\tau}_{l' , \alpha '})\phi _{m', \beta  '}(\br-\mathbf{R}_j'-\bm{\tau} _{m', \beta '}) \nonumber \\
&\quad \times \chi ^{\dagger}_\sigma \chi ^{\dagger}_{\sigma '}\chi _{\sigma '}\chi _{\sigma} .
\end{align}
Here $i$, $\alpha$, and $\sigma$ refer to Bravais lattice vectors, layer/sublattice index, and spin index. $\phi$ is Wannier function and $\chi$ is the two-component spinor wave function. We further assume that the "density-density" like interaction is dominant in the system, i.e., $V^{\alpha \beta l m \sigma , \alpha ' \beta ' l' m' \sigma '} _{ij,i'j'}\approx V^{\alpha \alpha l l \sigma , \alpha ' \alpha ' l' l' \sigma '} _{ii,i'i'}\equiv V_{i \sigma l \alpha  ,i' \sigma ' l' \alpha '}$, then the Coulomb interaction is simplified to
\begin{align}
\hat{V}_\text{ee}=&\frac{1}{2}\sum _{i i'}\sum _{\alpha \alpha ', l l'}\sum _{\sigma \sigma '}\hat{c}^{\dagger}_{i, \sigma l \alpha}\hat{c}^{\dagger}_{i', \sigma' l' \alpha} V_{i\sigma l \alpha, i' \sigma ' l' \alpha '}\hat{c}_{i', \sigma ' l' \alpha '}\hat{c}_{i, \sigma l \alpha} \nonumber \\
\approx&\frac{1}{2}\sum _{i l \alpha \neq i' l' \alpha '}\sum _{\sigma \sigma '}\hat{c}^{\dagger}_{i, \sigma l \alpha} \hat{c}^{\dagger}_{i', \sigma ' l' \alpha '} V_{i l \alpha,i' l' \alpha '}\hat{c}_{i', \sigma ' l' \alpha '}\hat{c}_{i, \sigma l \alpha} \nonumber 
\end{align}
Here we neglect on-site Coulomb interactions which is at least one order of magnitude weaker than long-ranged inter-site Coulomb interactions in the context of moir\'e superlattice \cite{zhang_prl2022}. Given that the electron density is low ($10^{12}$~cm$^{-2}$), i.e., a few electrons per supercell, the chance that two electrons meet at the same atomic site is very low. The Coulomb interactions between two electrons are mostly contributed by the inter-site ones. 

In order to model the screening effects to the $e$-$e$ Coulomb interactions, we introduce the Thomas-Fermi screening form of the intralayer Coulomb interactions, whose Fourier transform is expressed as 
\begin{equation}
    V_{ll} (\mathbf{q})=\frac{e^2}{2 \Omega_0 \epsilon_r \epsilon_0 \sqrt{q^2+\kappa^2}}
 \label{eq:V_thomasfermi}
\end{equation}
where $\Omega _0$ is the area of the superlattice's primitive cell and $\kappa = 1/400$\,\AA$^{-1}$. For the Coulomb interactions between electrons from different layers, we use
\begin{equation}
    V_{ll'}(\mathbf{q})=\frac{e^2}{2 \Omega_0 \epsilon_r \epsilon_0 q} e^{-q|l-l'|d_0}
    \label{eq:V_interlayer}
\end{equation}
with $l \neq l'$.

Since we are interested in the low-energy bands, the intersite Coulomb interactions can be divided into the intra-valley term and the inter-valley term. The intervalley term is neglected in our present study because of small moir\'e Brillouin zone. The intra-valley term $\hat{V}^{\text{intra}}$ can be expressed as
\begin{equation}
\hat{V}^{\rm{intra}}=\frac{1}{2N_s}\sum_{\alpha\alpha ', l l'}\sum_{\mu\mu ',\sigma\sigma '}\sum_{\bk \bk ' \bq} V_{l l'}(\bq)\,
\hat{c}^{\dagger}_{\sigma \mu l \alpha}(\bk+\bq) \hat{c}^{\dagger}_{\sigma' \mu ' l' \alpha '}(\bk ' - \bq) \hat{c}_{\sigma ' \mu ' l' \alpha '}(\bk ')\hat{c}_{\sigma \mu l \alpha}(\bk)\;,
\label{eq:h-intra}
\end{equation}
where $N_s$ is the total number of the superlattice's sites. 

The electron annihilation operator can be transformed from the original basis to the band basis:
\begin{equation}
\hat{c}_{\sigma\mu l \alpha}(\bk)\equiv \hat{c}_{\sigma\mu l \alpha \G}(\btk) =\sum_n C_{ \mu l \alpha \mathbf{G},n}(\btk)\,\hat{c}_{\sigma \mu,n}(\btk)\;,
\label{eq:transform}
\end{equation}
where $C_{\mu l \alpha \mathbf{G},n}(\btk)$ is the expansion coefficient in the $n$-th Bloch eigenstate at $\btk$ of valley $\mu$: 
\begin{equation}
\ket{\sigma \mu, n; \btk}=\sum_{l \alpha \mathbf{G}}C_{\mu l \alpha \mathbf{G},n}(\btk)\,\ket{ \sigma, \mu, l, \alpha, \mathbf{G}; \btk }\;.
\end{equation}
We note that the non-interacting Bloch functions are spin degenerate due to the separate spin rotational symmetry ($SU(2)\otimes SU(2)$ symmetry) of each valley. Using the transformation given in Eq.~(\ref{eq:transform}), the intra-valley Coulomb interaction can be written in the band basis
\begin{align}
\hat{V}^{\rm{intra}}&=\frac{1}{2N_s}\sum _{\btk \btk'\btq}\sum_{\substack{\mu\mu' \\ \sigma\sigma'\\l l'}}\sum_{\substack{nm\\ n'm'}} \left(\sum _{\mathbf{Q}}\,V_{ll'} (\mathbf{Q}+\btq)\,\Omega^{\mu l ,\mu' l'}_{nm,n'm'}(\btk,\btk',\btq,\mathbf{Q})\right) \nonumber \\
&\times \hat{c}^{\dagger}_{\sigma\mu,n}(\btk+\btq) \hat{c}^{\dagger}_{\sigma'\mu',n'}(\btk'-\btq) \hat{c}_{\sigma'\mu',m'}(\btk') \hat{c}_{\sigma\mu,m}(\btk)
\label{eq:Hintra-band}
\end{align}
where the form factor $\Omega ^{\mu  l,\mu' l'}_{nm,n'm'}$ are written respectively as
\begin{equation}
\Omega ^{\mu l ,\mu' l'}_{nm,n'm'}(\btk,\btk',\btq,\mathbf{Q})
=\sum _{\alpha\alpha'\mathbf{G}\mathbf{G}'}C^*_{\mu l \alpha\mathbf{G}+\mathbf{Q},n}(\btk+\btq) C^*_{\mu'l'\alpha'\mathbf{G}'-\mathbf{Q},n'}(\btk'-\btq)C_{\mu'l'\alpha'\mathbf{G}',m'}(\btk')C_{\mu l \alpha\mathbf{G},m}(\btk).
\end{equation}

We make Hartree-Fock approximation to Eq.~\eqref{eq:Hintra-band} such that the two-particle Hamiltonian is decomposed into a superposition of the  Hartree and Fock effective single-particle terms, where the Hartree term is expressed as
\begin{equation}
\begin{split}
\hat{V}_H^{\rm{intra}}=&\frac{1}{2N_s}\sum _{\btk \btk'}\sum _{\substack{\mu\mu'\\ \sigma\sigma'\\ ll'}}\sum_{\substack{nm\\ n'm'}}\left(\sum _{\mathbf{Q}} V_{ll'} (\mathbf{Q}) \, \Omega^{\mu l ,\mu' l'}_{nm,n'm'}(\btk,\btk',0,\mathbf{Q})\right)\\
&\times \left(\langle \hat{c}^{\dagger}_{\sigma\mu,n}(\btk)\hat{c}_{\sigma\mu,m}(\btk)\rangle \hat{c}^{\dagger}_{\sigma'\mu',n'}(\btk')\hat{c}_{\sigma'\mu',m'}(\btk') + \langle \hat{c}^{\dagger}_{\sigma'\mu',n'}(\btk')\hat{c}_{\sigma'\mu',m'}(\btk')\rangle \hat{c}^{\dagger}_{\sigma\mu,n}(\btk)\hat{c}_{\sigma\mu,m}(\btk)\right)
\end{split}
\label{eq:hartree}
\end{equation}

The Fock term is expressed as:
\begin{equation}
    \begin{split}
\hat{V}_F^{\rm{intra}}&=-\frac{1}{2N_s}\sum_{\btk \btk'}\sum _{\substack{\mu\mu'\\ \sigma\\ ll'}} \sum_{\substack{nm\\ n'm'}} \left( \sum_{\mathbf{Q}} V_{ll'} (\btk'-\btk+\mathbf{Q}) \, \Omega^{\mu l ,\mu' l'}_{nm,n'm'}(\btk, \btk ', \btk '-\btk,\mathbf{Q})\right)\\
&\times \left(\langle \hat{c}^{\dagger}_{\sigma\mu,n}(\btk')\hat{c}_{\sigma\mu',m'}(\btk')\rangle \hat{c}^{\dagger}_{\sigma\mu',n'}(\btk)\hat{c}_{\sigma\mu,m}(\btk) + \langle \hat{c}^{\dagger}_{\sigma\mu',n'}(\btk)\hat{c}_{\sigma\mu,m}(\btk)\rangle \hat{c}^{\dagger}_{\sigma\mu,n}(\btk')\hat{c}_{\sigma\mu',m'}(\btk')\right)\;.   
\end{split}
\label{eq:fock}
\end{equation}

After integrating out the fast modes from the remote bands, we obtain a low-energy effective model with cutoff $E_C^* \sim 0.15\,$eV. Practically, we keep three valence and three conduction bands per spin per valley ($n_{\rm{cut}}$=3), and project $e$-$e$ interactions to the ``non-interacting" wavefunctions of the renormalized low-energy effective model, which describes the effective low-energy single-particle behavior in the system. The Coulomb interaction effects between the remote-band electrons and the low-energy electrons have been taken into account by perturbative RG as discussed in Sec.~\III, which precisely lead to the renormalization of the continuum model parameters.
However, the $e$-$e$ interactions \textit{within the low-energy window} have not been included yet. This means that we can just normal order the $e$-$e$ interactions with respect to the vacuum of the truncated low-energy Hilbert space  $E_C^*$ ($n_{\textrm{cut}}=3$), which gives Eq.~\eqref{eq:Hintra-band} keeping in mind that the band indices refer to the those of the renormalized low-energy continuum model. The results presented in the main text (Figs.~2-6) are obtained with such a choice of normal ordering.

\section{More results on exact diagonalization calculations}

In this section, we provide more results on exact-diagonalization (ED) calculations associated with Figs.~3-6 in the main text considering $e$-$e$ Coulomb interactions being normal ordered with respect to the vacuum of the low-energy Hilbert space (see Eq.~\eqref{eq:Hintra-band}).

In our numerical implementations, the ED calculations at 2/3 and 1/3 electron doping are performed on a 27-site cluster generated from three sets of 9-site clusters \cite{Fu-27site-prb23,Bernevig-prx11}, and those at 3/5 and 2/5 electron doping are done on a 20-site (4$\times$5) cluster. The spectral-flow calculations of (2/3, 1/3) and (3/5, 2/5) electron fillings are done on 24-site (4$\times$6) and 20-site (4$\times$5) cluster, respectively. The Brillouin zone samplings for the ED calculations are shown in Supplementary Figure~\ref{fig:EDkmesh}. The Coulomb interactions in the ED calculations are identical to those adopted in the HF calculations in the previous section, which are then projected to the isolated flat band right below the chemical potential for HF states at filling 1. As discussed in the main text, there are two competing HF states, which give rise to two types of topologically distinct flat bands, and we have performed ED calculations by hole doping both types of flat bands and finding the global many-body ground state by comparing their many-body total energy.

As Fig.~5 in the main text showing the finer phase diagram for electron doping 2/3, we provide here the phase diagram at 3/5 electron filling (e.q. 2/5 hole filling) within a finer parameter space: $0.9\,\textrm{V/nm}\leq D\leq 1.0\,\textrm{V/nm}$ and $4\leq\epsilon_r\leq 6$ in Supplementary Figure~\ref{fig:fig7}(a). We see that there is a finite domain in the selected parameter space where FCI remains the global many-body ground state. In other words, FCI state is not only the most stable state while hole doping Chern-number-1 HF flat band but also has a total energy lower than that of any other many-body states generated by identically hole doping the topologically trivial HF flat band. More specifically, the system stays in composite-fermion FCI state for $D=0.9\,\rm{V/nm}$ and for $D=0.93\,\rm{V/nm}\!\leq\!D\!\leq\!0.95\,$V/nm (with interval of 0.1\,V/nm) with $\epsilon_r=5$. We have additionally discovered another type of CDW (CDW3) different from the two aforementioned types discussed in the main text, whose total reduced crystalline momenta of the three degenerate ground states are (1/2,0), (1/4,2/5) and (3/4,3/5). In other words, the two degenerate CDW states with crystalline wavevectors $\pm(1/4,2/5)$ are quasi-degenerate with that with wavevector (1/2,0), which may be an artifact due to finite size effects. In Supplementary Figure~\ref{fig:fig7}(b) we present the ``gap/spread" on a logarithmic scale, which indicates from some aspects the stability of the FCI or CDW states. Due to their energy degeneracy, the ``gap/spread" values should peak for the FCI and CDW states, as illustrated in Supplementary Figure~\ref{fig:fig7}(b). 

For the sake of completeness, we provide the many-body band structures of the global many-body ground states at electron doping 2/3, 1/3, 3/5 and 2/5 in Supplementary Figures~\ref{fig:EDband_2_3}-\ref{fig:EDband_2_5}, which correspond to the phase diagrams depicted in Figs.~3(a), 3(c), and Figs.~6(a), 6(c) of the main text, respectively. The band structures in Figs.~\ref{fig:EDband_2_3}-\ref{fig:EDband_2_5} are arranged in a sequential manner from left to right for $D$ field values ranging from 0.07 to 0.11 V/Å, and from bottom to top for $\epsilon_r$ values ranging from 4 to 8. This arrangement matches exactly to that in the phase diagrams presented in the main text.


We also present the many-body band structures of the global many-body ground states in the finer range of $D$ field values at 2/3 and 3/5 electron doping in Supplementary Figure~\ref{fig:EDrealGS_2_3} and Supplementary Figure~\ref{fig:EDrealGS_3_5}, which correspond to the finer phase diagrams in Fig.~5(a) of the main text and in Supplementary Figure~\ref{fig:fig7}, respectively. The arrangement of the band plots is also for the purpose to match exactly to that of the finer phase diagrams.


\begin{figure}[bth!]
\begin{center}
    \includegraphics[width=16cm]{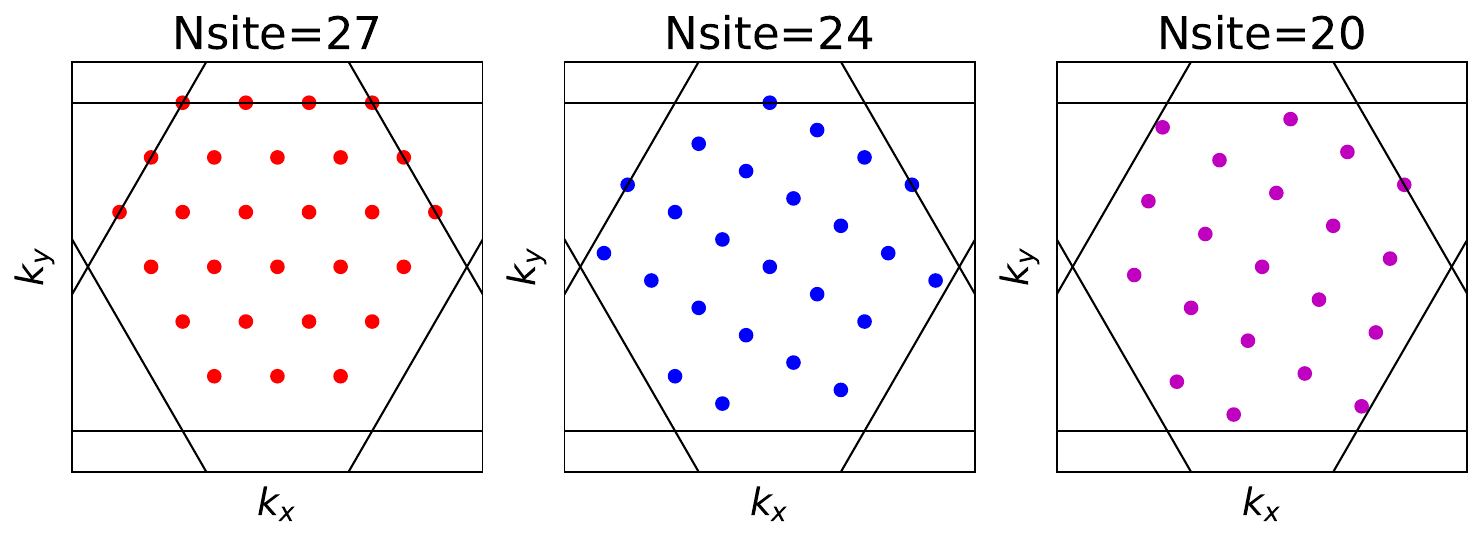}
\caption{
Brillouin zone samplings for the ED calculations.}
\label{fig:EDkmesh}
\end{center}
\end{figure}

\begin{figure}[bth!]
\begin{center}
    \includegraphics[width=16cm]{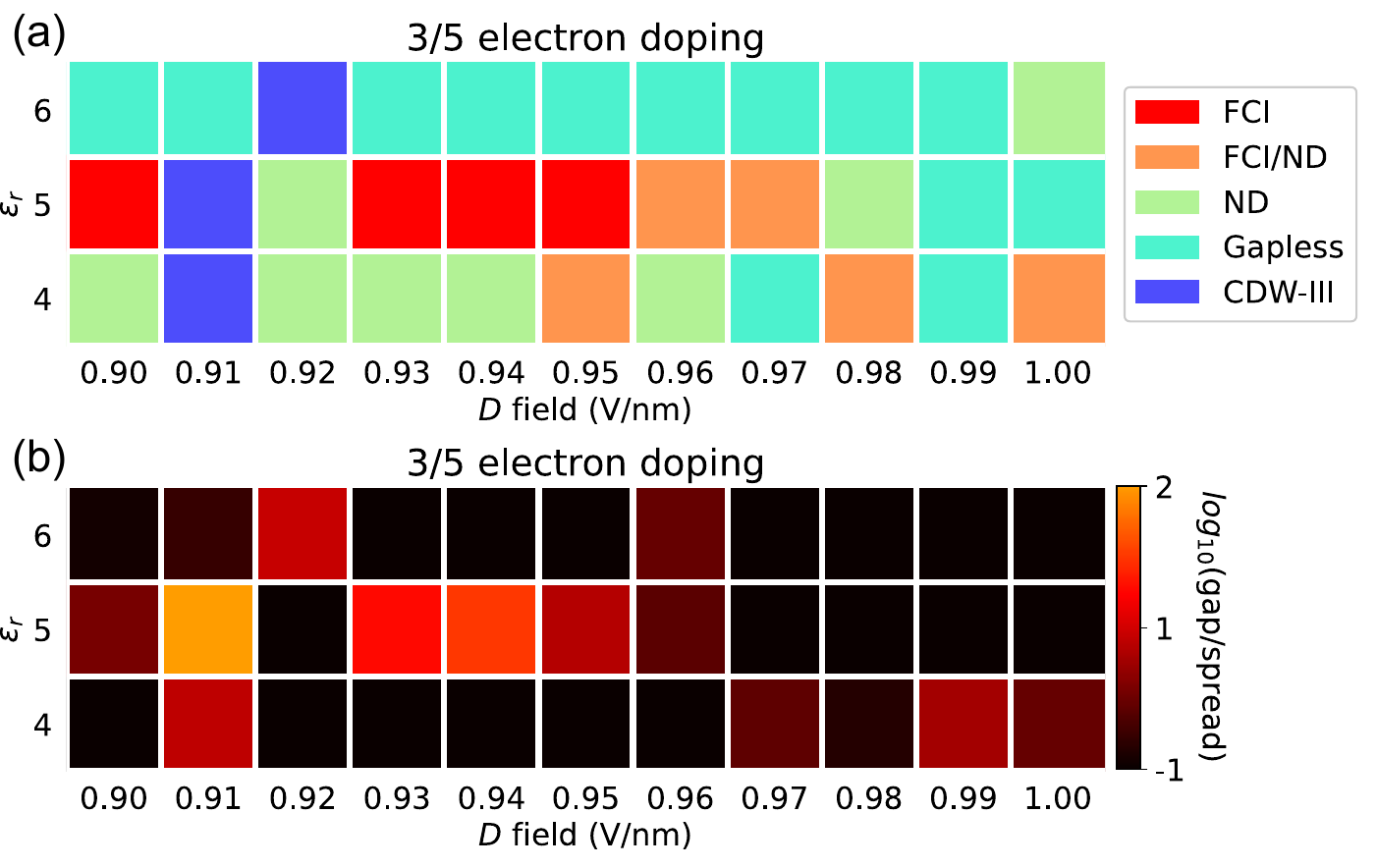}
\caption{(a) Finer phase diagrams in a smaller range of parameters $0.9\textrm{V/nm}\leq D\leq 1.0\,$V/nm and $4\leq\epsilon_r\leq 6$ at 3/5 electron filling. (b) The corresponding ``gap/spread" (see text) at 3/5 electron filling plotted on a logarithmic scale.}
\label{fig:fig7}
\end{center}
\end{figure}

\begin{figure}[bth!]
\begin{center}
    \includegraphics[width=7in]{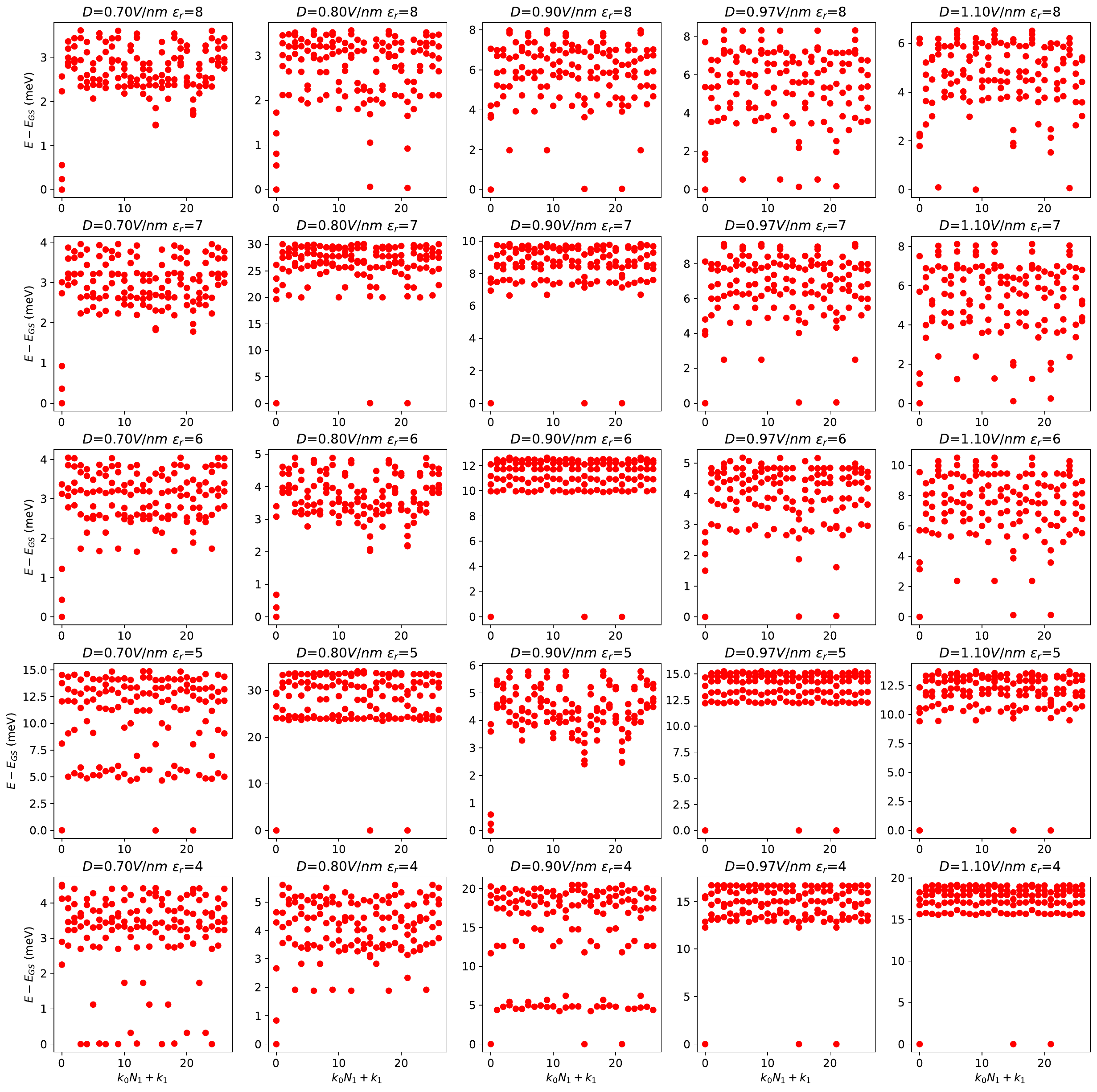}
\caption{
Global many-body ground states at 2/3 electron doping, corresponding to the phase diagram Fig.~3(a) in the main text.}
\label{fig:EDband_2_3}
\end{center}
\end{figure}

\begin{figure}[bth!]
\begin{center}
    \includegraphics[width=7in]{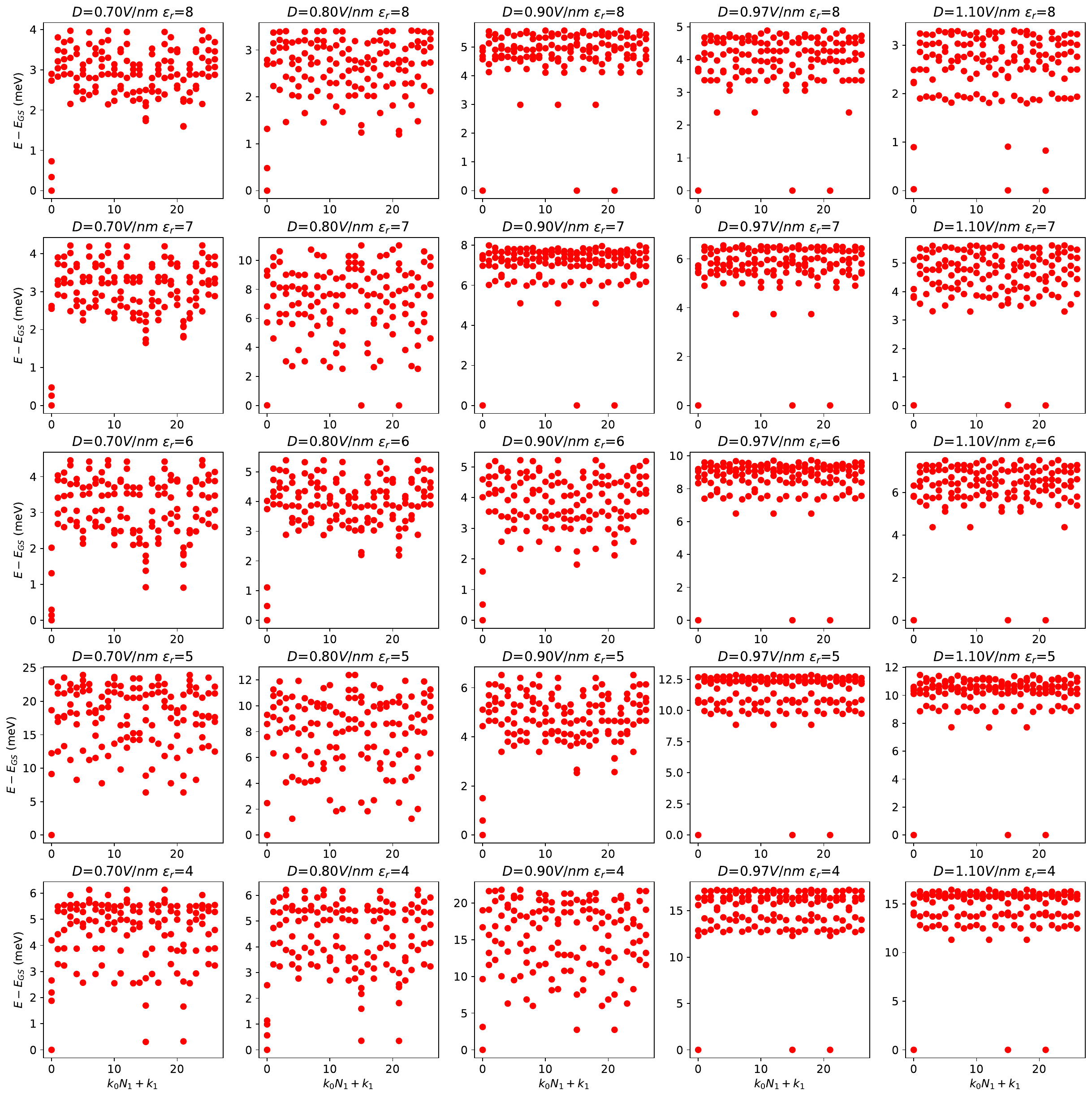}
\caption{
Global many-body ground states at 1/3 electron doping, corresponding to the phase diagram Fig.~3(c) in the main text.}
\label{fig:EDband_1_3}
\end{center}
\end{figure}

\begin{figure}[bth!]
\begin{center}
    \includegraphics[width=7in]{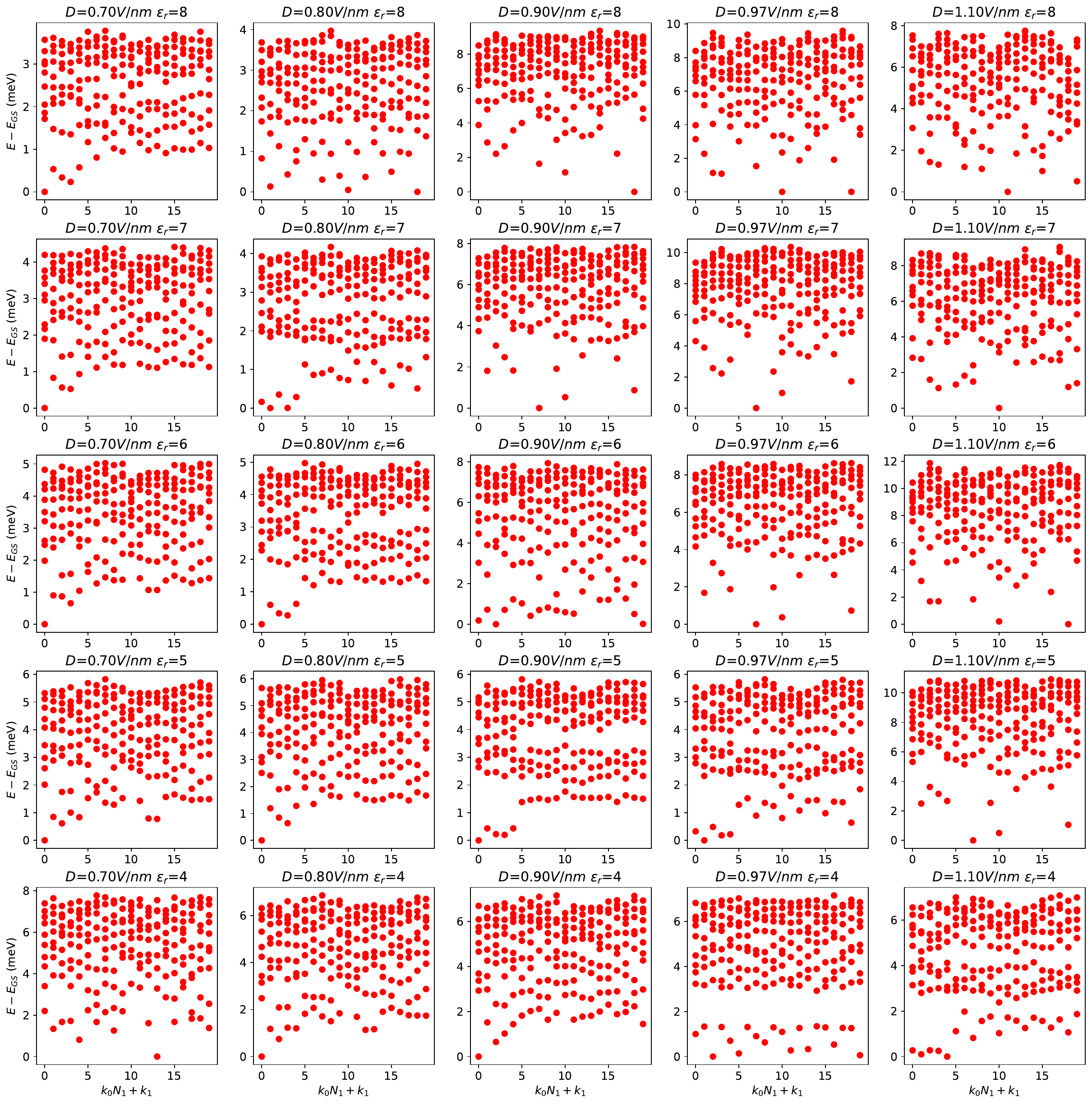}
\caption{
Global many-body ground states at 3/5 electron doping, corresponding to the phase diagram Fig.~6(a) in the main text.}
\label{fig:EDband_3_5}
\end{center}
\end{figure}

\begin{figure}[bth!]
\begin{center}
    \includegraphics[width=7in]{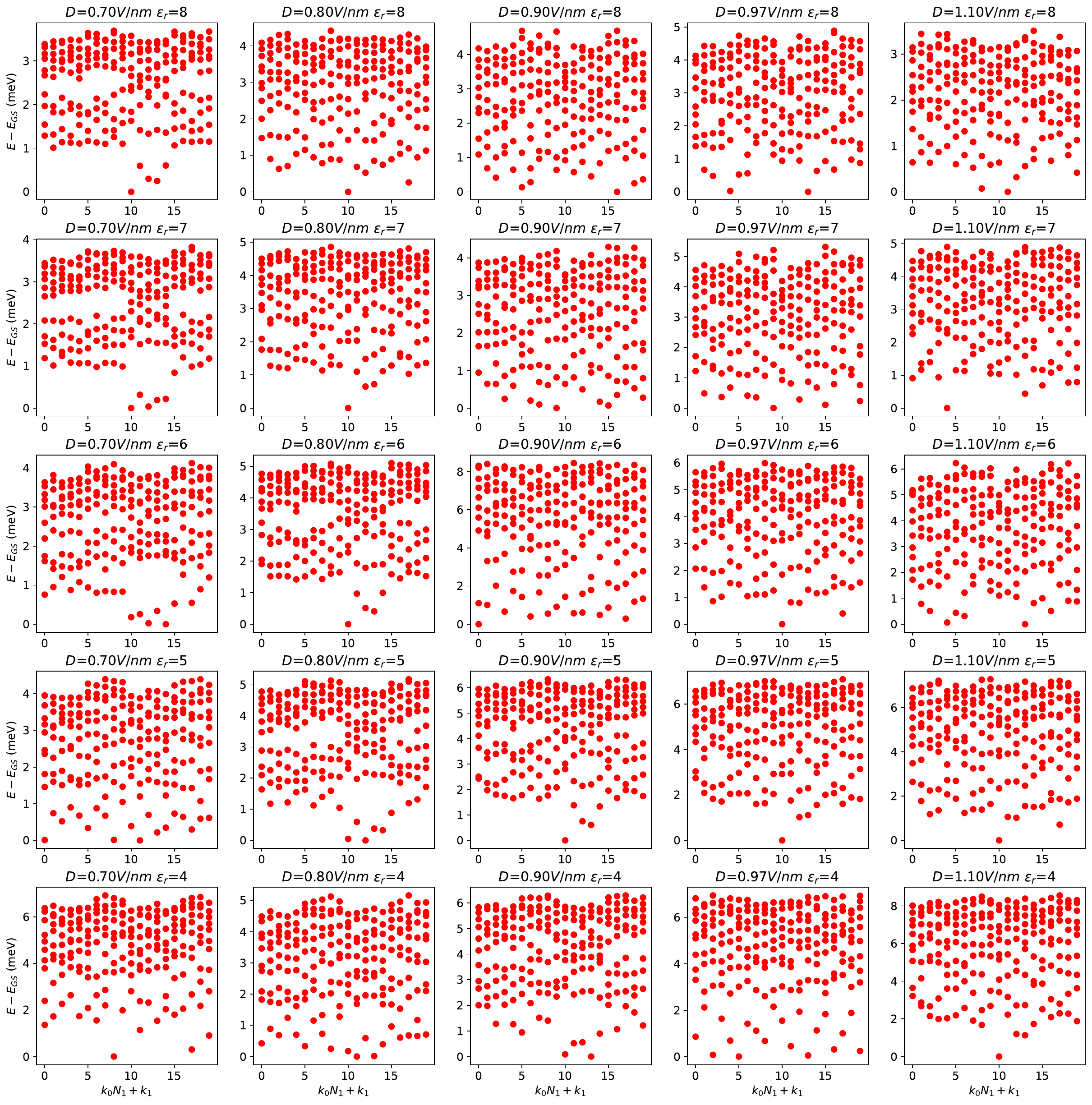}
\caption{
Global many-body ground states at 2/5 electron doping, corresponding to the phase diagram Fig.~6(c) in the main text.}
\label{fig:EDband_2_5}
\end{center}
\end{figure}

\begin{figure}[bth!]
\begin{center}
    \includegraphics[width=7in]{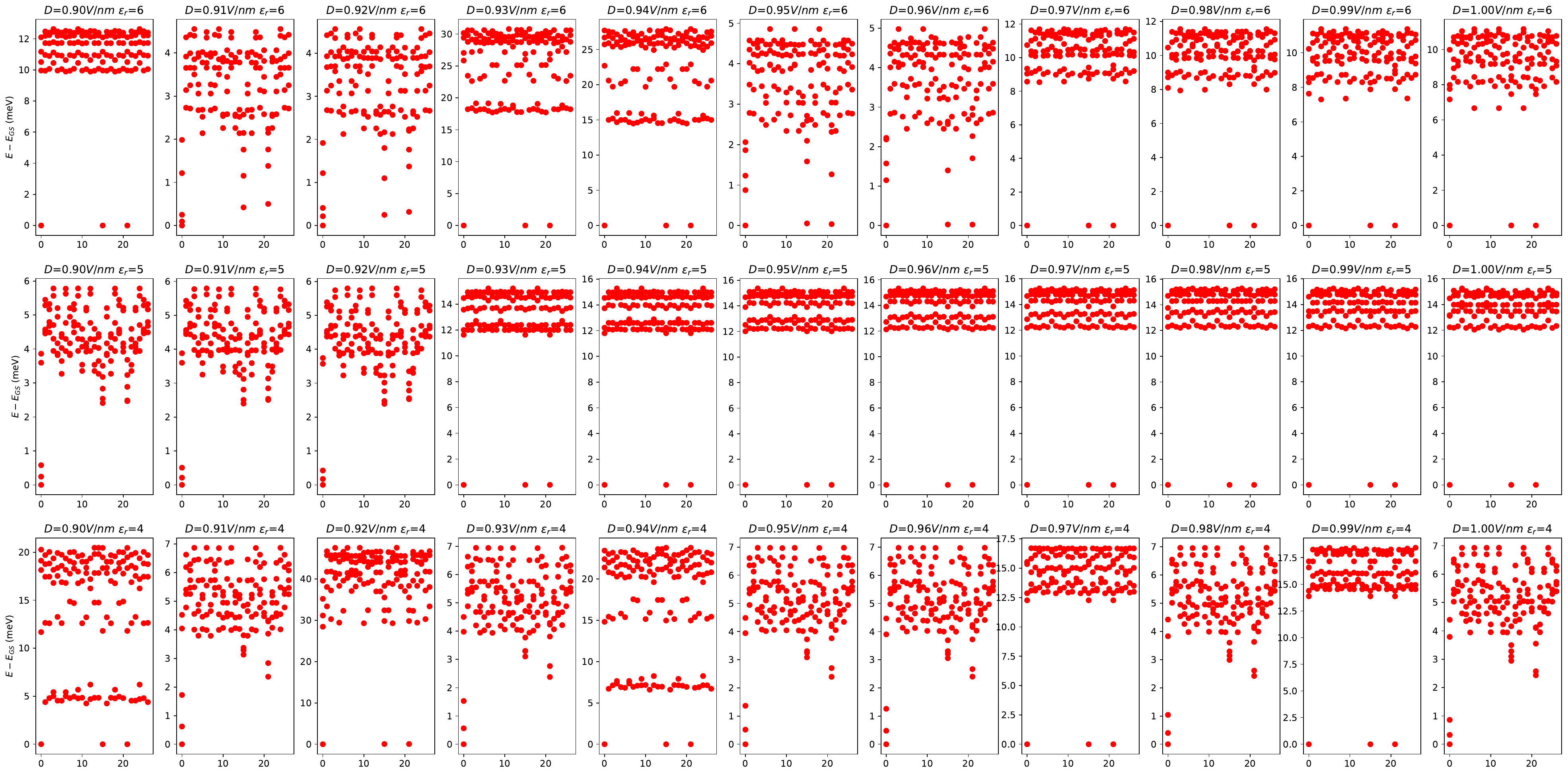}
\caption{
Global many-body ground states at 2/3 electron doping, corresponding to the finer phase diagrams Fig.~5(a) in the main text.}
\label{fig:EDrealGS_2_3}
\end{center}
\end{figure}

\begin{figure}[bth!]
\begin{center}
    \includegraphics[width=7in]{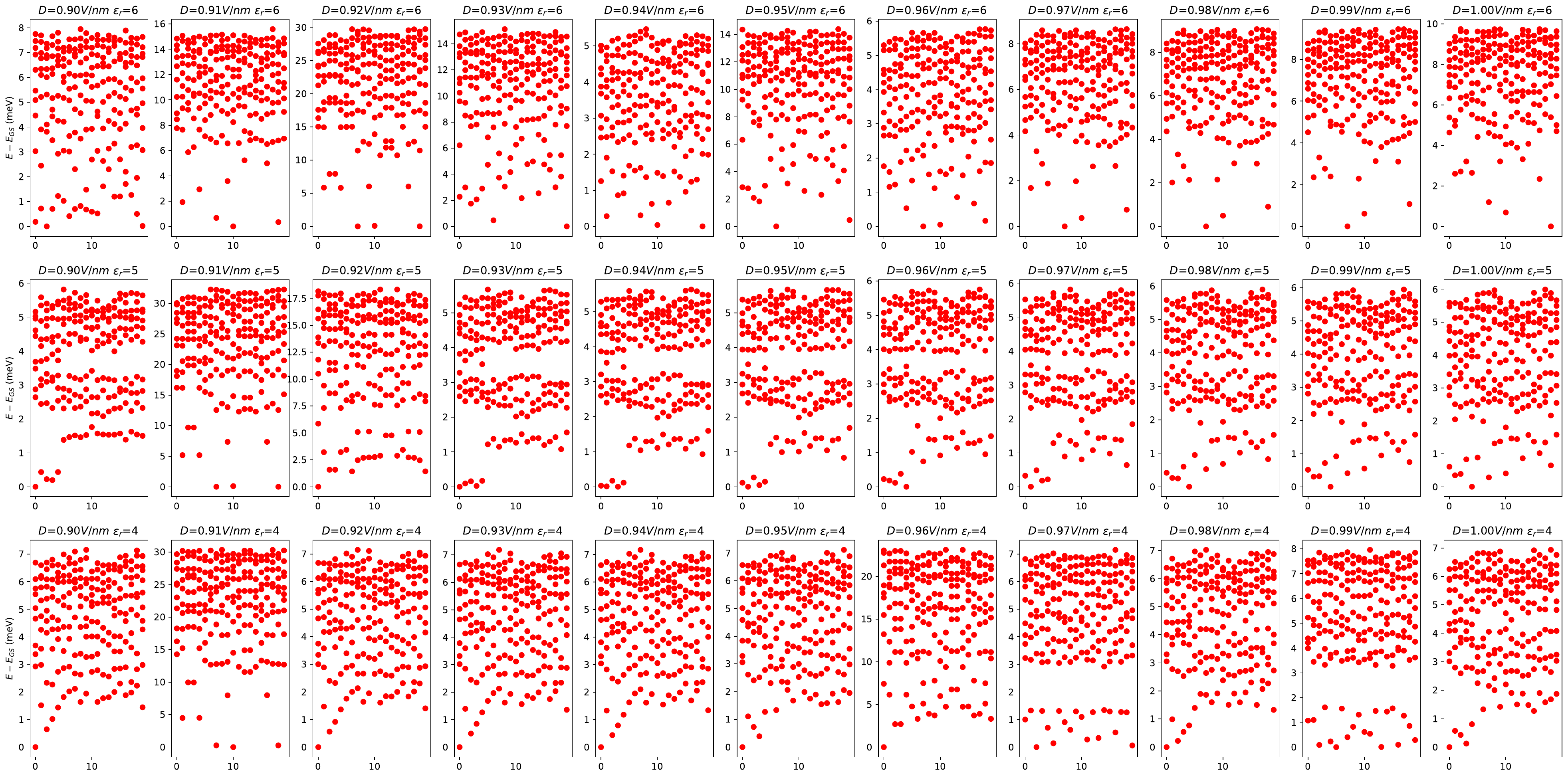}
\caption{
Global many-body ground states at 3/5 electron doping, corresponding to the finer phase diagrams Supplementary Figure~\ref{fig:fig7}.}
\label{fig:EDrealGS_3_5}
\end{center}
\end{figure}




\section{Lattice relaxations}

\begin{figure}[b]
\begin{center}
    \includegraphics[width=12cm]{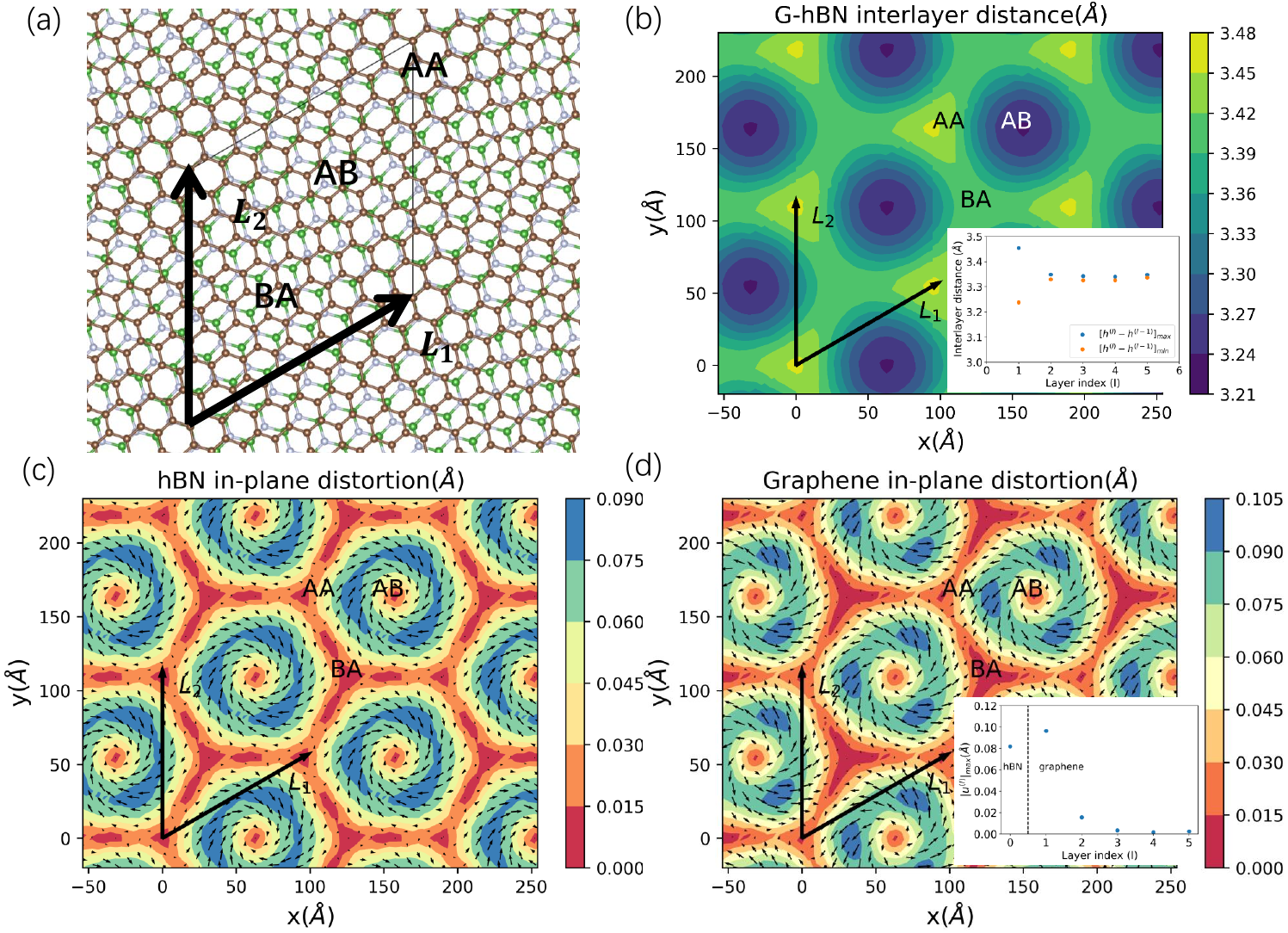}
\caption{
(a) The moir\'e supercell of hBN-pentalayer graphene. $L_{1}$ and $L_{2}$ denote the lattice vectors of the moir\'e supercell. $AA$, $AB$ and $BA$ points are the local commensurate stacking configurations. (b) The real space distribution of the interlayer distance between the hBN layer (0-th layer) and the adjacent graphene layer (1-st layer). The insert represents the maximum and the minimum value of the interlayer distance between the $l$-th layer and the $(l-1)$-th layer. (c) The real space distribution of the in-plane distortion field in the hBN layer (0-th layer). (d) The real space distribution of the in-plane distortion field in the adjacent graphene layer (1-st layer). The arrows depict the directions of the in-plane distortion. The colorbar represents the amplitude the displacement field. The insert of (d) shows the maximum amplitude of the in-plane distortion from the $l$-th layer. The influence of the lattice relaxation effects diminishes rapidly when distanced away from the interface.}
\label{fig:lattice}
\end{center}
\end{figure}

In this section, we study the lattice relaxation pattern and the influence of lattice relaxation effects on the electronic band structures of pentalayer graphene-hBN moir\'e superlattice. 

We start with a commensurate superlattice characterized by a twist angle of $0.77^{\circ}$ between pentalayer graphene and the hBN layer, with a corresponding lattice constant of the superlattice is $109.16\,$\AA, as schematically illustrated in Supplementary Figure~\ref{fig:lattice}(a). The local commensurate stacking configurations, denoted as $AA$ and $AB/BA$, are labeled in Supplementary Figure~\ref{fig:lattice}(a).
The structural relaxation is calculated utilizing Large-scale Atomic-Molecular Massively Parallel Simulation (LAMMPS) \cite{LAMMPS}. The interlayer interactions between adjacent layers are described using the Dispersion Interaction Random Phase (DRIP) potential, while the intralayer potentials are described by the Adaptive Intermolecular Reactive Bond Order (AIREBO) potential for the graphene layer and the Extended Tangent Plane (EXTEP) potential for the hBN layer \cite{G-HBN-MD}. The damped dynamics method is employed in the lattice relaxation calculations. We consider the in-plane distortions from the $l$-th layer, denoted as $\textbf{u}^{(l)}$, and the out-of-plane distortions from the $l$-th layer, denoted as $h^{(l)}$. The results of the lattice relaxation pattern are presented in Supplementary Figure~\ref{fig:lattice}(b)-(d). To be specific, we present the interlayer distance between hBN layer (0-th layer) and the adjacent graphene layer (1-st layer) in Supplementary Figure~\ref{fig:lattice}(b). The interlayer distance reaches a maximum  of $3.45\,$\AA\ at $AA$ point and a minimum  of $3.24\,$\AA\ at $AB$ point. The maximal value and the minimal value of interlayer distances between the $l$-th and ($l-1$)-th layer are presented in the inset. We obtain that the interlayer distance varies dramatically in the graphene/hBN interface, while it is roughly unchanged when moving far away from the interface. We present the real space distribution of the in-plane distortion in the hBN layer ($0$-th) layer in Supplementary Figure~\ref{fig:lattice}(c) and the in-plane distortion in the adjacent graphene layer ($1$-st layer) in Supplementary Figure~\ref{fig:lattice}(d). The arrows show the directions of in-plane components of lattice distortions and the colorbar represents the amplitudes of the distortions. The maximal amplitudes of in-plane distortions are approximately $0.09\,$\AA. The directions of  in-plane distortions exhibit an approximate inversion relationship between hBN layer and the adjacent graphene layer. Besides, the in-plane distortion forms a rotational field around the $AB$ region.  In the insert of Supplementary Figure~\ref{fig:lattice}(d), we present the maximal amplitudes of the in-plane distortions in the $l$-th layer, revealing a rapid decay in the pentalayer graphene as moving away from the interface. The tendency of the lattice relaxation pattern in hBN-pentalayer graphene  agrees well with previous researches \cite{G-HBN-MD,koshino-prb17,shengjun-yuan-npj2022}.
Despite the pronounced influence of lattice distortion on all atoms within the hBN and pentalayer graphene, it is noteworthy that such influence diminishes rapidly within the pentalayer graphene.

\begin{figure}[bth!]
\begin{center}
    \includegraphics[width=12cm]{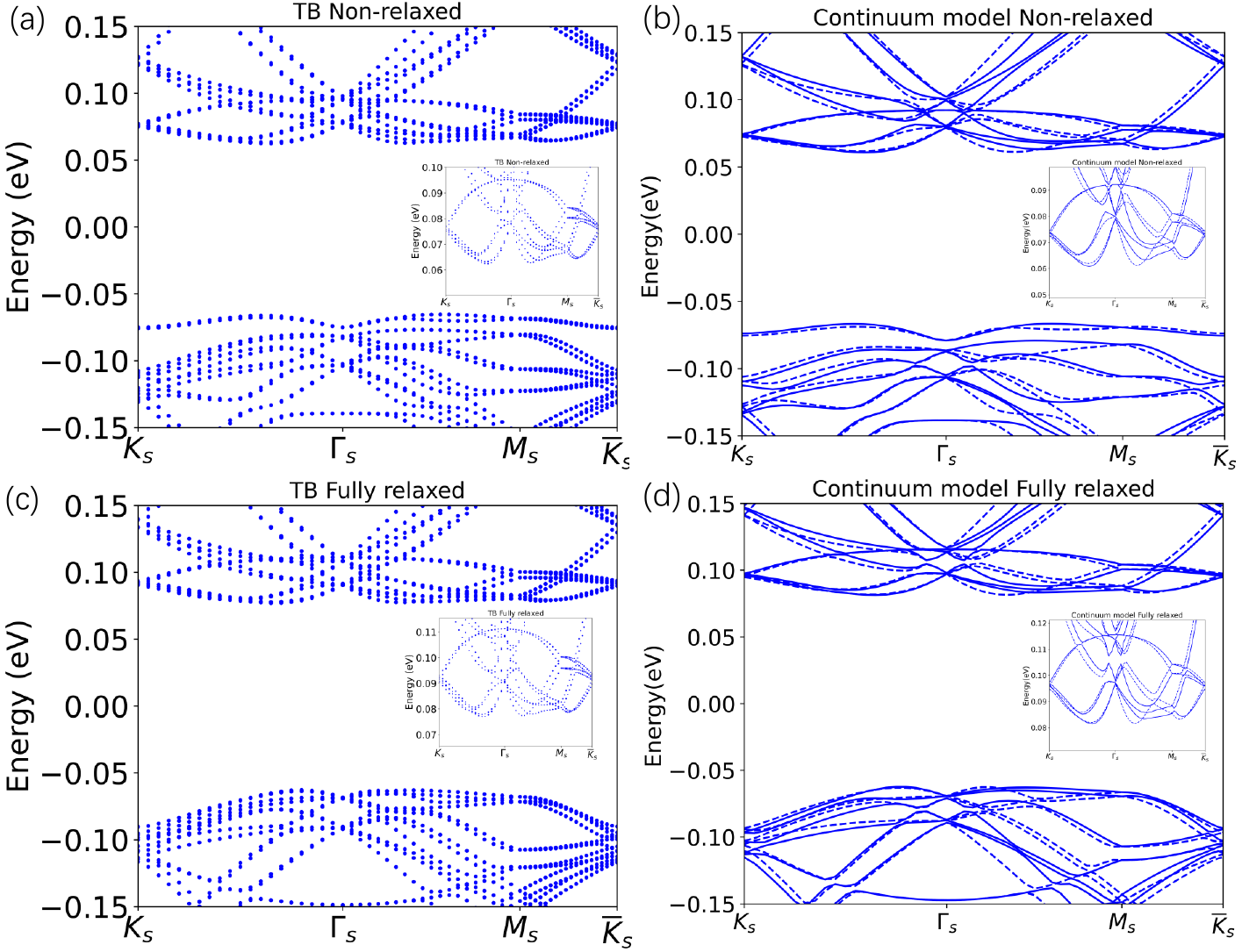}
\caption{
The band structure of hBN-pentalyer graphene with twist angle $\theta=0.77^{\circ}$ and $\Delta_D=65\,$meV. The band structures calculated by the TB model based on (a) the non-relaxed lattice structures and (c) the fully relaxed lattice structures. The band structures calculated by the effective continuum model based on (b) the non-relaxed lattice structures and (d) the fully relaxed lattice structures. In the insert, we preset a zoomed-in view of the lowest conduction band.}
\label{fig:bands}
\end{center}
\end{figure}

\begin{figure}[bth!]
\begin{center}
    \includegraphics[width=7in]{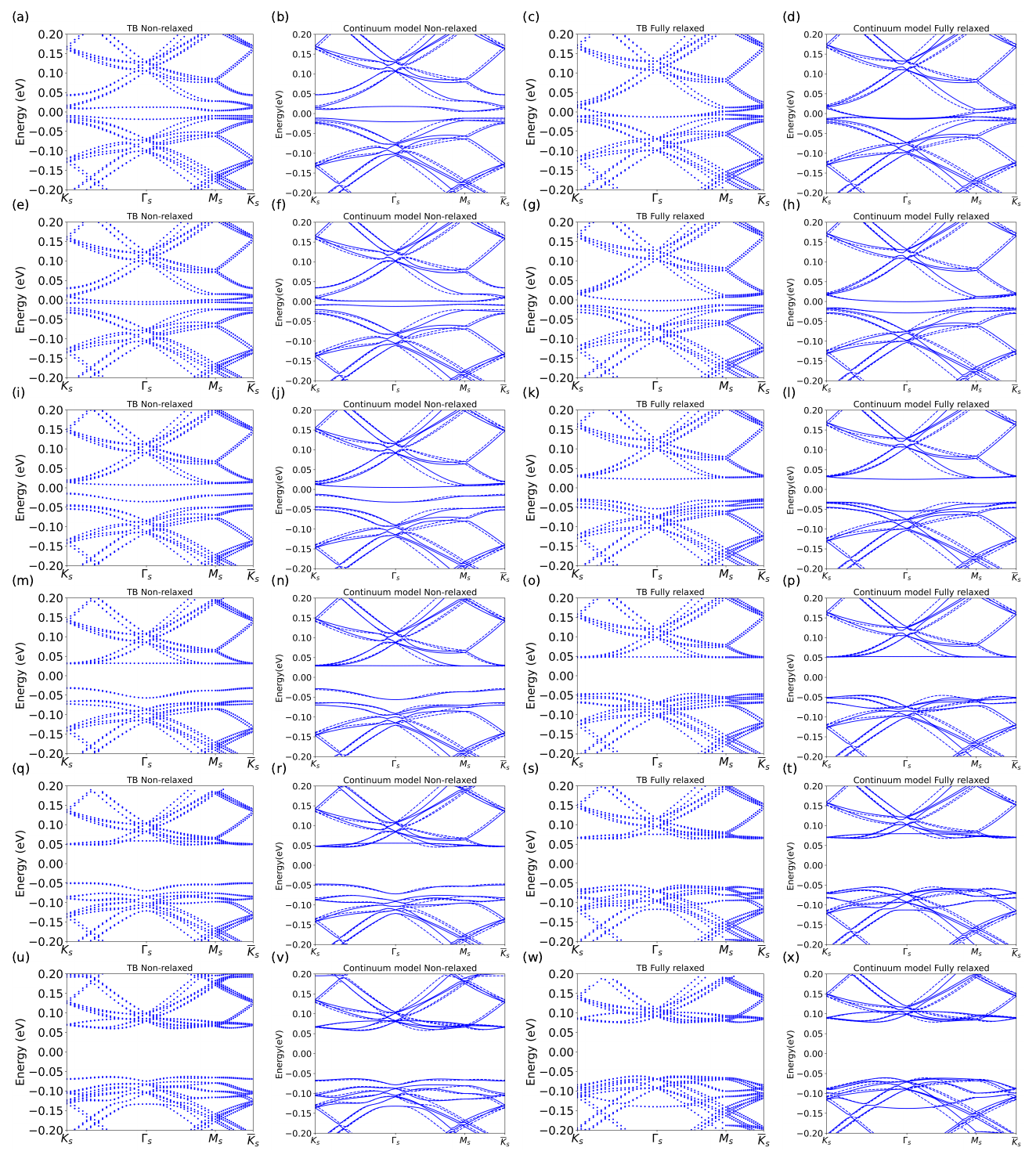}
\caption{
The band structures of hBN-pentalyer graphene with twist angle $\theta=0.77^{\circ}$ at different displacement fields. (a)-(d) $\Delta_{D}=0\,$meV, (e)-(h) $\Delta_{D}=6.7\,$meV, (i)-(l) $\Delta_{D}=20.1\,$meV, (m)-(p) $\Delta_D=33.5\,$meV, (q)-(t) $\Delta_D=47\,$meV, (u)-(x) $\Delta_D=60.3\,$meV.}
\label{fig:morebands}
\end{center}
\end{figure}

We continue to calculate the band structures of hBN-pentalayer graphene utilizing the atomistic Slater-Koster tight-binding (TB) model \cite{koshino-hBN-prb14} based on the fully relaxed lattice structure. To be specific, the hopping amplitude between two orbitals at different sites is expressed as:
\begin{equation}
-t(\textbf{R})=V_{pp\sigma}\left(\frac{\textbf{R}\cdot\textbf{e}_{z}}{\vert\textbf{R}\vert}\right)^{2}+V_{pp\pi}\left[1-\left(\frac{\textbf{R}\cdot\textbf{e}_{z}}{\vert\textbf{R}\vert}\right)^{2}\right],
\end{equation}
where $V_{pp\sigma}\!=\!V^{0}_{pp\sigma}e^{-(\vert\textbf{R}\vert-d_{0})/r_{0}}$ and $V_{pp\pi}\!=\!V^{0}_{pp\pi}e^{-(\vert\textbf{R}\vert-a_{0}/\sqrt{3})/r_{0}}$. $d_{0}\!=\!3.35\,$\AA\ is the interlayer distance between graphene layers. $r_{0}\!=\!0.184a_{0}$. $V^{0}_{pp\sigma}\!=\!0.48$ eV and $V^{0}_{pp\pi}=-2.7\,$eV. The on-site energies for different atoms are:
\begin{equation}
V_{\rm{C}}=0\ \mathrm{eV}, V_{\rm{B}}=3.34\ \mathrm{eV}, V_{\rm{N}}=-1.40\ \mathrm{eV}.
\end{equation}
We present the band structures of hBN-pentalayer graphene based on non-relaxed lattice structures in Supplementary Figure~\ref{fig:bands}(a) and that based on fully relaxed lattice structures in Supplementary Figure~\ref{fig:bands}(c), where a constant interlayer electrostatic potential energy drop $\Delta_D=65\,$meV has been applied.  In the inset, we present a zoomed-in view of the lowest conduction bands.

Next, we study the electronic properties of fully relaxed hBN-pentalyer graphene utilizing the effective continuum model. In previous section, we have derived the continuum Hamiltonian of hBN-pentalayer graphene, where the interlayer moir\'e hopping between the hBN layer and the graphene layer is treated as by second order perturbation, leading to a number of intralayer moir\'e potential terms acted on the bottom graphene layer cloest to hBN. Inspired by previous research \cite{koshino-tbg-epc-prb20,xie-tbg-phonon-prb23}, the lattice relaxation effects would change the position of each atoms and affect both the interlayer moir\'e potential and the intralayer Hamiltonian. For the intralayer Hamiltonian, the strain field acts as a pseudo vector potential coupled with the low-energy Dirac fermions. The pseudo vector potential induced by strain field in the $l$-th layer graphene $\textbf{A}^{(l)}=[A^{(l)}_{x},A^{(l)}_{y}]$ is given by:
\begin{align}
A^{(l)}_{x}=\frac{\beta\gamma_{0}}{e v_{F}}\frac{3}{4}\left(\frac{\partial u^{(l)}_{x}}{\partial x}-\frac{\partial u^{(l)}_{y}}{\partial y}\right),\ \ A^{(l)}_{y}=-\frac{\beta\gamma_{0}}{e v_{F}}\frac{3}{4}\left(\frac{\partial u^{(l)}_{x}}{\partial y}+\frac{\partial u^{(l)}_{y}}{\partial x}\right),
\end{align}
where $\beta=3.14$ and $\gamma_{0}=2.7\,$eV. The intralayer Hamiltonian is given by:
\begin{align}
h^{0,\mu,(l)}_{\rm{intra}, \rm{str}}(\textbf{k})=-\hbar v_{F}\left(\textbf{k}+\mu\frac{e}{\hbar}\textbf{A}^{(l)}\right)\cdot(\mu\sigma_{x},\sigma_{y}).
\end{align}
The Hamiltonian of pentalayer graphene is given by:
\begin{align}
H^{0,\mu}_{\rm{penta},\rm{str}}=\left[\begin{array}{ccccc}
				h^{0,\mu,(1)}_{\rm{intra}, \rm{str}} & (h^{0,\mu}_{\rm{inter}})^{\dagger} & 0&0&0 \\
				h^{0,\mu}_{\rm{inter}} & h^{0,\mu,(2)}_{\rm{intra}, \rm{str}}& (h^{0,\mu}_{\rm{inter}})^{\dagger}&0&0\\
				0&h^{0,\mu}_{\rm{inter}}&h^{0,\mu,(3)}_{\rm{intra}, \rm{str}}&(h^{0,\mu}_{\rm{inter}})^{\dagger}&0\\
				0&0&h^{0,\mu}_{\rm{inter}}&h^{0,\mu,(4)}_{\rm{intra}, \rm{str}}&(h^{0,\mu}_{\rm{inter}})^{\dagger}\\
				0&0& 0&h^{0,\mu}_{\rm{inter}}&h^{0,\mu,(5)}_{\rm{intra}, \rm{str}}
\end{array}\right]
\end{align}
where $h^{0,\mu}_{\rm{inter}}$ is the interlayer hopping within the pentalayer graphene, which to the leading order is unaffected by strain.
Since the hBN layer is an insulator with a huge band gap, we neglect the influence of the strain field on the hBN layer. For the interlayer moir\'e potential, the strain field would introduce additional momentum transfer and alter the hopping amplitude. To be specific, the moir\'e potential under strain field is given by:
\begin{align}
U^{\mu}_{X'X}(r)\approx&\sum_{j=1}^{3}\sum_{n_{1},n'_{1},\dots}t_{0}M^{j,\mu}_{X'X}\times\frac{[-i\textbf{Q}^{\mu}_{j}\cdot\textbf{u}^{(0)}_{\textbf{G}^{u}_{1}}]^{n_{1}}}{n_{1}!}\frac{[i\textbf{Q}^{\mu}_{j}\cdot\textbf{u}^{(1)}_{\textbf{G}^{u}_{1}}]^{n'_{1}}}{n'_{1}!}e^{i\mu\textbf{G}_{j}\cdot\textbf{r}}e^{i(n_{1}+n'_{1})\textbf{G}^{u}_{1}\cdot\textbf{r}}\dots+\nonumber\\
&\sum_{j=1}^{3}\sum_{n_{1},n'_{1},\dots}t_{1}h^{(0-1)}_{\textbf{G}^{h}_{1}}M^{j,\mu}_{X'X}\times\frac{[-i\textbf{Q}^{\mu}_{j}\cdot\textbf{u}^{(0)}_{\textbf{G}^{u}_{1}}]^{n_{1}}}{n_{1}!}\frac{[i\textbf{Q}^{\mu}_{j}\cdot\textbf{u}^{(1)}_{\textbf{G}^{u}_{1}}]^{n'_{1}}}{n'_{1}!}e^{i\mu\textbf{G}_{j}\cdot\textbf{r}}e^{i\textbf{G}^{h}_{1}\cdot\textbf{r}}e^{i(n_{1}+n'_{1})\textbf{G}^{u}_{1}\cdot\textbf{r}}\dots+\dots,
\end{align}
where $\{\textbf{G}^{u}_{1},\textbf{G}^{h}_{1},\dots\}$ are the moir\'e reciprocal lattice vectors, $h^{(0-1)}$ represents the interlayer distance between the graphene-hBN interface, $t_{0}=0.108\,$eV and $t_{1}\approx-0.265\,\mathrm{eV}/$\AA. $\textbf{Q}^{\mu}_{1}=\textbf{K}_{\mu}$, $\textbf{Q}^{\mu}_{2}=\textbf{K}_{\mu}+\mu\textbf{b}_{1}$ and $\textbf{Q}^{\mu}_{3}=\textbf{K}_{\mu}+\mu(\textbf{b}_{1}+\textbf{b}_{2})$. $M^{j}_{X'X}$ is given by:
\begin{align}
M^{1}=\left[\begin{array}{cc}
				1 & 1  \\
				1 & 1
\end{array}\right], M^{2}=\left[\begin{array}{cc}
				1 & e^{-i\mu\frac{2\pi}{3}}  \\
				e^{i\mu\frac{2\pi}{3}} & 1
\end{array}\right], M^{3}=\left[\begin{array}{cc}
				1 & e^{i\mu\frac{2\pi}{3}}  \\
				e^{-i\mu\frac{2\pi}{3}} & 1
\end{array}\right].
\end{align}

Then,we treat the hBN layer as the second order perturbation and obtain the effective Hamiltonian of fully-relaxed hBN-pentalayer graphene system. The result is $H^{0,\mu}_{s}=H^{0,\mu}_{\rm{penta},\rm{str}}+V^{s}_{\rm{hBN}}$, where the effective hBN term can be expressed as:
\begin{align}
V^{s}_{\rm{hBN}}=V_{\rm{hBN}}&\left(1+\sum_{n_{1},n'_{1},\dots}\frac{[-i\textbf{Q}^{\mu}_{j}\cdot\textbf{u}^{(0)}_{\textbf{G}^{u}_{1}}]^{n_{1}}}{n_{1}!}\frac{[i\textbf{Q}^{\mu}_{j}\cdot\textbf{u}^{(1)}_{\textbf{G}^{u}_{1}}]^{n'_{1}}}{n'_{1}!}e^{i(n_{1}+n'_{1})\textbf{G}^{u}_{1}\cdot\textbf{r}}\dots+H.c.+\right.\nonumber\\
&\left.\sum_{n_{1},n'_{1},\dots}\tilde{t}\,h^{(0-1)}_{\textbf{G}^{h}_{1}}\frac{[-i\textbf{Q}^{\mu}_{j}\cdot\textbf{u}^{(0)}_{\textbf{G}^{u}_{1}}]^{n_{1}}}{n_{1}!}\frac{[i\textbf{Q}^{\mu}_{j}\cdot\textbf{u}^{(1)}_{\textbf{G}^{u}_{1}}]^{n'_{1}}}{n'_{1}!}e^{i\textbf{G}^{h}_{1}\cdot\textbf{r}}e^{i(n_{1}+n'_{1})\textbf{G}^{u}_{1}\cdot\textbf{r}}\dots+H.c.\right),
\end{align}
where $\tilde{t}\approx6.03\,$. We present the band structure of hBN-pentalayer graphene calculated with the effective continuum model based on the non-relaxed lattice structures in Supplementary Figure~\ref{fig:bands}(b) and that based on the fully relaxed lattice structures in Supplementary Figure~\ref{fig:bands}(d), where a constant interlayer electrostatic potential energy drop $\Delta_D=65\,$meV has been applied. In the inset, we present a zoomed-in view of the lowset conduction band. The band structures calculated by the continuum model agree well with those calculated by the Slater-Koster tight-binding model.  We also note that, altough the lattice-relaxation effects on the valence moir\'e bands are non-negligible, the conduction moir\'e bands are barely affected. This is because under large positive $\Delta_D$, conduction band electrons are pushed to the top graphene layer that is far away from the hBN interface.

In Supplementary Figure~\ref{fig:morebands}, we present the band structures of hBN-pentalayer graphene under different interlayer electrostatic potential energy drop $\Delta_D$. First of all, we see nice agreements between the tight-binding band structures and the continuum-model ones. Second, we find that as $\Delta_D$ decreases from  $60.3\,$meV (Figs.~\ref{fig:morebands}(u)-(x)) to $0\,$meV (Figs.~\ref{fig:morebands}(a)-(d)), the lattice relaxation effects on the conduction moir\'e bands become stronger and stronger. For example, when $\Delta_D=0\,$meV (Figs.~\ref{fig:morebands}(a)-(d)), the gap between the conduction and valence moir\'e bands is greatly reduced by lattice relaxations.

\bibliography{tmg_tidy}

\begin{thebibliography}{109}%
\makeatletter
\providecommand \@ifxundefined [1]{%
 \@ifx{#1\undefined}
}%
\providecommand \@ifnum [1]{%
 \ifnum #1\expandafter \@firstoftwo
 \else \expandafter \@secondoftwo
 \fi
}%
\providecommand \@ifx [1]{%
 \ifx #1\expandafter \@firstoftwo
 \else \expandafter \@secondoftwo
 \fi
}%
\providecommand \natexlab [1]{#1}%
\providecommand \enquote  [1]{``#1''}%
\providecommand \bibnamefont  [1]{#1}%
\providecommand \bibfnamefont [1]{#1}%
\providecommand \citenamefont [1]{#1}%
\providecommand \href@noop [0]{\@secondoftwo}%
\providecommand \href [0]{\begingroup \@sanitize@url \@href}%
\providecommand \@href[1]{\@@startlink{#1}\@@href}%
\providecommand \@@href[1]{\endgroup#1\@@endlink}%
\providecommand \@sanitize@url [0]{\catcode `\\12\catcode `\$12\catcode
  `\&12\catcode `\#12\catcode `\^12\catcode `\_12\catcode `\%12\relax}%
\providecommand \@@startlink[1]{}%
\providecommand \@@endlink[0]{}%
\providecommand \url  [0]{\begingroup\@sanitize@url \@url }%
\providecommand \@url [1]{\endgroup\@href {#1}{\urlprefix }}%
\providecommand \urlprefix  [0]{URL }%
\providecommand \Eprint [0]{\href }%
\providecommand \doibase [0]{https://doi.org/}%
\providecommand \selectlanguage [0]{\@gobble}%
\providecommand \bibinfo  [0]{\@secondoftwo}%
\providecommand \bibfield  [0]{\@secondoftwo}%
\providecommand \translation [1]{[#1]}%
\providecommand \BibitemOpen [0]{}%
\providecommand \bibitemStop [0]{}%
\providecommand \bibitemNoStop [0]{.\EOS\space}%
\providecommand \EOS [0]{\spacefactor3000\relax}%
\providecommand \BibitemShut  [1]{\csname bibitem#1\endcsname}%
\let\auto@bib@innerbib\@empty
\bibitem [{\citenamefont {Park}\ \emph
  {et~al.}(2023{\natexlab{a}})\citenamefont {Park}, \citenamefont {Cai},
  \citenamefont {Anderson}, \citenamefont {Zhang}, \citenamefont {Zhu},
  \citenamefont {Liu}, \citenamefont {Wang}, \citenamefont {Holtzmann},
  \citenamefont {Hu}, \citenamefont {Liu}, \citenamefont {Taniguchi},
  \citenamefont {Watanabe}, \citenamefont {Chu}, \citenamefont {Cao},
  \citenamefont {Fu}, \citenamefont {Yao}, \citenamefont {Chang}, \citenamefont
  {Cobden}, \citenamefont {Xiao},\ and\ \citenamefont {Xu}}]{fqah-nature23}%
  \BibitemOpen
  \bibfield  {author} {\bibinfo {author} {\bibfnamefont {H.}~\bibnamefont
  {Park}}, \bibinfo {author} {\bibfnamefont {J.}~\bibnamefont {Cai}}, \bibinfo
  {author} {\bibfnamefont {E.}~\bibnamefont {Anderson}}, \bibinfo {author}
  {\bibfnamefont {Y.}~\bibnamefont {Zhang}}, \bibinfo {author} {\bibfnamefont
  {J.}~\bibnamefont {Zhu}}, \bibinfo {author} {\bibfnamefont {X.}~\bibnamefont
  {Liu}}, \bibinfo {author} {\bibfnamefont {C.}~\bibnamefont {Wang}}, \bibinfo
  {author} {\bibfnamefont {W.}~\bibnamefont {Holtzmann}}, \bibinfo {author}
  {\bibfnamefont {C.}~\bibnamefont {Hu}}, \bibinfo {author} {\bibfnamefont
  {Z.}~\bibnamefont {Liu}}, \bibinfo {author} {\bibfnamefont {T.}~\bibnamefont
  {Taniguchi}}, \bibinfo {author} {\bibfnamefont {K.}~\bibnamefont {Watanabe}},
  \bibinfo {author} {\bibfnamefont {J.-H.}\ \bibnamefont {Chu}}, \bibinfo
  {author} {\bibfnamefont {T.}~\bibnamefont {Cao}}, \bibinfo {author}
  {\bibfnamefont {L.}~\bibnamefont {Fu}}, \bibinfo {author} {\bibfnamefont
  {W.}~\bibnamefont {Yao}}, \bibinfo {author} {\bibfnamefont {C.-Z.}\
  \bibnamefont {Chang}}, \bibinfo {author} {\bibfnamefont {D.}~\bibnamefont
  {Cobden}}, \bibinfo {author} {\bibfnamefont {D.}~\bibnamefont {Xiao}},\ and\
  \bibinfo {author} {\bibfnamefont {X.}~\bibnamefont {Xu}},\ }\href
  {https://doi.org/10.1038/s41586-023-06536-0} {\bibfield  {journal} {\bibinfo
  {journal} {Nature}\ }\textbf {\bibinfo {volume} {622}},\ \bibinfo {pages}
  {74} (\bibinfo {year} {2023}{\natexlab{a}})}\BibitemShut {NoStop}%
\bibitem [{\citenamefont {Xu}\ \emph {et~al.}(2023{\natexlab{a}})\citenamefont
  {Xu}, \citenamefont {Sun}, \citenamefont {Jia}, \citenamefont {Liu},
  \citenamefont {Xu}, \citenamefont {Li}, \citenamefont {Gu}, \citenamefont
  {Watanabe}, \citenamefont {Taniguchi}, \citenamefont {Tong}, \citenamefont
  {Jia}, \citenamefont {Shi}, \citenamefont {Jiang}, \citenamefont {Zhang},
  \citenamefont {Liu},\ and\ \citenamefont {Li}}]{fqah-prx23}%
  \BibitemOpen
  \bibfield  {author} {\bibinfo {author} {\bibfnamefont {F.}~\bibnamefont
  {Xu}}, \bibinfo {author} {\bibfnamefont {Z.}~\bibnamefont {Sun}}, \bibinfo
  {author} {\bibfnamefont {T.}~\bibnamefont {Jia}}, \bibinfo {author}
  {\bibfnamefont {C.}~\bibnamefont {Liu}}, \bibinfo {author} {\bibfnamefont
  {C.}~\bibnamefont {Xu}}, \bibinfo {author} {\bibfnamefont {C.}~\bibnamefont
  {Li}}, \bibinfo {author} {\bibfnamefont {Y.}~\bibnamefont {Gu}}, \bibinfo
  {author} {\bibfnamefont {K.}~\bibnamefont {Watanabe}}, \bibinfo {author}
  {\bibfnamefont {T.}~\bibnamefont {Taniguchi}}, \bibinfo {author}
  {\bibfnamefont {B.}~\bibnamefont {Tong}}, \bibinfo {author} {\bibfnamefont
  {J.}~\bibnamefont {Jia}}, \bibinfo {author} {\bibfnamefont {Z.}~\bibnamefont
  {Shi}}, \bibinfo {author} {\bibfnamefont {S.}~\bibnamefont {Jiang}}, \bibinfo
  {author} {\bibfnamefont {Y.}~\bibnamefont {Zhang}}, \bibinfo {author}
  {\bibfnamefont {X.}~\bibnamefont {Liu}},\ and\ \bibinfo {author}
  {\bibfnamefont {T.}~\bibnamefont {Li}},\ }\href
  {https://doi.org/10.1103/PhysRevX.13.031037} {\bibfield  {journal} {\bibinfo
  {journal} {Phys. Rev. X}\ }\textbf {\bibinfo {volume} {13}},\ \bibinfo
  {pages} {031037} (\bibinfo {year} {2023}{\natexlab{a}})}\BibitemShut
  {NoStop}%
\bibitem [{\citenamefont {Cai}\ \emph {et~al.}(2023)\citenamefont {Cai},
  \citenamefont {Anderson}, \citenamefont {Wang}, \citenamefont {Zhang},
  \citenamefont {Liu}, \citenamefont {Holtzmann}, \citenamefont {Zhang},
  \citenamefont {Fan}, \citenamefont {Taniguchi}, \citenamefont {Watanabe},
  \citenamefont {Ran}, \citenamefont {Cao}, \citenamefont {Fu}, \citenamefont
  {Xiao}, \citenamefont {Yao},\ and\ \citenamefont
  {Xu}}]{fqah-optics-xu-nature23}%
  \BibitemOpen
  \bibfield  {author} {\bibinfo {author} {\bibfnamefont {J.}~\bibnamefont
  {Cai}}, \bibinfo {author} {\bibfnamefont {E.}~\bibnamefont {Anderson}},
  \bibinfo {author} {\bibfnamefont {C.}~\bibnamefont {Wang}}, \bibinfo {author}
  {\bibfnamefont {X.}~\bibnamefont {Zhang}}, \bibinfo {author} {\bibfnamefont
  {X.}~\bibnamefont {Liu}}, \bibinfo {author} {\bibfnamefont {W.}~\bibnamefont
  {Holtzmann}}, \bibinfo {author} {\bibfnamefont {Y.}~\bibnamefont {Zhang}},
  \bibinfo {author} {\bibfnamefont {F.}~\bibnamefont {Fan}}, \bibinfo {author}
  {\bibfnamefont {T.}~\bibnamefont {Taniguchi}}, \bibinfo {author}
  {\bibfnamefont {K.}~\bibnamefont {Watanabe}}, \bibinfo {author}
  {\bibfnamefont {Y.}~\bibnamefont {Ran}}, \bibinfo {author} {\bibfnamefont
  {T.}~\bibnamefont {Cao}}, \bibinfo {author} {\bibfnamefont {L.}~\bibnamefont
  {Fu}}, \bibinfo {author} {\bibfnamefont {D.}~\bibnamefont {Xiao}}, \bibinfo
  {author} {\bibfnamefont {W.}~\bibnamefont {Yao}},\ and\ \bibinfo {author}
  {\bibfnamefont {X.}~\bibnamefont {Xu}},\ }\href
  {https://doi.org/10.1038/s41586-023-06289-w} {\bibfield  {journal} {\bibinfo
  {journal} {Nature}\ }\textbf {\bibinfo {volume} {622}},\ \bibinfo {pages}
  {63} (\bibinfo {year} {2023})}\BibitemShut {NoStop}%
\bibitem [{\citenamefont {Zeng}\ \emph {et~al.}(2023)\citenamefont {Zeng},
  \citenamefont {Xia}, \citenamefont {Kang}, \citenamefont {Zhu}, \citenamefont
  {Kn{\"u}ppel}, \citenamefont {Vaswani}, \citenamefont {Watanabe},
  \citenamefont {Taniguchi}, \citenamefont {Mak},\ and\ \citenamefont
  {Shan}}]{fqah-mak-nature23}%
  \BibitemOpen
  \bibfield  {author} {\bibinfo {author} {\bibfnamefont {Y.}~\bibnamefont
  {Zeng}}, \bibinfo {author} {\bibfnamefont {Z.}~\bibnamefont {Xia}}, \bibinfo
  {author} {\bibfnamefont {K.}~\bibnamefont {Kang}}, \bibinfo {author}
  {\bibfnamefont {J.}~\bibnamefont {Zhu}}, \bibinfo {author} {\bibfnamefont
  {P.}~\bibnamefont {Kn{\"u}ppel}}, \bibinfo {author} {\bibfnamefont
  {C.}~\bibnamefont {Vaswani}}, \bibinfo {author} {\bibfnamefont
  {K.}~\bibnamefont {Watanabe}}, \bibinfo {author} {\bibfnamefont
  {T.}~\bibnamefont {Taniguchi}}, \bibinfo {author} {\bibfnamefont {K.~F.}\
  \bibnamefont {Mak}},\ and\ \bibinfo {author} {\bibfnamefont {J.}~\bibnamefont
  {Shan}},\ }\href {https://doi.org/10.1038/s41586-023-06452-3} {\bibfield
  {journal} {\bibinfo  {journal} {Nature}\ }\textbf {\bibinfo {volume} {622}},\
  \bibinfo {pages} {69} (\bibinfo {year} {2023})}\BibitemShut {NoStop}%
\bibitem [{\citenamefont {Lu}\ \emph {et~al.}(2023{\natexlab{a}})\citenamefont
  {Lu}, \citenamefont {Han}, \citenamefont {Yao}, \citenamefont {Reddy},
  \citenamefont {Yang}, \citenamefont {Seo}, \citenamefont {Watanabe},
  \citenamefont {Taniguchi}, \citenamefont {Fu},\ and\ \citenamefont
  {Ju}}]{fqah-ju-arxiv23}%
  \BibitemOpen
  \bibfield  {author} {\bibinfo {author} {\bibfnamefont {Z.}~\bibnamefont
  {Lu}}, \bibinfo {author} {\bibfnamefont {T.}~\bibnamefont {Han}}, \bibinfo
  {author} {\bibfnamefont {Y.}~\bibnamefont {Yao}}, \bibinfo {author}
  {\bibfnamefont {A.~P.}\ \bibnamefont {Reddy}}, \bibinfo {author}
  {\bibfnamefont {J.}~\bibnamefont {Yang}}, \bibinfo {author} {\bibfnamefont
  {J.}~\bibnamefont {Seo}}, \bibinfo {author} {\bibfnamefont {K.}~\bibnamefont
  {Watanabe}}, \bibinfo {author} {\bibfnamefont {T.}~\bibnamefont {Taniguchi}},
  \bibinfo {author} {\bibfnamefont {L.}~\bibnamefont {Fu}},\ and\ \bibinfo
  {author} {\bibfnamefont {L.}~\bibnamefont {Ju}},\ }\href@noop {} {\bibfield
  {journal} {\bibinfo  {journal} {arXiv preprint arXiv:2309.17436}\ } (\bibinfo
  {year} {2023}{\natexlab{a}})}\BibitemShut {NoStop}%
\bibitem [{\citenamefont {Regnault}\ and\ \citenamefont
  {Bernevig}(2011{\natexlab{a}})}]{fci-prx11}%
  \BibitemOpen
  \bibfield  {author} {\bibinfo {author} {\bibfnamefont {N.}~\bibnamefont
  {Regnault}}\ and\ \bibinfo {author} {\bibfnamefont {B.~A.}\ \bibnamefont
  {Bernevig}},\ }\href {https://doi.org/10.1103/PhysRevX.1.021014} {\bibfield
  {journal} {\bibinfo  {journal} {Phys. Rev. X}\ }\textbf {\bibinfo {volume}
  {1}},\ \bibinfo {pages} {021014} (\bibinfo {year}
  {2011}{\natexlab{a}})}\BibitemShut {NoStop}%
\bibitem [{\citenamefont {Sheng}\ \emph {et~al.}(2011)\citenamefont {Sheng},
  \citenamefont {Gu}, \citenamefont {Sun},\ and\ \citenamefont
  {Sheng}}]{sheng-fci-nc11}%
  \BibitemOpen
  \bibfield  {author} {\bibinfo {author} {\bibfnamefont {D.~N.}\ \bibnamefont
  {Sheng}}, \bibinfo {author} {\bibfnamefont {Z.-C.}\ \bibnamefont {Gu}},
  \bibinfo {author} {\bibfnamefont {K.}~\bibnamefont {Sun}},\ and\ \bibinfo
  {author} {\bibfnamefont {L.}~\bibnamefont {Sheng}},\ }\href
  {https://doi.org/10.1038/ncomms1380} {\bibfield  {journal} {\bibinfo
  {journal} {Nature Communications}\ }\textbf {\bibinfo {volume} {2}},\
  \bibinfo {pages} {389} (\bibinfo {year} {2011})}\BibitemShut {NoStop}%
\bibitem [{\citenamefont {Neupert}\ \emph {et~al.}(2011)\citenamefont
  {Neupert}, \citenamefont {Santos}, \citenamefont {Chamon},\ and\
  \citenamefont {Mudry}}]{murdy-fci-prl11}%
  \BibitemOpen
  \bibfield  {author} {\bibinfo {author} {\bibfnamefont {T.}~\bibnamefont
  {Neupert}}, \bibinfo {author} {\bibfnamefont {L.}~\bibnamefont {Santos}},
  \bibinfo {author} {\bibfnamefont {C.}~\bibnamefont {Chamon}},\ and\ \bibinfo
  {author} {\bibfnamefont {C.}~\bibnamefont {Mudry}},\ }\href
  {https://doi.org/10.1103/PhysRevLett.106.236804} {\bibfield  {journal}
  {\bibinfo  {journal} {Phys. Rev. Lett.}\ }\textbf {\bibinfo {volume} {106}},\
  \bibinfo {pages} {236804} (\bibinfo {year} {2011})}\BibitemShut {NoStop}%
\bibitem [{\citenamefont {Tang}\ \emph {et~al.}(2011)\citenamefont {Tang},
  \citenamefont {Mei},\ and\ \citenamefont {Wen}}]{wen-kagome-prl11}%
  \BibitemOpen
  \bibfield  {author} {\bibinfo {author} {\bibfnamefont {E.}~\bibnamefont
  {Tang}}, \bibinfo {author} {\bibfnamefont {J.-W.}\ \bibnamefont {Mei}},\ and\
  \bibinfo {author} {\bibfnamefont {X.-G.}\ \bibnamefont {Wen}},\ }\href
  {https://doi.org/10.1103/PhysRevLett.106.236802} {\bibfield  {journal}
  {\bibinfo  {journal} {Phys. Rev. Lett.}\ }\textbf {\bibinfo {volume} {106}},\
  \bibinfo {pages} {236802} (\bibinfo {year} {2011})}\BibitemShut {NoStop}%
\bibitem [{\citenamefont {Sun}\ \emph {et~al.}(2011)\citenamefont {Sun},
  \citenamefont {Gu}, \citenamefont {Katsura},\ and\ \citenamefont
  {Das~Sarma}}]{sarma-flatchern-prl11}%
  \BibitemOpen
  \bibfield  {author} {\bibinfo {author} {\bibfnamefont {K.}~\bibnamefont
  {Sun}}, \bibinfo {author} {\bibfnamefont {Z.}~\bibnamefont {Gu}}, \bibinfo
  {author} {\bibfnamefont {H.}~\bibnamefont {Katsura}},\ and\ \bibinfo {author}
  {\bibfnamefont {S.}~\bibnamefont {Das~Sarma}},\ }\href
  {https://doi.org/10.1103/PhysRevLett.106.236803} {\bibfield  {journal}
  {\bibinfo  {journal} {Phys. Rev. Lett.}\ }\textbf {\bibinfo {volume} {106}},\
  \bibinfo {pages} {236803} (\bibinfo {year} {2011})}\BibitemShut {NoStop}%
\bibitem [{\citenamefont {M\"oller}\ and\ \citenamefont
  {Cooper}(2009)}]{cooper-fci-prl09}%
  \BibitemOpen
  \bibfield  {author} {\bibinfo {author} {\bibfnamefont {G.}~\bibnamefont
  {M\"oller}}\ and\ \bibinfo {author} {\bibfnamefont {N.~R.}\ \bibnamefont
  {Cooper}},\ }\href {https://doi.org/10.1103/PhysRevLett.103.105303}
  {\bibfield  {journal} {\bibinfo  {journal} {Phys. Rev. Lett.}\ }\textbf
  {\bibinfo {volume} {103}},\ \bibinfo {pages} {105303} (\bibinfo {year}
  {2009})}\BibitemShut {NoStop}%
\bibitem [{\citenamefont {Tsui}\ \emph {et~al.}(1982)\citenamefont {Tsui},
  \citenamefont {Stormer},\ and\ \citenamefont {Gossard}}]{fqhe-prl82}%
  \BibitemOpen
  \bibfield  {author} {\bibinfo {author} {\bibfnamefont {D.~C.}\ \bibnamefont
  {Tsui}}, \bibinfo {author} {\bibfnamefont {H.~L.}\ \bibnamefont {Stormer}},\
  and\ \bibinfo {author} {\bibfnamefont {A.~C.}\ \bibnamefont {Gossard}},\
  }\href {https://doi.org/10.1103/PhysRevLett.48.1559} {\bibfield  {journal}
  {\bibinfo  {journal} {Phys. Rev. Lett.}\ }\textbf {\bibinfo {volume} {48}},\
  \bibinfo {pages} {1559} (\bibinfo {year} {1982})}\BibitemShut {NoStop}%
\bibitem [{\citenamefont {Laughlin}(1983)}]{laughlin-prl83}%
  \BibitemOpen
  \bibfield  {author} {\bibinfo {author} {\bibfnamefont {R.~B.}\ \bibnamefont
  {Laughlin}},\ }\href {https://doi.org/10.1103/PhysRevLett.50.1395} {\bibfield
   {journal} {\bibinfo  {journal} {Physical Review Letters}\ }\textbf {\bibinfo
  {volume} {50}},\ \bibinfo {pages} {1395} (\bibinfo {year}
  {1983})}\BibitemShut {NoStop}%
\bibitem [{\citenamefont {Jain}(1989)}]{jain-prl89}%
  \BibitemOpen
  \bibfield  {author} {\bibinfo {author} {\bibfnamefont {J.~K.}\ \bibnamefont
  {Jain}},\ }\href@noop {} {\bibfield  {journal} {\bibinfo  {journal} {Physical
  review letters}\ }\textbf {\bibinfo {volume} {63}},\ \bibinfo {pages} {199}
  (\bibinfo {year} {1989})}\BibitemShut {NoStop}%
\bibitem [{\citenamefont {Moore}\ and\ \citenamefont
  {Read}(1991)}]{moore-read}%
  \BibitemOpen
  \bibfield  {author} {\bibinfo {author} {\bibfnamefont {G.}~\bibnamefont
  {Moore}}\ and\ \bibinfo {author} {\bibfnamefont {N.}~\bibnamefont {Read}},\
  }\href@noop {} {\bibfield  {journal} {\bibinfo  {journal} {Nuclear Physics
  B}\ }\textbf {\bibinfo {volume} {360}},\ \bibinfo {pages} {362} (\bibinfo
  {year} {1991})}\BibitemShut {NoStop}%
\bibitem [{\citenamefont {Stormer}\ \emph {et~al.}(1999)\citenamefont
  {Stormer}, \citenamefont {Tsui},\ and\ \citenamefont {Gossard}}]{fqhe-rmp99}%
  \BibitemOpen
  \bibfield  {author} {\bibinfo {author} {\bibfnamefont {H.~L.}\ \bibnamefont
  {Stormer}}, \bibinfo {author} {\bibfnamefont {D.~C.}\ \bibnamefont {Tsui}},\
  and\ \bibinfo {author} {\bibfnamefont {A.~C.}\ \bibnamefont {Gossard}},\
  }\href@noop {} {\bibfield  {journal} {\bibinfo  {journal} {Reviews of Modern
  Physics}\ }\textbf {\bibinfo {volume} {71}},\ \bibinfo {pages} {S298}
  (\bibinfo {year} {1999})}\BibitemShut {NoStop}%
\bibitem [{\citenamefont {Cage}\ \emph {et~al.}(2012)\citenamefont {Cage},
  \citenamefont {Klitzing}, \citenamefont {Chang}, \citenamefont {Duncan},
  \citenamefont {Haldane}, \citenamefont {Laughlin}, \citenamefont {Pruisken},\
  and\ \citenamefont {Thouless}}]{qhe-book-2012}%
  \BibitemOpen
  \bibfield  {author} {\bibinfo {author} {\bibfnamefont {M.~E.}\ \bibnamefont
  {Cage}}, \bibinfo {author} {\bibfnamefont {K.}~\bibnamefont {Klitzing}},
  \bibinfo {author} {\bibfnamefont {A.}~\bibnamefont {Chang}}, \bibinfo
  {author} {\bibfnamefont {F.}~\bibnamefont {Duncan}}, \bibinfo {author}
  {\bibfnamefont {M.}~\bibnamefont {Haldane}}, \bibinfo {author} {\bibfnamefont
  {R.~B.}\ \bibnamefont {Laughlin}}, \bibinfo {author} {\bibfnamefont
  {A.}~\bibnamefont {Pruisken}},\ and\ \bibinfo {author} {\bibfnamefont
  {D.}~\bibnamefont {Thouless}},\ }\href@noop {} {\emph {\bibinfo {title} {The
  quantum Hall effect}}}\ (\bibinfo  {publisher} {Springer Science \& Business
  Media},\ \bibinfo {year} {2012})\BibitemShut {NoStop}%
\bibitem [{\citenamefont {Wang}\ \emph {et~al.}(2011)\citenamefont {Wang},
  \citenamefont {Gu}, \citenamefont {Gong},\ and\ \citenamefont
  {Sheng}}]{fqh-boson-prl11}%
  \BibitemOpen
  \bibfield  {author} {\bibinfo {author} {\bibfnamefont {Y.-F.}\ \bibnamefont
  {Wang}}, \bibinfo {author} {\bibfnamefont {Z.-C.}\ \bibnamefont {Gu}},
  \bibinfo {author} {\bibfnamefont {C.-D.}\ \bibnamefont {Gong}},\ and\
  \bibinfo {author} {\bibfnamefont {D.~N.}\ \bibnamefont {Sheng}},\ }\href
  {https://doi.org/10.1103/PhysRevLett.107.146803} {\bibfield  {journal}
  {\bibinfo  {journal} {Phys. Rev. Lett.}\ }\textbf {\bibinfo {volume} {107}},\
  \bibinfo {pages} {146803} (\bibinfo {year} {2011})}\BibitemShut {NoStop}%
\bibitem [{\citenamefont {Wu}\ \emph {et~al.}(2012)\citenamefont {Wu},
  \citenamefont {Bernevig},\ and\ \citenamefont {Regnault}}]{fci-zoo-prb12}%
  \BibitemOpen
  \bibfield  {author} {\bibinfo {author} {\bibfnamefont {Y.-L.}\ \bibnamefont
  {Wu}}, \bibinfo {author} {\bibfnamefont {B.~A.}\ \bibnamefont {Bernevig}},\
  and\ \bibinfo {author} {\bibfnamefont {N.}~\bibnamefont {Regnault}},\ }\href
  {https://doi.org/10.1103/PhysRevB.85.075116} {\bibfield  {journal} {\bibinfo
  {journal} {Phys. Rev. B}\ }\textbf {\bibinfo {volume} {85}},\ \bibinfo
  {pages} {075116} (\bibinfo {year} {2012})}\BibitemShut {NoStop}%
\bibitem [{\citenamefont {Liu}\ \emph {et~al.}(2012)\citenamefont {Liu},
  \citenamefont {Bergholtz}, \citenamefont {Fan},\ and\ \citenamefont
  {L\"auchli}}]{liuzhao-fci-prl12}%
  \BibitemOpen
  \bibfield  {author} {\bibinfo {author} {\bibfnamefont {Z.}~\bibnamefont
  {Liu}}, \bibinfo {author} {\bibfnamefont {E.~J.}\ \bibnamefont {Bergholtz}},
  \bibinfo {author} {\bibfnamefont {H.}~\bibnamefont {Fan}},\ and\ \bibinfo
  {author} {\bibfnamefont {A.~M.}\ \bibnamefont {L\"auchli}},\ }\href
  {https://doi.org/10.1103/PhysRevLett.109.186805} {\bibfield  {journal}
  {\bibinfo  {journal} {Phys. Rev. Lett.}\ }\textbf {\bibinfo {volume} {109}},\
  \bibinfo {pages} {186805} (\bibinfo {year} {2012})}\BibitemShut {NoStop}%
\bibitem [{\citenamefont {Venderbos}\ \emph {et~al.}(2012)\citenamefont
  {Venderbos}, \citenamefont {Kourtis}, \citenamefont {van~den Brink},\ and\
  \citenamefont {Daghofer}}]{vanderbos-fci-prl12}%
  \BibitemOpen
  \bibfield  {author} {\bibinfo {author} {\bibfnamefont {J.~W.~F.}\
  \bibnamefont {Venderbos}}, \bibinfo {author} {\bibfnamefont {S.}~\bibnamefont
  {Kourtis}}, \bibinfo {author} {\bibfnamefont {J.}~\bibnamefont {van~den
  Brink}},\ and\ \bibinfo {author} {\bibfnamefont {M.}~\bibnamefont
  {Daghofer}},\ }\href {https://doi.org/10.1103/PhysRevLett.108.126405}
  {\bibfield  {journal} {\bibinfo  {journal} {Phys. Rev. Lett.}\ }\textbf
  {\bibinfo {volume} {108}},\ \bibinfo {pages} {126405} (\bibinfo {year}
  {2012})}\BibitemShut {NoStop}%
\bibitem [{\citenamefont {Liu}\ \emph {et~al.}(2013)\citenamefont {Liu},
  \citenamefont {Repellin}, \citenamefont {Bernevig},\ and\ \citenamefont
  {Regnault}}]{liu-fci-prb13}%
  \BibitemOpen
  \bibfield  {author} {\bibinfo {author} {\bibfnamefont {T.}~\bibnamefont
  {Liu}}, \bibinfo {author} {\bibfnamefont {C.}~\bibnamefont {Repellin}},
  \bibinfo {author} {\bibfnamefont {B.~A.}\ \bibnamefont {Bernevig}},\ and\
  \bibinfo {author} {\bibfnamefont {N.}~\bibnamefont {Regnault}},\ }\href
  {https://doi.org/10.1103/PhysRevB.87.205136} {\bibfield  {journal} {\bibinfo
  {journal} {Phys. Rev. B}\ }\textbf {\bibinfo {volume} {87}},\ \bibinfo
  {pages} {205136} (\bibinfo {year} {2013})}\BibitemShut {NoStop}%
\bibitem [{\citenamefont {Hu}\ \emph {et~al.}(2011)\citenamefont {Hu},
  \citenamefont {Kargarian},\ and\ \citenamefont {Fiete}}]{fiete-ruby-prb11}%
  \BibitemOpen
  \bibfield  {author} {\bibinfo {author} {\bibfnamefont {X.}~\bibnamefont
  {Hu}}, \bibinfo {author} {\bibfnamefont {M.}~\bibnamefont {Kargarian}},\ and\
  \bibinfo {author} {\bibfnamefont {G.~A.}\ \bibnamefont {Fiete}},\ }\href
  {https://doi.org/10.1103/PhysRevB.84.155116} {\bibfield  {journal} {\bibinfo
  {journal} {Phys. Rev. B}\ }\textbf {\bibinfo {volume} {84}},\ \bibinfo
  {pages} {155116} (\bibinfo {year} {2011})}\BibitemShut {NoStop}%
\bibitem [{\citenamefont {Roy}(2014)}]{roy-prb14}%
  \BibitemOpen
  \bibfield  {author} {\bibinfo {author} {\bibfnamefont {R.}~\bibnamefont
  {Roy}},\ }\href {https://doi.org/10.1103/PhysRevB.90.165139} {\bibfield
  {journal} {\bibinfo  {journal} {Phys. Rev. B}\ }\textbf {\bibinfo {volume}
  {90}},\ \bibinfo {pages} {165139} (\bibinfo {year} {2014})}\BibitemShut
  {NoStop}%
\bibitem [{\citenamefont {Claassen}\ \emph {et~al.}(2015)\citenamefont
  {Claassen}, \citenamefont {Lee}, \citenamefont {Thomale}, \citenamefont
  {Qi},\ and\ \citenamefont {Devereaux}}]{claassen-prl15}%
  \BibitemOpen
  \bibfield  {author} {\bibinfo {author} {\bibfnamefont {M.}~\bibnamefont
  {Claassen}}, \bibinfo {author} {\bibfnamefont {C.~H.}\ \bibnamefont {Lee}},
  \bibinfo {author} {\bibfnamefont {R.}~\bibnamefont {Thomale}}, \bibinfo
  {author} {\bibfnamefont {X.-L.}\ \bibnamefont {Qi}},\ and\ \bibinfo {author}
  {\bibfnamefont {T.~P.}\ \bibnamefont {Devereaux}},\ }\href
  {https://doi.org/10.1103/PhysRevLett.114.236802} {\bibfield  {journal}
  {\bibinfo  {journal} {Phys. Rev. Lett.}\ }\textbf {\bibinfo {volume} {114}},\
  \bibinfo {pages} {236802} (\bibinfo {year} {2015})}\BibitemShut {NoStop}%
\bibitem [{\citenamefont {Wang}\ \emph
  {et~al.}(2021{\natexlab{a}})\citenamefont {Wang}, \citenamefont {Cano},
  \citenamefont {Millis}, \citenamefont {Liu},\ and\ \citenamefont
  {Yang}}]{wangjie-prl21}%
  \BibitemOpen
  \bibfield  {author} {\bibinfo {author} {\bibfnamefont {J.}~\bibnamefont
  {Wang}}, \bibinfo {author} {\bibfnamefont {J.}~\bibnamefont {Cano}}, \bibinfo
  {author} {\bibfnamefont {A.~J.}\ \bibnamefont {Millis}}, \bibinfo {author}
  {\bibfnamefont {Z.}~\bibnamefont {Liu}},\ and\ \bibinfo {author}
  {\bibfnamefont {B.}~\bibnamefont {Yang}},\ }\href
  {https://doi.org/10.1103/PhysRevLett.127.246403} {\bibfield  {journal}
  {\bibinfo  {journal} {Phys. Rev. Lett.}\ }\textbf {\bibinfo {volume} {127}},\
  \bibinfo {pages} {246403} (\bibinfo {year} {2021}{\natexlab{a}})}\BibitemShut
  {NoStop}%
\bibitem [{\citenamefont {Ledwith}\ \emph
  {et~al.}(2022{\natexlab{a}})\citenamefont {Ledwith}, \citenamefont
  {Vishwanath},\ and\ \citenamefont {Parker}}]{ledwith-vortex-arxiv22}%
  \BibitemOpen
  \bibfield  {author} {\bibinfo {author} {\bibfnamefont {P.~J.}\ \bibnamefont
  {Ledwith}}, \bibinfo {author} {\bibfnamefont {A.}~\bibnamefont
  {Vishwanath}},\ and\ \bibinfo {author} {\bibfnamefont {D.~E.}\ \bibnamefont
  {Parker}},\ }\href@noop {} {\bibinfo {title} {Vortexability: A unifying
  criterion for ideal fractional chern insulators}} (\bibinfo {year}
  {2022}{\natexlab{a}}),\ \Eprint {https://arxiv.org/abs/2209.15023}
  {arXiv:2209.15023 [cond-mat.str-el]} \BibitemShut {NoStop}%
\bibitem [{\citenamefont {Bistritzer}\ and\ \citenamefont
  {MacDonald}(2011)}]{macdonald-pnas11}%
  \BibitemOpen
  \bibfield  {author} {\bibinfo {author} {\bibfnamefont {R.}~\bibnamefont
  {Bistritzer}}\ and\ \bibinfo {author} {\bibfnamefont {A.~H.}\ \bibnamefont
  {MacDonald}},\ }\href@noop {} {\bibfield  {journal} {\bibinfo  {journal}
  {Proceedings of the National Academy of Sciences}\ }\textbf {\bibinfo
  {volume} {108}},\ \bibinfo {pages} {12233} (\bibinfo {year}
  {2011})}\BibitemShut {NoStop}%
\bibitem [{\citenamefont {Song}\ \emph {et~al.}(2019)\citenamefont {Song},
  \citenamefont {Wang}, \citenamefont {Shi}, \citenamefont {Li}, \citenamefont
  {Fang},\ and\ \citenamefont {Bernevig}}]{song-tbg-prl19}%
  \BibitemOpen
  \bibfield  {author} {\bibinfo {author} {\bibfnamefont {Z.}~\bibnamefont
  {Song}}, \bibinfo {author} {\bibfnamefont {Z.}~\bibnamefont {Wang}}, \bibinfo
  {author} {\bibfnamefont {W.}~\bibnamefont {Shi}}, \bibinfo {author}
  {\bibfnamefont {G.}~\bibnamefont {Li}}, \bibinfo {author} {\bibfnamefont
  {C.}~\bibnamefont {Fang}},\ and\ \bibinfo {author} {\bibfnamefont {B.~A.}\
  \bibnamefont {Bernevig}},\ }\href
  {https://doi.org/10.1103/PhysRevLett.123.036401} {\bibfield  {journal}
  {\bibinfo  {journal} {Phys. Rev. Lett.}\ }\textbf {\bibinfo {volume} {123}},\
  \bibinfo {pages} {036401} (\bibinfo {year} {2019})}\BibitemShut {NoStop}%
\bibitem [{\citenamefont {Tarnopolsky}\ \emph {et~al.}(2019)\citenamefont
  {Tarnopolsky}, \citenamefont {Kruchkov},\ and\ \citenamefont
  {Vishwanath}}]{origin-magic-angle-prl19}%
  \BibitemOpen
  \bibfield  {author} {\bibinfo {author} {\bibfnamefont {G.}~\bibnamefont
  {Tarnopolsky}}, \bibinfo {author} {\bibfnamefont {A.~J.}\ \bibnamefont
  {Kruchkov}},\ and\ \bibinfo {author} {\bibfnamefont {A.}~\bibnamefont
  {Vishwanath}},\ }\href {https://doi.org/10.1103/PhysRevLett.122.106405}
  {\bibfield  {journal} {\bibinfo  {journal} {Phys. Rev. Lett.}\ }\textbf
  {\bibinfo {volume} {122}},\ \bibinfo {pages} {106405} (\bibinfo {year}
  {2019})}\BibitemShut {NoStop}%
\bibitem [{\citenamefont {Liu}\ \emph {et~al.}(2019{\natexlab{a}})\citenamefont
  {Liu}, \citenamefont {Liu},\ and\ \citenamefont {Dai}}]{jpliu-prb19}%
  \BibitemOpen
  \bibfield  {author} {\bibinfo {author} {\bibfnamefont {J.}~\bibnamefont
  {Liu}}, \bibinfo {author} {\bibfnamefont {J.}~\bibnamefont {Liu}},\ and\
  \bibinfo {author} {\bibfnamefont {X.}~\bibnamefont {Dai}},\ }\href
  {https://doi.org/10.1103/PhysRevB.99.155415} {\bibfield  {journal} {\bibinfo
  {journal} {Phys. Rev. B}\ }\textbf {\bibinfo {volume} {99}},\ \bibinfo
  {pages} {155415} (\bibinfo {year} {2019}{\natexlab{a}})}\BibitemShut
  {NoStop}%
\bibitem [{\citenamefont {Ahn}\ \emph {et~al.}(2019)\citenamefont {Ahn},
  \citenamefont {Park},\ and\ \citenamefont {Yang}}]{yang-tbg-prx19}%
  \BibitemOpen
  \bibfield  {author} {\bibinfo {author} {\bibfnamefont {J.}~\bibnamefont
  {Ahn}}, \bibinfo {author} {\bibfnamefont {S.}~\bibnamefont {Park}},\ and\
  \bibinfo {author} {\bibfnamefont {B.-J.}\ \bibnamefont {Yang}},\ }\href
  {https://doi.org/10.1103/PhysRevX.9.021013} {\bibfield  {journal} {\bibinfo
  {journal} {Phys. Rev. X}\ }\textbf {\bibinfo {volume} {9}},\ \bibinfo {pages}
  {021013} (\bibinfo {year} {2019})}\BibitemShut {NoStop}%
\bibitem [{\citenamefont {Po}\ \emph {et~al.}(2019)\citenamefont {Po},
  \citenamefont {Zou}, \citenamefont {Senthil},\ and\ \citenamefont
  {Vishwanath}}]{po-tbg-prb19}%
  \BibitemOpen
  \bibfield  {author} {\bibinfo {author} {\bibfnamefont {H.~C.}\ \bibnamefont
  {Po}}, \bibinfo {author} {\bibfnamefont {L.}~\bibnamefont {Zou}}, \bibinfo
  {author} {\bibfnamefont {T.}~\bibnamefont {Senthil}},\ and\ \bibinfo {author}
  {\bibfnamefont {A.}~\bibnamefont {Vishwanath}},\ }\href
  {https://doi.org/10.1103/PhysRevB.99.195455} {\bibfield  {journal} {\bibinfo
  {journal} {Phys. Rev. B}\ }\textbf {\bibinfo {volume} {99}},\ \bibinfo
  {pages} {195455} (\bibinfo {year} {2019})}\BibitemShut {NoStop}%
\bibitem [{\citenamefont {Bultinck}\ \emph {et~al.}(2020)\citenamefont
  {Bultinck}, \citenamefont {Chatterjee},\ and\ \citenamefont
  {Zaletel}}]{zaletel-tbg-2019}%
  \BibitemOpen
  \bibfield  {author} {\bibinfo {author} {\bibfnamefont {N.}~\bibnamefont
  {Bultinck}}, \bibinfo {author} {\bibfnamefont {S.}~\bibnamefont
  {Chatterjee}},\ and\ \bibinfo {author} {\bibfnamefont {M.~P.}\ \bibnamefont
  {Zaletel}},\ }\href {https://doi.org/10.1103/PhysRevLett.124.166601}
  {\bibfield  {journal} {\bibinfo  {journal} {Phys. Rev. Lett.}\ }\textbf
  {\bibinfo {volume} {124}},\ \bibinfo {pages} {166601} (\bibinfo {year}
  {2020})}\BibitemShut {NoStop}%
\bibitem [{\citenamefont {Wang}\ \emph
  {et~al.}(2021{\natexlab{b}})\citenamefont {Wang}, \citenamefont {Zheng},
  \citenamefont {Millis},\ and\ \citenamefont {Cano}}]{wang2021chiral}%
  \BibitemOpen
  \bibfield  {author} {\bibinfo {author} {\bibfnamefont {J.}~\bibnamefont
  {Wang}}, \bibinfo {author} {\bibfnamefont {Y.}~\bibnamefont {Zheng}},
  \bibinfo {author} {\bibfnamefont {A.~J.}\ \bibnamefont {Millis}},\ and\
  \bibinfo {author} {\bibfnamefont {J.}~\bibnamefont {Cano}},\ }\href@noop {}
  {\bibfield  {journal} {\bibinfo  {journal} {Physical Review Research}\
  }\textbf {\bibinfo {volume} {3}},\ \bibinfo {pages} {023155} (\bibinfo {year}
  {2021}{\natexlab{b}})}\BibitemShut {NoStop}%
\bibitem [{\citenamefont {Song}\ \emph {et~al.}(2021)\citenamefont {Song},
  \citenamefont {Lian}, \citenamefont {Regnault},\ and\ \citenamefont
  {Bernevig}}]{song-tbg-ii-prb21}%
  \BibitemOpen
  \bibfield  {author} {\bibinfo {author} {\bibfnamefont {Z.-D.}\ \bibnamefont
  {Song}}, \bibinfo {author} {\bibfnamefont {B.}~\bibnamefont {Lian}}, \bibinfo
  {author} {\bibfnamefont {N.}~\bibnamefont {Regnault}},\ and\ \bibinfo
  {author} {\bibfnamefont {B.~A.}\ \bibnamefont {Bernevig}},\ }\href
  {https://doi.org/10.1103/PhysRevB.103.205412} {\bibfield  {journal} {\bibinfo
   {journal} {Phys. Rev. B}\ }\textbf {\bibinfo {volume} {103}},\ \bibinfo
  {pages} {205412} (\bibinfo {year} {2021})}\BibitemShut {NoStop}%
\bibitem [{\citenamefont {Song}\ and\ \citenamefont
  {Bernevig}(2022)}]{song-heavyfermion-prl22}%
  \BibitemOpen
  \bibfield  {author} {\bibinfo {author} {\bibfnamefont {Z.-D.}\ \bibnamefont
  {Song}}\ and\ \bibinfo {author} {\bibfnamefont {B.~A.}\ \bibnamefont
  {Bernevig}},\ }\href {https://doi.org/10.1103/PhysRevLett.129.047601}
  {\bibfield  {journal} {\bibinfo  {journal} {Phys. Rev. Lett.}\ }\textbf
  {\bibinfo {volume} {129}},\ \bibinfo {pages} {047601} (\bibinfo {year}
  {2022})}\BibitemShut {NoStop}%
\bibitem [{\citenamefont {Shi}\ and\ \citenamefont
  {Dai}(2022)}]{shi-dai-heavy-prb22}%
  \BibitemOpen
  \bibfield  {author} {\bibinfo {author} {\bibfnamefont {H.}~\bibnamefont
  {Shi}}\ and\ \bibinfo {author} {\bibfnamefont {X.}~\bibnamefont {Dai}},\
  }\href {https://doi.org/10.1103/PhysRevB.106.245129} {\bibfield  {journal}
  {\bibinfo  {journal} {Phys. Rev. B}\ }\textbf {\bibinfo {volume} {106}},\
  \bibinfo {pages} {245129} (\bibinfo {year} {2022})}\BibitemShut {NoStop}%
\bibitem [{\citenamefont {Zhang}\ \emph {et~al.}(2019)\citenamefont {Zhang},
  \citenamefont {Mao},\ and\ \citenamefont {Senthil}}]{senthil-tbg-prr19}%
  \BibitemOpen
  \bibfield  {author} {\bibinfo {author} {\bibfnamefont {Y.-H.}\ \bibnamefont
  {Zhang}}, \bibinfo {author} {\bibfnamefont {D.}~\bibnamefont {Mao}},\ and\
  \bibinfo {author} {\bibfnamefont {T.}~\bibnamefont {Senthil}},\ }\href
  {https://doi.org/10.1103/PhysRevResearch.1.033126} {\bibfield  {journal}
  {\bibinfo  {journal} {Phys. Rev. Research}\ }\textbf {\bibinfo {volume}
  {1}},\ \bibinfo {pages} {033126} (\bibinfo {year} {2019})}\BibitemShut
  {NoStop}%
\bibitem [{\citenamefont {Chittari}\ \emph {et~al.}(2019)\citenamefont
  {Chittari}, \citenamefont {Chen}, \citenamefont {Zhang}, \citenamefont
  {Wang},\ and\ \citenamefont {Jung}}]{jung-hbn-trilayer-prl19}%
  \BibitemOpen
  \bibfield  {author} {\bibinfo {author} {\bibfnamefont {B.~L.}\ \bibnamefont
  {Chittari}}, \bibinfo {author} {\bibfnamefont {G.}~\bibnamefont {Chen}},
  \bibinfo {author} {\bibfnamefont {Y.}~\bibnamefont {Zhang}}, \bibinfo
  {author} {\bibfnamefont {F.}~\bibnamefont {Wang}},\ and\ \bibinfo {author}
  {\bibfnamefont {J.}~\bibnamefont {Jung}},\ }\href
  {https://doi.org/10.1103/PhysRevLett.122.016401} {\bibfield  {journal}
  {\bibinfo  {journal} {Phys. Rev. Lett.}\ }\textbf {\bibinfo {volume} {122}},\
  \bibinfo {pages} {016401} (\bibinfo {year} {2019})}\BibitemShut {NoStop}%
\bibitem [{\citenamefont {Park}\ \emph
  {et~al.}(2023{\natexlab{b}})\citenamefont {Park}, \citenamefont {Kim},
  \citenamefont {Chittari},\ and\ \citenamefont
  {Jung}}]{jung-multilayer-arxiv23}%
  \BibitemOpen
  \bibfield  {author} {\bibinfo {author} {\bibfnamefont {Y.}~\bibnamefont
  {Park}}, \bibinfo {author} {\bibfnamefont {Y.}~\bibnamefont {Kim}}, \bibinfo
  {author} {\bibfnamefont {B.~L.}\ \bibnamefont {Chittari}},\ and\ \bibinfo
  {author} {\bibfnamefont {J.}~\bibnamefont {Jung}},\ }\href@noop {} {\bibinfo
  {title} {Topological flat bands in rhombohedral tetralayer and multilayer
  graphene on hexagonal boron nitride moire superlattices}} (\bibinfo {year}
  {2023}{\natexlab{b}}),\ \Eprint {https://arxiv.org/abs/2304.12874}
  {arXiv:2304.12874 [cond-mat.mes-hall]} \BibitemShut {NoStop}%
\bibitem [{\citenamefont {Liu}\ \emph {et~al.}(2019{\natexlab{b}})\citenamefont
  {Liu}, \citenamefont {Ma}, \citenamefont {Gao},\ and\ \citenamefont
  {Dai}}]{jpliu-prx19}%
  \BibitemOpen
  \bibfield  {author} {\bibinfo {author} {\bibfnamefont {J.}~\bibnamefont
  {Liu}}, \bibinfo {author} {\bibfnamefont {Z.}~\bibnamefont {Ma}}, \bibinfo
  {author} {\bibfnamefont {J.}~\bibnamefont {Gao}},\ and\ \bibinfo {author}
  {\bibfnamefont {X.}~\bibnamefont {Dai}},\ }\href
  {https://doi.org/10.1103/PhysRevX.9.031021} {\bibfield  {journal} {\bibinfo
  {journal} {Phys. Rev. X}\ }\textbf {\bibinfo {volume} {9}},\ \bibinfo {pages}
  {031021} (\bibinfo {year} {2019}{\natexlab{b}})}\BibitemShut {NoStop}%
\bibitem [{\citenamefont {Koshino}(2019)}]{koshino-tdbg-prb19}%
  \BibitemOpen
  \bibfield  {author} {\bibinfo {author} {\bibfnamefont {M.}~\bibnamefont
  {Koshino}},\ }\href {https://doi.org/10.1103/PhysRevB.99.235406} {\bibfield
  {journal} {\bibinfo  {journal} {Phys. Rev. B}\ }\textbf {\bibinfo {volume}
  {99}},\ \bibinfo {pages} {235406} (\bibinfo {year} {2019})}\BibitemShut
  {NoStop}%
\bibitem [{\citenamefont {Lee}\ \emph {et~al.}(2019)\citenamefont {Lee},
  \citenamefont {Khalaf}, \citenamefont {Liu}, \citenamefont {Liu},
  \citenamefont {Hao}, \citenamefont {Kim},\ and\ \citenamefont
  {Vishwanath}}]{lee-tdbg-nc19}%
  \BibitemOpen
  \bibfield  {author} {\bibinfo {author} {\bibfnamefont {J.~Y.}\ \bibnamefont
  {Lee}}, \bibinfo {author} {\bibfnamefont {E.}~\bibnamefont {Khalaf}},
  \bibinfo {author} {\bibfnamefont {S.}~\bibnamefont {Liu}}, \bibinfo {author}
  {\bibfnamefont {X.}~\bibnamefont {Liu}}, \bibinfo {author} {\bibfnamefont
  {Z.}~\bibnamefont {Hao}}, \bibinfo {author} {\bibfnamefont {P.}~\bibnamefont
  {Kim}},\ and\ \bibinfo {author} {\bibfnamefont {A.}~\bibnamefont
  {Vishwanath}},\ }\href {https://doi.org/10.1038/s41467-019-12981-1}
  {\bibfield  {journal} {\bibinfo  {journal} {Nature Communications}\ }\textbf
  {\bibinfo {volume} {10}},\ \bibinfo {pages} {5333} (\bibinfo {year}
  {2019})}\BibitemShut {NoStop}%
\bibitem [{\citenamefont {Haddadi}\ \emph {et~al.}(2020)\citenamefont
  {Haddadi}, \citenamefont {Wu}, \citenamefont {Kruchkov},\ and\ \citenamefont
  {Yazyev}}]{quansheng-tdbg-nanoletter20}%
  \BibitemOpen
  \bibfield  {author} {\bibinfo {author} {\bibfnamefont {F.}~\bibnamefont
  {Haddadi}}, \bibinfo {author} {\bibfnamefont {Q.}~\bibnamefont {Wu}},
  \bibinfo {author} {\bibfnamefont {A.~J.}\ \bibnamefont {Kruchkov}},\ and\
  \bibinfo {author} {\bibfnamefont {O.~V.}\ \bibnamefont {Yazyev}},\ }\href
  {https://doi.org/10.1021/acs.nanolett.9b05117} {\bibfield  {journal}
  {\bibinfo  {journal} {Nano Letters}\ }\textbf {\bibinfo {volume} {20}},\
  \bibinfo {pages} {2410} (\bibinfo {year} {2020})}\BibitemShut {NoStop}%
\bibitem [{\citenamefont {Ledwith}\ \emph
  {et~al.}(2022{\natexlab{b}})\citenamefont {Ledwith}, \citenamefont
  {Vishwanath},\ and\ \citenamefont {Khalaf}}]{eslam-tmg-prl22}%
  \BibitemOpen
  \bibfield  {author} {\bibinfo {author} {\bibfnamefont {P.~J.}\ \bibnamefont
  {Ledwith}}, \bibinfo {author} {\bibfnamefont {A.}~\bibnamefont
  {Vishwanath}},\ and\ \bibinfo {author} {\bibfnamefont {E.}~\bibnamefont
  {Khalaf}},\ }\href {https://doi.org/10.1103/PhysRevLett.128.176404}
  {\bibfield  {journal} {\bibinfo  {journal} {Phys. Rev. Lett.}\ }\textbf
  {\bibinfo {volume} {128}},\ \bibinfo {pages} {176404} (\bibinfo {year}
  {2022}{\natexlab{b}})}\BibitemShut {NoStop}%
\bibitem [{\citenamefont {Wang}\ and\ \citenamefont
  {Liu}(2022)}]{wang-tmg-prl22}%
  \BibitemOpen
  \bibfield  {author} {\bibinfo {author} {\bibfnamefont {J.}~\bibnamefont
  {Wang}}\ and\ \bibinfo {author} {\bibfnamefont {Z.}~\bibnamefont {Liu}},\
  }\href {https://doi.org/10.1103/PhysRevLett.128.176403} {\bibfield  {journal}
  {\bibinfo  {journal} {Phys. Rev. Lett.}\ }\textbf {\bibinfo {volume} {128}},\
  \bibinfo {pages} {176403} (\bibinfo {year} {2022})}\BibitemShut {NoStop}%
\bibitem [{\citenamefont {Xie}\ \emph {et~al.}(2022{\natexlab{a}})\citenamefont
  {Xie}, \citenamefont {Peng}, \citenamefont {Zhang},\ and\ \citenamefont
  {Liu}}]{xie-atmg-npj22}%
  \BibitemOpen
  \bibfield  {author} {\bibinfo {author} {\bibfnamefont {B.}~\bibnamefont
  {Xie}}, \bibinfo {author} {\bibfnamefont {R.}~\bibnamefont {Peng}}, \bibinfo
  {author} {\bibfnamefont {S.}~\bibnamefont {Zhang}},\ and\ \bibinfo {author}
  {\bibfnamefont {J.}~\bibnamefont {Liu}},\ }\href
  {https://doi.org/10.1038/s41524-022-00789-5} {\bibfield  {journal} {\bibinfo
  {journal} {npj Computational Materials}\ }\textbf {\bibinfo {volume} {8}},\
  \bibinfo {pages} {110} (\bibinfo {year} {2022}{\natexlab{a}})}\BibitemShut
  {NoStop}%
\bibitem [{\citenamefont {Ledwith}\ \emph {et~al.}(2021)\citenamefont
  {Ledwith}, \citenamefont {Khalaf}, \citenamefont {Zhu}, \citenamefont {Carr},
  \citenamefont {Kaxiras},\ and\ \citenamefont
  {Vishwanath}}]{ashvin-atmg-arxiv21}%
  \BibitemOpen
  \bibfield  {author} {\bibinfo {author} {\bibfnamefont {P.~J.}\ \bibnamefont
  {Ledwith}}, \bibinfo {author} {\bibfnamefont {E.}~\bibnamefont {Khalaf}},
  \bibinfo {author} {\bibfnamefont {Z.}~\bibnamefont {Zhu}}, \bibinfo {author}
  {\bibfnamefont {S.}~\bibnamefont {Carr}}, \bibinfo {author} {\bibfnamefont
  {E.}~\bibnamefont {Kaxiras}},\ and\ \bibinfo {author} {\bibfnamefont
  {A.}~\bibnamefont {Vishwanath}},\ }\href
  {https://doi.org/10.48550/ARXIV.2111.11060} {\bibinfo {title} {Tb or not tb?
  contrasting properties of twisted bilayer graphene and the alternating twist
  $n$-layer structures ($n=3, 4, 5, \dots$)}} (\bibinfo {year}
  {2021})\BibitemShut {NoStop}%
\bibitem [{\citenamefont {Ma}\ \emph {et~al.}(2021)\citenamefont {Ma},
  \citenamefont {Li}, \citenamefont {Zheng}, \citenamefont {Xiao},
  \citenamefont {Jiang}, \citenamefont {Gao},\ and\ \citenamefont
  {Xie}}]{ma-sb2021}%
  \BibitemOpen
  \bibfield  {author} {\bibinfo {author} {\bibfnamefont {Z.}~\bibnamefont
  {Ma}}, \bibinfo {author} {\bibfnamefont {S.}~\bibnamefont {Li}}, \bibinfo
  {author} {\bibfnamefont {Y.-W.}\ \bibnamefont {Zheng}}, \bibinfo {author}
  {\bibfnamefont {M.-M.}\ \bibnamefont {Xiao}}, \bibinfo {author}
  {\bibfnamefont {H.}~\bibnamefont {Jiang}}, \bibinfo {author} {\bibfnamefont
  {J.-H.}\ \bibnamefont {Gao}},\ and\ \bibinfo {author} {\bibfnamefont
  {X.}~\bibnamefont {Xie}},\ }\href
  {https://doi.org/https://doi.org/10.1016/j.scib.2020.10.004} {\bibfield
  {journal} {\bibinfo  {journal} {Science Bulletin}\ }\textbf {\bibinfo
  {volume} {66}},\ \bibinfo {pages} {18} (\bibinfo {year} {2021})}\BibitemShut
  {NoStop}%
\bibitem [{\citenamefont {Zhang}\ \emph {et~al.}(2023)\citenamefont {Zhang},
  \citenamefont {Xie}, \citenamefont {Wu}, \citenamefont {Liu},\ and\
  \citenamefont {Yazyev}}]{zhang-chiral-nl23}%
  \BibitemOpen
  \bibfield  {author} {\bibinfo {author} {\bibfnamefont {S.}~\bibnamefont
  {Zhang}}, \bibinfo {author} {\bibfnamefont {B.}~\bibnamefont {Xie}}, \bibinfo
  {author} {\bibfnamefont {Q.}~\bibnamefont {Wu}}, \bibinfo {author}
  {\bibfnamefont {J.}~\bibnamefont {Liu}},\ and\ \bibinfo {author}
  {\bibfnamefont {O.~V.}\ \bibnamefont {Yazyev}},\ }\href
  {https://doi.org/10.1021/acs.nanolett.3c00275} {\bibfield  {journal}
  {\bibinfo  {journal} {Nano Letters}\ }\textbf {\bibinfo {volume} {23}},\
  \bibinfo {pages} {2921} (\bibinfo {year} {2023})}\BibitemShut {NoStop}%
\bibitem [{\citenamefont {Wu}\ \emph {et~al.}(2019)\citenamefont {Wu},
  \citenamefont {Lovorn}, \citenamefont {Tutuc}, \citenamefont {Martin},\ and\
  \citenamefont {MacDonald}}]{wu-tmd-prl19}%
  \BibitemOpen
  \bibfield  {author} {\bibinfo {author} {\bibfnamefont {F.}~\bibnamefont
  {Wu}}, \bibinfo {author} {\bibfnamefont {T.}~\bibnamefont {Lovorn}}, \bibinfo
  {author} {\bibfnamefont {E.}~\bibnamefont {Tutuc}}, \bibinfo {author}
  {\bibfnamefont {I.}~\bibnamefont {Martin}},\ and\ \bibinfo {author}
  {\bibfnamefont {A.~H.}\ \bibnamefont {MacDonald}},\ }\href
  {https://doi.org/10.1103/PhysRevLett.122.086402} {\bibfield  {journal}
  {\bibinfo  {journal} {Phys. Rev. Lett.}\ }\textbf {\bibinfo {volume} {122}},\
  \bibinfo {pages} {086402} (\bibinfo {year} {2019})}\BibitemShut {NoStop}%
\bibitem [{\citenamefont {Morales-Durán}\ \emph {et~al.}(2023)\citenamefont
  {Morales-Durán}, \citenamefont {Wei},\ and\ \citenamefont
  {MacDonald}}]{macdonald-tmd-ll-arxiv23}%
  \BibitemOpen
  \bibfield  {author} {\bibinfo {author} {\bibfnamefont {N.}~\bibnamefont
  {Morales-Durán}}, \bibinfo {author} {\bibfnamefont {N.}~\bibnamefont
  {Wei}},\ and\ \bibinfo {author} {\bibfnamefont {A.~H.}\ \bibnamefont
  {MacDonald}},\ }\href@noop {} {\bibinfo {title} {Magic angles and fractional
  chern insulators in twisted homobilayer tmds}} (\bibinfo {year} {2023}),\
  \Eprint {https://arxiv.org/abs/2308.03143} {arXiv:2308.03143
  [cond-mat.str-el]} \BibitemShut {NoStop}%
\bibitem [{\citenamefont {Su}\ \emph {et~al.}(2022)\citenamefont {Su},
  \citenamefont {Li}, \citenamefont {Zhang}, \citenamefont {Sun},\ and\
  \citenamefont {Lin}}]{lin-prr22}%
  \BibitemOpen
  \bibfield  {author} {\bibinfo {author} {\bibfnamefont {Y.}~\bibnamefont
  {Su}}, \bibinfo {author} {\bibfnamefont {H.}~\bibnamefont {Li}}, \bibinfo
  {author} {\bibfnamefont {C.}~\bibnamefont {Zhang}}, \bibinfo {author}
  {\bibfnamefont {K.}~\bibnamefont {Sun}},\ and\ \bibinfo {author}
  {\bibfnamefont {S.-Z.}\ \bibnamefont {Lin}},\ }\href
  {https://doi.org/10.1103/PhysRevResearch.4.L032024} {\bibfield  {journal}
  {\bibinfo  {journal} {Phys. Rev. Res.}\ }\textbf {\bibinfo {volume} {4}},\
  \bibinfo {pages} {L032024} (\bibinfo {year} {2022})}\BibitemShut {NoStop}%
\bibitem [{\citenamefont {Reddy}\ \emph
  {et~al.}(2023{\natexlab{a}})\citenamefont {Reddy}, \citenamefont {Alsallom},
  \citenamefont {Zhang}, \citenamefont {Devakul},\ and\ \citenamefont
  {Fu}}]{liangfu-fqah-tmd-prb23}%
  \BibitemOpen
  \bibfield  {author} {\bibinfo {author} {\bibfnamefont {A.~P.}\ \bibnamefont
  {Reddy}}, \bibinfo {author} {\bibfnamefont {F.}~\bibnamefont {Alsallom}},
  \bibinfo {author} {\bibfnamefont {Y.}~\bibnamefont {Zhang}}, \bibinfo
  {author} {\bibfnamefont {T.}~\bibnamefont {Devakul}},\ and\ \bibinfo {author}
  {\bibfnamefont {L.}~\bibnamefont {Fu}},\ }\href
  {https://doi.org/10.1103/PhysRevB.108.085117} {\bibfield  {journal} {\bibinfo
   {journal} {Phys. Rev. B}\ }\textbf {\bibinfo {volume} {108}},\ \bibinfo
  {pages} {085117} (\bibinfo {year} {2023}{\natexlab{a}})}\BibitemShut
  {NoStop}%
\bibitem [{\citenamefont {Crépel}\ \emph {et~al.}(2023)\citenamefont
  {Crépel}, \citenamefont {Regnault},\ and\ \citenamefont
  {Queiroz}}]{nicolas-tmd-chiral-arxiv23}%
  \BibitemOpen
  \bibfield  {author} {\bibinfo {author} {\bibfnamefont {V.}~\bibnamefont
  {Crépel}}, \bibinfo {author} {\bibfnamefont {N.}~\bibnamefont {Regnault}},\
  and\ \bibinfo {author} {\bibfnamefont {R.}~\bibnamefont {Queiroz}},\
  }\href@noop {} {\bibinfo {title} {The chiral limits of moir\'e
  semiconductors: origin of flat bands and topology in twisted transition metal
  dichalcogenides homobilayers}} (\bibinfo {year} {2023}),\ \Eprint
  {https://arxiv.org/abs/2305.10477} {arXiv:2305.10477 [cond-mat.mes-hall]}
  \BibitemShut {NoStop}%
\bibitem [{\citenamefont {Wang}\ \emph {et~al.}(2023)\citenamefont {Wang},
  \citenamefont {Zhang}, \citenamefont {Liu}, \citenamefont {He}, \citenamefont
  {Xu}, \citenamefont {Ran}, \citenamefont {Cao},\ and\ \citenamefont
  {Xiao}}]{xiao-fqah-arxiv23}%
  \BibitemOpen
  \bibfield  {author} {\bibinfo {author} {\bibfnamefont {C.}~\bibnamefont
  {Wang}}, \bibinfo {author} {\bibfnamefont {X.-W.}\ \bibnamefont {Zhang}},
  \bibinfo {author} {\bibfnamefont {X.}~\bibnamefont {Liu}}, \bibinfo {author}
  {\bibfnamefont {Y.}~\bibnamefont {He}}, \bibinfo {author} {\bibfnamefont
  {X.}~\bibnamefont {Xu}}, \bibinfo {author} {\bibfnamefont {Y.}~\bibnamefont
  {Ran}}, \bibinfo {author} {\bibfnamefont {T.}~\bibnamefont {Cao}},\ and\
  \bibinfo {author} {\bibfnamefont {D.}~\bibnamefont {Xiao}},\ }\href@noop {}
  {\bibinfo {title} {Fractional chern insulator in twisted bilayer mote$_2$}}
  (\bibinfo {year} {2023}),\ \Eprint {https://arxiv.org/abs/2304.11864}
  {arXiv:2304.11864 [cond-mat.str-el]} \BibitemShut {NoStop}%
\bibitem [{\citenamefont {Xu}\ \emph {et~al.}(2023{\natexlab{b}})\citenamefont
  {Xu}, \citenamefont {Li}, \citenamefont {Xu}, \citenamefont {Bi},\ and\
  \citenamefont {Zhang}}]{zhangyang-fqah-tmd-arxiv23}%
  \BibitemOpen
  \bibfield  {author} {\bibinfo {author} {\bibfnamefont {C.}~\bibnamefont
  {Xu}}, \bibinfo {author} {\bibfnamefont {J.}~\bibnamefont {Li}}, \bibinfo
  {author} {\bibfnamefont {Y.}~\bibnamefont {Xu}}, \bibinfo {author}
  {\bibfnamefont {Z.}~\bibnamefont {Bi}},\ and\ \bibinfo {author}
  {\bibfnamefont {Y.}~\bibnamefont {Zhang}},\ }\href@noop {} {\bibinfo {title}
  {Maximally localized wannier orbitals, interaction models and fractional
  quantum anomalous hall effect in twisted bilayer mote2}} (\bibinfo {year}
  {2023}{\natexlab{b}}),\ \Eprint {https://arxiv.org/abs/2308.09697}
  {arXiv:2308.09697 [cond-mat.str-el]} \BibitemShut {NoStop}%
\bibitem [{\citenamefont {Fan}\ \emph {et~al.}(2023)\citenamefont {Fan},
  \citenamefont {Xiao},\ and\ \citenamefont {Yao}}]{yao-orbitalchern-arxiv23}%
  \BibitemOpen
  \bibfield  {author} {\bibinfo {author} {\bibfnamefont {F.-R.}\ \bibnamefont
  {Fan}}, \bibinfo {author} {\bibfnamefont {C.}~\bibnamefont {Xiao}},\ and\
  \bibinfo {author} {\bibfnamefont {W.}~\bibnamefont {Yao}},\ }\href@noop {}
  {\bibinfo {title} {Altermagnetic orbital chern insulator in twisted
  mote$_{2}$}} (\bibinfo {year} {2023}),\ \Eprint
  {https://arxiv.org/abs/2308.11454} {arXiv:2308.11454 [cond-mat.str-el]}
  \BibitemShut {NoStop}%
\bibitem [{\citenamefont {Xie}\ \emph {et~al.}(2022{\natexlab{b}})\citenamefont
  {Xie}, \citenamefont {Zhang}, \citenamefont {Hu}, \citenamefont {Mak},\ and\
  \citenamefont {Law}}]{law-tmd-prl22}%
  \BibitemOpen
  \bibfield  {author} {\bibinfo {author} {\bibfnamefont {Y.-M.}\ \bibnamefont
  {Xie}}, \bibinfo {author} {\bibfnamefont {C.-P.}\ \bibnamefont {Zhang}},
  \bibinfo {author} {\bibfnamefont {J.-X.}\ \bibnamefont {Hu}}, \bibinfo
  {author} {\bibfnamefont {K.~F.}\ \bibnamefont {Mak}},\ and\ \bibinfo {author}
  {\bibfnamefont {K.~T.}\ \bibnamefont {Law}},\ }\href
  {https://doi.org/10.1103/PhysRevLett.128.026402} {\bibfield  {journal}
  {\bibinfo  {journal} {Phys. Rev. Lett.}\ }\textbf {\bibinfo {volume} {128}},\
  \bibinfo {pages} {026402} (\bibinfo {year} {2022}{\natexlab{b}})}\BibitemShut
  {NoStop}%
\bibitem [{\citenamefont {Serlin}\ \emph {et~al.}(2019)\citenamefont {Serlin},
  \citenamefont {Tschirhart}, \citenamefont {Polshyn}, \citenamefont {Zhang},
  \citenamefont {Zhu}, \citenamefont {Watanabe}, \citenamefont {Taniguchi},
  \citenamefont {Balents},\ and\ \citenamefont {Young}}]{young-tbg-science19}%
  \BibitemOpen
  \bibfield  {author} {\bibinfo {author} {\bibfnamefont {M.}~\bibnamefont
  {Serlin}}, \bibinfo {author} {\bibfnamefont {C.}~\bibnamefont {Tschirhart}},
  \bibinfo {author} {\bibfnamefont {H.}~\bibnamefont {Polshyn}}, \bibinfo
  {author} {\bibfnamefont {Y.}~\bibnamefont {Zhang}}, \bibinfo {author}
  {\bibfnamefont {J.}~\bibnamefont {Zhu}}, \bibinfo {author} {\bibfnamefont
  {K.}~\bibnamefont {Watanabe}}, \bibinfo {author} {\bibfnamefont
  {T.}~\bibnamefont {Taniguchi}}, \bibinfo {author} {\bibfnamefont
  {L.}~\bibnamefont {Balents}},\ and\ \bibinfo {author} {\bibfnamefont
  {A.}~\bibnamefont {Young}},\ }\href@noop {} {\bibfield  {journal} {\bibinfo
  {journal} {Science}\ } (\bibinfo {year} {2019})}\BibitemShut {NoStop}%
\bibitem [{\citenamefont {Li}\ \emph {et~al.}(2021)\citenamefont {Li},
  \citenamefont {Jiang}, \citenamefont {Shen}, \citenamefont {Zhang},
  \citenamefont {Li}, \citenamefont {Tao}, \citenamefont {Devakul},
  \citenamefont {Watanabe}, \citenamefont {Taniguchi}, \citenamefont {Fu},
  \citenamefont {Shan},\ and\ \citenamefont {Mak}}]{Mak-mote2-wse2-nature2021}%
  \BibitemOpen
  \bibfield  {author} {\bibinfo {author} {\bibfnamefont {T.}~\bibnamefont
  {Li}}, \bibinfo {author} {\bibfnamefont {S.}~\bibnamefont {Jiang}}, \bibinfo
  {author} {\bibfnamefont {B.}~\bibnamefont {Shen}}, \bibinfo {author}
  {\bibfnamefont {Y.}~\bibnamefont {Zhang}}, \bibinfo {author} {\bibfnamefont
  {L.}~\bibnamefont {Li}}, \bibinfo {author} {\bibfnamefont {Z.}~\bibnamefont
  {Tao}}, \bibinfo {author} {\bibfnamefont {T.}~\bibnamefont {Devakul}},
  \bibinfo {author} {\bibfnamefont {K.}~\bibnamefont {Watanabe}}, \bibinfo
  {author} {\bibfnamefont {T.}~\bibnamefont {Taniguchi}}, \bibinfo {author}
  {\bibfnamefont {L.}~\bibnamefont {Fu}}, \bibinfo {author} {\bibfnamefont
  {J.}~\bibnamefont {Shan}},\ and\ \bibinfo {author} {\bibfnamefont {K.~F.}\
  \bibnamefont {Mak}},\ }\href {https://doi.org/10.1038/s41586-021-04171-1}
  {\bibfield  {journal} {\bibinfo  {journal} {Nature}\ }\textbf {\bibinfo
  {volume} {600}},\ \bibinfo {pages} {641} (\bibinfo {year}
  {2021})}\BibitemShut {NoStop}%
\bibitem [{\citenamefont {Polshyn}\ \emph {et~al.}(2020)\citenamefont
  {Polshyn}, \citenamefont {Zhu}, \citenamefont {Kumar}, \citenamefont {Zhang},
  \citenamefont {Yang}, \citenamefont {Tschirhart}, \citenamefont {Serlin},
  \citenamefont {Watanabe}, \citenamefont {Taniguchi}, \citenamefont
  {MacDonald} \emph {et~al.}}]{young-monobi-nature20}%
  \BibitemOpen
  \bibfield  {author} {\bibinfo {author} {\bibfnamefont {H.}~\bibnamefont
  {Polshyn}}, \bibinfo {author} {\bibfnamefont {J.}~\bibnamefont {Zhu}},
  \bibinfo {author} {\bibfnamefont {M.}~\bibnamefont {Kumar}}, \bibinfo
  {author} {\bibfnamefont {Y.}~\bibnamefont {Zhang}}, \bibinfo {author}
  {\bibfnamefont {F.}~\bibnamefont {Yang}}, \bibinfo {author} {\bibfnamefont
  {C.}~\bibnamefont {Tschirhart}}, \bibinfo {author} {\bibfnamefont
  {M.}~\bibnamefont {Serlin}}, \bibinfo {author} {\bibfnamefont
  {K.}~\bibnamefont {Watanabe}}, \bibinfo {author} {\bibfnamefont
  {T.}~\bibnamefont {Taniguchi}}, \bibinfo {author} {\bibfnamefont
  {A.}~\bibnamefont {MacDonald}}, \emph {et~al.},\ }\href@noop {} {\bibfield
  {journal} {\bibinfo  {journal} {Nature}\ ,\ \bibinfo {pages} {1}} (\bibinfo
  {year} {2020})}\BibitemShut {NoStop}%
\bibitem [{\citenamefont {Chen}\ \emph {et~al.}(2019)\citenamefont {Chen},
  \citenamefont {Sharpe}, \citenamefont {Gallagher}, \citenamefont {Rosen},
  \citenamefont {Fox}, \citenamefont {Jiang}, \citenamefont {Lyu},
  \citenamefont {Li}, \citenamefont {Watanabe}, \citenamefont {Taniguchi},
  \citenamefont {Jung}, \citenamefont {Shi}, \citenamefont {Goldhaber-Gordon},
  \citenamefont {Zhang},\ and\ \citenamefont
  {Wang}}]{chen-hbn-trilayer-nature19}%
  \BibitemOpen
  \bibfield  {author} {\bibinfo {author} {\bibfnamefont {G.}~\bibnamefont
  {Chen}}, \bibinfo {author} {\bibfnamefont {A.~L.}\ \bibnamefont {Sharpe}},
  \bibinfo {author} {\bibfnamefont {P.}~\bibnamefont {Gallagher}}, \bibinfo
  {author} {\bibfnamefont {I.~T.}\ \bibnamefont {Rosen}}, \bibinfo {author}
  {\bibfnamefont {E.~J.}\ \bibnamefont {Fox}}, \bibinfo {author} {\bibfnamefont
  {L.}~\bibnamefont {Jiang}}, \bibinfo {author} {\bibfnamefont
  {B.}~\bibnamefont {Lyu}}, \bibinfo {author} {\bibfnamefont {H.}~\bibnamefont
  {Li}}, \bibinfo {author} {\bibfnamefont {K.}~\bibnamefont {Watanabe}},
  \bibinfo {author} {\bibfnamefont {T.}~\bibnamefont {Taniguchi}}, \bibinfo
  {author} {\bibfnamefont {J.}~\bibnamefont {Jung}}, \bibinfo {author}
  {\bibfnamefont {Z.}~\bibnamefont {Shi}}, \bibinfo {author} {\bibfnamefont
  {D.}~\bibnamefont {Goldhaber-Gordon}}, \bibinfo {author} {\bibfnamefont
  {Y.}~\bibnamefont {Zhang}},\ and\ \bibinfo {author} {\bibfnamefont
  {F.}~\bibnamefont {Wang}},\ }\href
  {https://doi.org/10.1038/s41586-019-1393-y} {\bibfield  {journal} {\bibinfo
  {journal} {Nature}\ }\textbf {\bibinfo {volume} {572}},\ \bibinfo {pages}
  {215} (\bibinfo {year} {2019})}\BibitemShut {NoStop}%
\bibitem [{\citenamefont {Wu}\ \emph {et~al.}(2021)\citenamefont {Wu},
  \citenamefont {Zhang}, \citenamefont {Watanabe}, \citenamefont {Taniguchi},\
  and\ \citenamefont {Andrei}}]{andrei-tbg-chern-arxiv20}%
  \BibitemOpen
  \bibfield  {author} {\bibinfo {author} {\bibfnamefont {S.}~\bibnamefont
  {Wu}}, \bibinfo {author} {\bibfnamefont {Z.}~\bibnamefont {Zhang}}, \bibinfo
  {author} {\bibfnamefont {K.}~\bibnamefont {Watanabe}}, \bibinfo {author}
  {\bibfnamefont {T.}~\bibnamefont {Taniguchi}},\ and\ \bibinfo {author}
  {\bibfnamefont {E.~Y.}\ \bibnamefont {Andrei}},\ }\href@noop {} {\bibfield
  {journal} {\bibinfo  {journal} {Nature Materials}\ } (\bibinfo {year}
  {2021})}\BibitemShut {NoStop}%
\bibitem [{\citenamefont {Nuckolls}\ \emph {et~al.}(2020)\citenamefont
  {Nuckolls}, \citenamefont {Oh}, \citenamefont {Wong}, \citenamefont {Lian},
  \citenamefont {Watanabe}, \citenamefont {Taniguchi}, \citenamefont
  {Bernevig},\ and\ \citenamefont {Yazdani}}]{yazdani-tbg-chern-arxiv20}%
  \BibitemOpen
  \bibfield  {author} {\bibinfo {author} {\bibfnamefont {K.~P.}\ \bibnamefont
  {Nuckolls}}, \bibinfo {author} {\bibfnamefont {M.}~\bibnamefont {Oh}},
  \bibinfo {author} {\bibfnamefont {D.}~\bibnamefont {Wong}}, \bibinfo {author}
  {\bibfnamefont {B.}~\bibnamefont {Lian}}, \bibinfo {author} {\bibfnamefont
  {K.}~\bibnamefont {Watanabe}}, \bibinfo {author} {\bibfnamefont
  {T.}~\bibnamefont {Taniguchi}}, \bibinfo {author} {\bibfnamefont {B.~A.}\
  \bibnamefont {Bernevig}},\ and\ \bibinfo {author} {\bibfnamefont
  {A.}~\bibnamefont {Yazdani}},\ }\href
  {https://doi.org/10.1038/s41586-020-3028-8} {\bibfield  {journal} {\bibinfo
  {journal} {Nature}\ }\textbf {\bibinfo {volume} {588}},\ \bibinfo {pages}
  {610} (\bibinfo {year} {2020})}\BibitemShut {NoStop}%
\bibitem [{\citenamefont {Pierce}\ \emph
  {et~al.}(2021{\natexlab{a}})\citenamefont {Pierce}, \citenamefont {Xie},
  \citenamefont {Park}, \citenamefont {Khalaf}, \citenamefont {Lee},
  \citenamefont {Cao}, \citenamefont {Parker}, \citenamefont {Forrester},
  \citenamefont {Chen}, \citenamefont {Watanabe} \emph
  {et~al.}}]{pablo-tbg-chern-arxiv21}%
  \BibitemOpen
  \bibfield  {author} {\bibinfo {author} {\bibfnamefont {A.~T.}\ \bibnamefont
  {Pierce}}, \bibinfo {author} {\bibfnamefont {Y.}~\bibnamefont {Xie}},
  \bibinfo {author} {\bibfnamefont {J.~M.}\ \bibnamefont {Park}}, \bibinfo
  {author} {\bibfnamefont {E.}~\bibnamefont {Khalaf}}, \bibinfo {author}
  {\bibfnamefont {S.~H.}\ \bibnamefont {Lee}}, \bibinfo {author} {\bibfnamefont
  {Y.}~\bibnamefont {Cao}}, \bibinfo {author} {\bibfnamefont {D.~E.}\
  \bibnamefont {Parker}}, \bibinfo {author} {\bibfnamefont {P.~R.}\
  \bibnamefont {Forrester}}, \bibinfo {author} {\bibfnamefont {S.}~\bibnamefont
  {Chen}}, \bibinfo {author} {\bibfnamefont {K.}~\bibnamefont {Watanabe}},
  \emph {et~al.},\ }\href@noop {} {\bibfield  {journal} {\bibinfo  {journal}
  {Nature Physics}\ }\textbf {\bibinfo {volume} {17}},\ \bibinfo {pages} {1210}
  (\bibinfo {year} {2021}{\natexlab{a}})}\BibitemShut {NoStop}%
\bibitem [{\citenamefont {Pierce}\ \emph
  {et~al.}(2021{\natexlab{b}})\citenamefont {Pierce}, \citenamefont {Xie},
  \citenamefont {Park}, \citenamefont {Khalaf}, \citenamefont {Lee},
  \citenamefont {Cao}, \citenamefont {Parker}, \citenamefont {Forrester},
  \citenamefont {Chen}, \citenamefont {Watanabe}, \citenamefont {Taniguchi},
  \citenamefont {Vishwanath}, \citenamefont {Jarillo-Herrero},\ and\
  \citenamefont {Yacoby}}]{yacoby-ashvin-chern-tbg-natphys21}%
  \BibitemOpen
  \bibfield  {author} {\bibinfo {author} {\bibfnamefont {A.~T.}\ \bibnamefont
  {Pierce}}, \bibinfo {author} {\bibfnamefont {Y.}~\bibnamefont {Xie}},
  \bibinfo {author} {\bibfnamefont {J.~M.}\ \bibnamefont {Park}}, \bibinfo
  {author} {\bibfnamefont {E.}~\bibnamefont {Khalaf}}, \bibinfo {author}
  {\bibfnamefont {S.~H.}\ \bibnamefont {Lee}}, \bibinfo {author} {\bibfnamefont
  {Y.}~\bibnamefont {Cao}}, \bibinfo {author} {\bibfnamefont {D.~E.}\
  \bibnamefont {Parker}}, \bibinfo {author} {\bibfnamefont {P.~R.}\
  \bibnamefont {Forrester}}, \bibinfo {author} {\bibfnamefont {S.}~\bibnamefont
  {Chen}}, \bibinfo {author} {\bibfnamefont {K.}~\bibnamefont {Watanabe}},
  \bibinfo {author} {\bibfnamefont {T.}~\bibnamefont {Taniguchi}}, \bibinfo
  {author} {\bibfnamefont {A.}~\bibnamefont {Vishwanath}}, \bibinfo {author}
  {\bibfnamefont {P.}~\bibnamefont {Jarillo-Herrero}},\ and\ \bibinfo {author}
  {\bibfnamefont {A.}~\bibnamefont {Yacoby}},\ }\href
  {https://doi.org/10.1038/s41567-021-01347-4} {\bibfield  {journal} {\bibinfo
  {journal} {Nature Physics}\ }\textbf {\bibinfo {volume} {17}},\ \bibinfo
  {pages} {1210} (\bibinfo {year} {2021}{\natexlab{b}})}\BibitemShut {NoStop}%
\bibitem [{\citenamefont {Das}\ \emph {et~al.}(2021)\citenamefont {Das},
  \citenamefont {Lu}, \citenamefont {Herzog-Arbeitman}, \citenamefont {Song},
  \citenamefont {Watanabe}, \citenamefont {Taniguchi}, \citenamefont
  {Bernevig},\ and\ \citenamefont {Efetov}}]{efetov-tbg-chern-arxiv20}%
  \BibitemOpen
  \bibfield  {author} {\bibinfo {author} {\bibfnamefont {I.}~\bibnamefont
  {Das}}, \bibinfo {author} {\bibfnamefont {X.}~\bibnamefont {Lu}}, \bibinfo
  {author} {\bibfnamefont {J.}~\bibnamefont {Herzog-Arbeitman}}, \bibinfo
  {author} {\bibfnamefont {Z.-D.}\ \bibnamefont {Song}}, \bibinfo {author}
  {\bibfnamefont {K.}~\bibnamefont {Watanabe}}, \bibinfo {author}
  {\bibfnamefont {T.}~\bibnamefont {Taniguchi}}, \bibinfo {author}
  {\bibfnamefont {B.~A.}\ \bibnamefont {Bernevig}},\ and\ \bibinfo {author}
  {\bibfnamefont {D.~K.}\ \bibnamefont {Efetov}},\ }\href
  {https://doi.org/10.1038/s41567-021-01186-3} {\bibfield  {journal} {\bibinfo
  {journal} {Nature Physics}\ }\textbf {\bibinfo {volume} {17}},\ \bibinfo
  {pages} {710} (\bibinfo {year} {2021})}\BibitemShut {NoStop}%
\bibitem [{\citenamefont {Ledwith}\ \emph {et~al.}(2020)\citenamefont
  {Ledwith}, \citenamefont {Tarnopolsky}, \citenamefont {Khalaf},\ and\
  \citenamefont {Vishwanath}}]{ashvin-fci-tbg-prr20}%
  \BibitemOpen
  \bibfield  {author} {\bibinfo {author} {\bibfnamefont {P.~J.}\ \bibnamefont
  {Ledwith}}, \bibinfo {author} {\bibfnamefont {G.}~\bibnamefont
  {Tarnopolsky}}, \bibinfo {author} {\bibfnamefont {E.}~\bibnamefont
  {Khalaf}},\ and\ \bibinfo {author} {\bibfnamefont {A.}~\bibnamefont
  {Vishwanath}},\ }\href {https://doi.org/10.1103/PhysRevResearch.2.023237}
  {\bibfield  {journal} {\bibinfo  {journal} {Phys. Rev. Res.}\ }\textbf
  {\bibinfo {volume} {2}},\ \bibinfo {pages} {023237} (\bibinfo {year}
  {2020})}\BibitemShut {NoStop}%
\bibitem [{\citenamefont {Parker}\ \emph {et~al.}(2021)\citenamefont {Parker},
  \citenamefont {Ledwith}, \citenamefont {Khalaf}, \citenamefont {Soejima},
  \citenamefont {Hauschild}, \citenamefont {Xie}, \citenamefont {Pierce},
  \citenamefont {Zaletel}, \citenamefont {Yacoby},\ and\ \citenamefont
  {Vishwanath}}]{ashvin-fci-tbg-arxiv21}%
  \BibitemOpen
  \bibfield  {author} {\bibinfo {author} {\bibfnamefont {D.}~\bibnamefont
  {Parker}}, \bibinfo {author} {\bibfnamefont {P.}~\bibnamefont {Ledwith}},
  \bibinfo {author} {\bibfnamefont {E.}~\bibnamefont {Khalaf}}, \bibinfo
  {author} {\bibfnamefont {T.}~\bibnamefont {Soejima}}, \bibinfo {author}
  {\bibfnamefont {J.}~\bibnamefont {Hauschild}}, \bibinfo {author}
  {\bibfnamefont {Y.}~\bibnamefont {Xie}}, \bibinfo {author} {\bibfnamefont
  {A.}~\bibnamefont {Pierce}}, \bibinfo {author} {\bibfnamefont {M.~P.}\
  \bibnamefont {Zaletel}}, \bibinfo {author} {\bibfnamefont {A.}~\bibnamefont
  {Yacoby}},\ and\ \bibinfo {author} {\bibfnamefont {A.}~\bibnamefont
  {Vishwanath}},\ }\href@noop {} {\bibinfo {title} {Field-tuned and zero-field
  fractional chern insulators in magic angle graphene}} (\bibinfo {year}
  {2021}),\ \Eprint {https://arxiv.org/abs/2112.13837} {arXiv:2112.13837
  [cond-mat.str-el]} \BibitemShut {NoStop}%
\bibitem [{\citenamefont {Spanton}\ \emph {et~al.}(2018)\citenamefont
  {Spanton}, \citenamefont {Zibrov}, \citenamefont {Zhou}, \citenamefont
  {Taniguchi}, \citenamefont {Watanabe}, \citenamefont {Zaletel},\ and\
  \citenamefont {Young}}]{young-science18}%
  \BibitemOpen
  \bibfield  {author} {\bibinfo {author} {\bibfnamefont {E.~M.}\ \bibnamefont
  {Spanton}}, \bibinfo {author} {\bibfnamefont {A.~A.}\ \bibnamefont {Zibrov}},
  \bibinfo {author} {\bibfnamefont {H.}~\bibnamefont {Zhou}}, \bibinfo {author}
  {\bibfnamefont {T.}~\bibnamefont {Taniguchi}}, \bibinfo {author}
  {\bibfnamefont {K.}~\bibnamefont {Watanabe}}, \bibinfo {author}
  {\bibfnamefont {M.~P.}\ \bibnamefont {Zaletel}},\ and\ \bibinfo {author}
  {\bibfnamefont {A.~F.}\ \bibnamefont {Young}},\ }\href
  {https://doi.org/10.1126/science.aan8458} {\bibfield  {journal} {\bibinfo
  {journal} {Science}\ }\textbf {\bibinfo {volume} {360}},\ \bibinfo {pages}
  {62} (\bibinfo {year} {2018})},\ \Eprint
  {https://arxiv.org/abs/https://www.science.org/doi/pdf/10.1126/science.aan8458}
  {https://www.science.org/doi/pdf/10.1126/science.aan8458} \BibitemShut
  {NoStop}%
\bibitem [{\citenamefont {Xie}\ \emph {et~al.}(2021{\natexlab{a}})\citenamefont
  {Xie}, \citenamefont {Pierce}, \citenamefont {Park}, \citenamefont {Parker},
  \citenamefont {Khalaf}, \citenamefont {Ledwith}, \citenamefont {Cao},
  \citenamefont {Lee}, \citenamefont {Chen}, \citenamefont {Forrester},
  \citenamefont {Watanabe}, \citenamefont {Taniguchi}, \citenamefont
  {Vishwanath}, \citenamefont {Jarillo-Herrero},\ and\ \citenamefont
  {Yacoby}}]{xie-tbgfci-nature21}%
  \BibitemOpen
  \bibfield  {author} {\bibinfo {author} {\bibfnamefont {Y.}~\bibnamefont
  {Xie}}, \bibinfo {author} {\bibfnamefont {A.~T.}\ \bibnamefont {Pierce}},
  \bibinfo {author} {\bibfnamefont {J.~M.}\ \bibnamefont {Park}}, \bibinfo
  {author} {\bibfnamefont {D.~E.}\ \bibnamefont {Parker}}, \bibinfo {author}
  {\bibfnamefont {E.}~\bibnamefont {Khalaf}}, \bibinfo {author} {\bibfnamefont
  {P.}~\bibnamefont {Ledwith}}, \bibinfo {author} {\bibfnamefont
  {Y.}~\bibnamefont {Cao}}, \bibinfo {author} {\bibfnamefont {S.~H.}\
  \bibnamefont {Lee}}, \bibinfo {author} {\bibfnamefont {S.}~\bibnamefont
  {Chen}}, \bibinfo {author} {\bibfnamefont {P.~R.}\ \bibnamefont {Forrester}},
  \bibinfo {author} {\bibfnamefont {K.}~\bibnamefont {Watanabe}}, \bibinfo
  {author} {\bibfnamefont {T.}~\bibnamefont {Taniguchi}}, \bibinfo {author}
  {\bibfnamefont {A.}~\bibnamefont {Vishwanath}}, \bibinfo {author}
  {\bibfnamefont {P.}~\bibnamefont {Jarillo-Herrero}},\ and\ \bibinfo {author}
  {\bibfnamefont {A.}~\bibnamefont {Yacoby}},\ }\href
  {https://doi.org/10.1038/s41586-021-04002-3} {\bibfield  {journal} {\bibinfo
  {journal} {Nature}\ }\textbf {\bibinfo {volume} {600}},\ \bibinfo {pages}
  {439} (\bibinfo {year} {2021}{\natexlab{a}})}\BibitemShut {NoStop}%
\bibitem [{\citenamefont {Xie}\ \emph {et~al.}(2021{\natexlab{b}})\citenamefont
  {Xie}, \citenamefont {Pierce}, \citenamefont {Park}, \citenamefont {Parker},
  \citenamefont {Khalaf}, \citenamefont {Ledwith}, \citenamefont {Cao},
  \citenamefont {Lee}, \citenamefont {Chen}, \citenamefont {Forrester},
  \citenamefont {Watanabe}, \citenamefont {Taniguchi}, \citenamefont
  {Vishwanath}, \citenamefont {Jarillo-Herrero},\ and\ \citenamefont
  {Yacoby}}]{yacoby-tbg-fqah-arxiv21}%
  \BibitemOpen
  \bibfield  {author} {\bibinfo {author} {\bibfnamefont {Y.}~\bibnamefont
  {Xie}}, \bibinfo {author} {\bibfnamefont {A.~T.}\ \bibnamefont {Pierce}},
  \bibinfo {author} {\bibfnamefont {J.~M.}\ \bibnamefont {Park}}, \bibinfo
  {author} {\bibfnamefont {D.~E.}\ \bibnamefont {Parker}}, \bibinfo {author}
  {\bibfnamefont {E.}~\bibnamefont {Khalaf}}, \bibinfo {author} {\bibfnamefont
  {P.}~\bibnamefont {Ledwith}}, \bibinfo {author} {\bibfnamefont
  {Y.}~\bibnamefont {Cao}}, \bibinfo {author} {\bibfnamefont {S.~H.}\
  \bibnamefont {Lee}}, \bibinfo {author} {\bibfnamefont {S.}~\bibnamefont
  {Chen}}, \bibinfo {author} {\bibfnamefont {P.~R.}\ \bibnamefont {Forrester}},
  \bibinfo {author} {\bibfnamefont {K.}~\bibnamefont {Watanabe}}, \bibinfo
  {author} {\bibfnamefont {T.}~\bibnamefont {Taniguchi}}, \bibinfo {author}
  {\bibfnamefont {A.}~\bibnamefont {Vishwanath}}, \bibinfo {author}
  {\bibfnamefont {P.}~\bibnamefont {Jarillo-Herrero}},\ and\ \bibinfo {author}
  {\bibfnamefont {A.}~\bibnamefont {Yacoby}},\ }\href
  {https://doi.org/10.1038/s41586-021-04002-3} {\bibfield  {journal} {\bibinfo
  {journal} {Nature}\ }\textbf {\bibinfo {volume} {600}},\ \bibinfo {pages}
  {439} (\bibinfo {year} {2021}{\natexlab{b}})},\ \Eprint
  {https://arxiv.org/abs/arXiv:2107.10854} {arXiv:arXiv:2107.10854
  [cond-mat.mes-hall]} \BibitemShut {NoStop}%
\bibitem [{\citenamefont {Moon}\ and\ \citenamefont
  {Koshino}(2013)}]{moon-tbg-prb13}%
  \BibitemOpen
  \bibfield  {author} {\bibinfo {author} {\bibfnamefont {P.}~\bibnamefont
  {Moon}}\ and\ \bibinfo {author} {\bibfnamefont {M.}~\bibnamefont {Koshino}},\
  }\href@noop {} {\bibfield  {journal} {\bibinfo  {journal} {Physical Review
  B}\ }\textbf {\bibinfo {volume} {87}},\ \bibinfo {pages} {205404} (\bibinfo
  {year} {2013})}\BibitemShut {NoStop}%
\bibitem [{sup()}]{supp_info}%
  \BibitemOpen
  \href@noop {} {}\bibinfo {note} {See Supplemental Information for: (a)
  details of the continuum model, (b) details of the renormalization group
  calculations, (c) detailed formalism for the Hartree-Fock calculations, (d)
  more results from the exact-diagonalization calculations, (e) lattice
  relaxations and their effects on electronic band structures.}\BibitemShut
  {Stop}%
\bibitem [{\citenamefont {Moon}\ and\ \citenamefont
  {Koshino}(2014{\natexlab{a}})}]{moon-prb14}%
  \BibitemOpen
  \bibfield  {author} {\bibinfo {author} {\bibfnamefont {P.}~\bibnamefont
  {Moon}}\ and\ \bibinfo {author} {\bibfnamefont {M.}~\bibnamefont {Koshino}},\
  }\href {https://doi.org/10.1103/PhysRevB.90.155406} {\bibfield  {journal}
  {\bibinfo  {journal} {Phys. Rev. B}\ }\textbf {\bibinfo {volume} {90}},\
  \bibinfo {pages} {155406} (\bibinfo {year} {2014}{\natexlab{a}})}\BibitemShut
  {NoStop}%
\bibitem [{\citenamefont {Jang}\ \emph
  {et~al.}(2023{\natexlab{a}})\citenamefont {Jang}, \citenamefont {Park},
  \citenamefont {Jung},\ and\ \citenamefont {Min}}]{min-prb23}%
  \BibitemOpen
  \bibfield  {author} {\bibinfo {author} {\bibfnamefont {Y.}~\bibnamefont
  {Jang}}, \bibinfo {author} {\bibfnamefont {Y.}~\bibnamefont {Park}}, \bibinfo
  {author} {\bibfnamefont {J.}~\bibnamefont {Jung}},\ and\ \bibinfo {author}
  {\bibfnamefont {H.}~\bibnamefont {Min}},\ }\href
  {https://doi.org/10.1103/PhysRevB.108.L041101} {\bibfield  {journal}
  {\bibinfo  {journal} {Phys. Rev. B}\ }\textbf {\bibinfo {volume} {108}},\
  \bibinfo {pages} {L041101} (\bibinfo {year}
  {2023}{\natexlab{a}})}\BibitemShut {NoStop}%
\bibitem [{\citenamefont {Gonzalez}\ \emph {et~al.}(1994)\citenamefont
  {Gonzalez}, \citenamefont {Guinea},\ and\ \citenamefont
  {H.}}]{gonzalez_nuclphysb1993}%
  \BibitemOpen
  \bibfield  {author} {\bibinfo {author} {\bibfnamefont {J.}~\bibnamefont
  {Gonzalez}}, \bibinfo {author} {\bibfnamefont {F.}~\bibnamefont {Guinea}},\
  and\ \bibinfo {author} {\bibfnamefont {V.~M.~A.}\ \bibnamefont {H.}},\ }\href
  {https://doi.org/https://doi.org/10.1016/0550-3213(94)90410-3} {\bibfield
  {journal} {\bibinfo  {journal} {Nuclear Physics B}\ }\textbf {\bibinfo
  {volume} {424}},\ \bibinfo {pages} {595} (\bibinfo {year}
  {1994})}\BibitemShut {NoStop}%
\bibitem [{\citenamefont {Kotov}\ \emph {et~al.}(2012)\citenamefont {Kotov},
  \citenamefont {Uchoa}, \citenamefont {Pereira}, \citenamefont {Guinea},\ and\
  \citenamefont {Castro~Neto}}]{kotov_rmp2012}%
  \BibitemOpen
  \bibfield  {author} {\bibinfo {author} {\bibfnamefont {V.~N.}\ \bibnamefont
  {Kotov}}, \bibinfo {author} {\bibfnamefont {B.}~\bibnamefont {Uchoa}},
  \bibinfo {author} {\bibfnamefont {V.~M.}\ \bibnamefont {Pereira}}, \bibinfo
  {author} {\bibfnamefont {F.}~\bibnamefont {Guinea}},\ and\ \bibinfo {author}
  {\bibfnamefont {A.~H.}\ \bibnamefont {Castro~Neto}},\ }\href
  {https://doi.org/10.1103/RevModPhys.84.1067} {\bibfield  {journal} {\bibinfo
  {journal} {Rev. Mod. Phys.}\ }\textbf {\bibinfo {volume} {84}},\ \bibinfo
  {pages} {1067} (\bibinfo {year} {2012})}\BibitemShut {NoStop}%
\bibitem [{\citenamefont {Castro~Neto}\ \emph {et~al.}(2009)\citenamefont
  {Castro~Neto}, \citenamefont {Guinea}, \citenamefont {Peres}, \citenamefont
  {Novoselov},\ and\ \citenamefont {Geim}}]{castroneto_rmp2009}%
  \BibitemOpen
  \bibfield  {author} {\bibinfo {author} {\bibfnamefont {A.~H.}\ \bibnamefont
  {Castro~Neto}}, \bibinfo {author} {\bibfnamefont {F.}~\bibnamefont {Guinea}},
  \bibinfo {author} {\bibfnamefont {N.~M.~R.}\ \bibnamefont {Peres}}, \bibinfo
  {author} {\bibfnamefont {K.~S.}\ \bibnamefont {Novoselov}},\ and\ \bibinfo
  {author} {\bibfnamefont {A.~K.}\ \bibnamefont {Geim}},\ }\href
  {https://doi.org/10.1103/RevModPhys.81.109} {\bibfield  {journal} {\bibinfo
  {journal} {Rev. Mod. Phys.}\ }\textbf {\bibinfo {volume} {81}},\ \bibinfo
  {pages} {109} (\bibinfo {year} {2009})}\BibitemShut {NoStop}%
\bibitem [{\citenamefont {Nair}\ \emph {et~al.}(2008)\citenamefont {Nair},
  \citenamefont {Blake}, \citenamefont {Grigorenko}, \citenamefont {Novoselov},
  \citenamefont {Booth}, \citenamefont {Stauber}, \citenamefont {Peres},\ and\
  \citenamefont {Geim}}]{nair_science2008}%
  \BibitemOpen
  \bibfield  {author} {\bibinfo {author} {\bibfnamefont {R.~R.}\ \bibnamefont
  {Nair}}, \bibinfo {author} {\bibfnamefont {P.}~\bibnamefont {Blake}},
  \bibinfo {author} {\bibfnamefont {A.~N.}\ \bibnamefont {Grigorenko}},
  \bibinfo {author} {\bibfnamefont {K.~S.}\ \bibnamefont {Novoselov}}, \bibinfo
  {author} {\bibfnamefont {T.~J.}\ \bibnamefont {Booth}}, \bibinfo {author}
  {\bibfnamefont {T.}~\bibnamefont {Stauber}}, \bibinfo {author} {\bibfnamefont
  {N.~M.}\ \bibnamefont {Peres}},\ and\ \bibinfo {author} {\bibfnamefont
  {A.~K.}\ \bibnamefont {Geim}},\ }\href@noop {} {\bibfield  {journal}
  {\bibinfo  {journal} {Science}\ }\textbf {\bibinfo {volume} {320}},\ \bibinfo
  {pages} {1308} (\bibinfo {year} {2008})}\BibitemShut {NoStop}%
\bibitem [{\citenamefont {Vafek}\ and\ \citenamefont
  {Kang}(2020{\natexlab{a}})}]{kang-rg-prl20}%
  \BibitemOpen
  \bibfield  {author} {\bibinfo {author} {\bibfnamefont {O.}~\bibnamefont
  {Vafek}}\ and\ \bibinfo {author} {\bibfnamefont {J.}~\bibnamefont {Kang}},\
  }\href {https://doi.org/10.1103/PhysRevLett.125.257602} {\bibfield  {journal}
  {\bibinfo  {journal} {Phys. Rev. Lett.}\ }\textbf {\bibinfo {volume} {125}},\
  \bibinfo {pages} {257602} (\bibinfo {year} {2020}{\natexlab{a}})}\BibitemShut
  {NoStop}%
\bibitem [{\citenamefont {Lu}\ \emph {et~al.}(2023{\natexlab{b}})\citenamefont
  {Lu}, \citenamefont {Zhang}, \citenamefont {Wang}, \citenamefont {Gao},
  \citenamefont {Yang}, \citenamefont {Guo}, \citenamefont {Gao}, \citenamefont
  {Ye}, \citenamefont {Han},\ and\ \citenamefont {Liu}}]{lu-nc23}%
  \BibitemOpen
  \bibfield  {author} {\bibinfo {author} {\bibfnamefont {X.}~\bibnamefont
  {Lu}}, \bibinfo {author} {\bibfnamefont {S.}~\bibnamefont {Zhang}}, \bibinfo
  {author} {\bibfnamefont {Y.}~\bibnamefont {Wang}}, \bibinfo {author}
  {\bibfnamefont {X.}~\bibnamefont {Gao}}, \bibinfo {author} {\bibfnamefont
  {K.}~\bibnamefont {Yang}}, \bibinfo {author} {\bibfnamefont {Z.}~\bibnamefont
  {Guo}}, \bibinfo {author} {\bibfnamefont {Y.}~\bibnamefont {Gao}}, \bibinfo
  {author} {\bibfnamefont {Y.}~\bibnamefont {Ye}}, \bibinfo {author}
  {\bibfnamefont {Z.}~\bibnamefont {Han}},\ and\ \bibinfo {author}
  {\bibfnamefont {J.}~\bibnamefont {Liu}},\ }\href
  {https://doi.org/10.1038/s41467-023-41293-8} {\bibfield  {journal} {\bibinfo
  {journal} {Nature Communications}\ }\textbf {\bibinfo {volume} {14}},\
  \bibinfo {pages} {5550} (\bibinfo {year} {2023}{\natexlab{b}})}\BibitemShut
  {NoStop}%
\bibitem [{\citenamefont {Dong}\ \emph
  {et~al.}(2023{\natexlab{a}})\citenamefont {Dong}, \citenamefont {Patri},\
  and\ \citenamefont {Senthil}}]{senthil-fqah-arxiv23}%
  \BibitemOpen
  \bibfield  {author} {\bibinfo {author} {\bibfnamefont {Z.}~\bibnamefont
  {Dong}}, \bibinfo {author} {\bibfnamefont {A.~S.}\ \bibnamefont {Patri}},\
  and\ \bibinfo {author} {\bibfnamefont {T.}~\bibnamefont {Senthil}},\
  }\href@noop {} {\bibinfo {title} {Theory of fractional quantum anomalous hall
  phases in pentalayer rhombohedral graphene moir\'e structures}} (\bibinfo
  {year} {2023}{\natexlab{a}}),\ \Eprint {https://arxiv.org/abs/2311.03445}
  {arXiv:2311.03445 [cond-mat.str-el]} \BibitemShut {NoStop}%
\bibitem [{\citenamefont {Zhou}\ \emph {et~al.}(2023)\citenamefont {Zhou},
  \citenamefont {Yang},\ and\ \citenamefont {Zhang}}]{zhang-fqah-arxiv23}%
  \BibitemOpen
  \bibfield  {author} {\bibinfo {author} {\bibfnamefont {B.}~\bibnamefont
  {Zhou}}, \bibinfo {author} {\bibfnamefont {H.}~\bibnamefont {Yang}},\ and\
  \bibinfo {author} {\bibfnamefont {Y.-H.}\ \bibnamefont {Zhang}},\ }\href@noop
  {} {\bibinfo {title} {Fractional quantum anomalous hall effects in
  rhombohedral multilayer graphene in the moir\'eless limit and in coulomb
  imprinted superlattice}} (\bibinfo {year} {2023}),\ \Eprint
  {https://arxiv.org/abs/2311.04217} {arXiv:2311.04217 [cond-mat.str-el]}
  \BibitemShut {NoStop}%
\bibitem [{\citenamefont {Dong}\ \emph
  {et~al.}(2023{\natexlab{b}})\citenamefont {Dong}, \citenamefont {Wang},
  \citenamefont {Wang}, \citenamefont {Soejima}, \citenamefont {Zaletel},
  \citenamefont {Vishwanath},\ and\ \citenamefont
  {Parker}}]{ashvin-fqah-arxiv23}%
  \BibitemOpen
  \bibfield  {author} {\bibinfo {author} {\bibfnamefont {J.}~\bibnamefont
  {Dong}}, \bibinfo {author} {\bibfnamefont {T.}~\bibnamefont {Wang}}, \bibinfo
  {author} {\bibfnamefont {T.}~\bibnamefont {Wang}}, \bibinfo {author}
  {\bibfnamefont {T.}~\bibnamefont {Soejima}}, \bibinfo {author} {\bibfnamefont
  {M.~P.}\ \bibnamefont {Zaletel}}, \bibinfo {author} {\bibfnamefont
  {A.}~\bibnamefont {Vishwanath}},\ and\ \bibinfo {author} {\bibfnamefont
  {D.~E.}\ \bibnamefont {Parker}},\ }\href@noop {} {\bibinfo {title} {Anomalous
  hall crystals in rhombohedral multilayer graphene i: Interaction-driven chern
  bands and fractional quantum hall states at zero magnetic field}} (\bibinfo
  {year} {2023}{\natexlab{b}}),\ \Eprint {https://arxiv.org/abs/2311.05568}
  {arXiv:2311.05568 [cond-mat.str-el]} \BibitemShut {NoStop}%
\bibitem [{\citenamefont {Herzog-Arbeitman}\ \emph {et~al.}(2023)\citenamefont
  {Herzog-Arbeitman}, \citenamefont {Wang}, \citenamefont {Liu}, \citenamefont
  {Tam}, \citenamefont {Qi}, \citenamefont {Jia}, \citenamefont {Efetov},
  \citenamefont {Vafek}, \citenamefont {Regnault}, \citenamefont {Weng},
  \citenamefont {Wu}, \citenamefont {Bernevig},\ and\ \citenamefont
  {Yu}}]{bernevig-fci-plg-arxiv23}%
  \BibitemOpen
  \bibfield  {author} {\bibinfo {author} {\bibfnamefont {J.}~\bibnamefont
  {Herzog-Arbeitman}}, \bibinfo {author} {\bibfnamefont {Y.}~\bibnamefont
  {Wang}}, \bibinfo {author} {\bibfnamefont {J.}~\bibnamefont {Liu}}, \bibinfo
  {author} {\bibfnamefont {P.~M.}\ \bibnamefont {Tam}}, \bibinfo {author}
  {\bibfnamefont {Z.}~\bibnamefont {Qi}}, \bibinfo {author} {\bibfnamefont
  {Y.}~\bibnamefont {Jia}}, \bibinfo {author} {\bibfnamefont {D.~K.}\
  \bibnamefont {Efetov}}, \bibinfo {author} {\bibfnamefont {O.}~\bibnamefont
  {Vafek}}, \bibinfo {author} {\bibfnamefont {N.}~\bibnamefont {Regnault}},
  \bibinfo {author} {\bibfnamefont {H.}~\bibnamefont {Weng}}, \bibinfo {author}
  {\bibfnamefont {Q.}~\bibnamefont {Wu}}, \bibinfo {author} {\bibfnamefont
  {B.~A.}\ \bibnamefont {Bernevig}},\ and\ \bibinfo {author} {\bibfnamefont
  {J.}~\bibnamefont {Yu}},\ }\href@noop {} {\bibinfo {title} {Moir\'e
  fractional chern insulators ii: First-principles calculations and continuum
  models of rhombohedral graphene superlattices}} (\bibinfo {year} {2023}),\
  \Eprint {https://arxiv.org/abs/2311.12920} {arXiv:2311.12920
  [cond-mat.mes-hall]} \BibitemShut {NoStop}%
\bibitem [{\citenamefont {Moon}\ and\ \citenamefont
  {Koshino}(2014{\natexlab{b}})}]{moon-hBNgr-prb-2014}%
  \BibitemOpen
  \bibfield  {author} {\bibinfo {author} {\bibfnamefont {P.}~\bibnamefont
  {Moon}}\ and\ \bibinfo {author} {\bibfnamefont {M.}~\bibnamefont {Koshino}},\
  }\href {https://doi.org/10.1103/PhysRevB.90.155406} {\bibfield  {journal}
  {\bibinfo  {journal} {Phys. Rev. B}\ }\textbf {\bibinfo {volume} {90}},\
  \bibinfo {pages} {155406} (\bibinfo {year} {2014}{\natexlab{b}})}\BibitemShut
  {NoStop}%
\bibitem [{\citenamefont {Lu}\ \emph {et~al.}(2023{\natexlab{c}})\citenamefont
  {Lu}, \citenamefont {Han}, \citenamefont {Yao}, \citenamefont {Reddy},
  \citenamefont {Yang}, \citenamefont {Seo}, \citenamefont {Watanabe},
  \citenamefont {Taniguchi}, \citenamefont {Fu},\ and\ \citenamefont
  {Ju}}]{lu-fqahe_pGr-arxiv-2023}%
  \BibitemOpen
  \bibfield  {author} {\bibinfo {author} {\bibfnamefont {Z.}~\bibnamefont
  {Lu}}, \bibinfo {author} {\bibfnamefont {T.}~\bibnamefont {Han}}, \bibinfo
  {author} {\bibfnamefont {Y.}~\bibnamefont {Yao}}, \bibinfo {author}
  {\bibfnamefont {A.~P.}\ \bibnamefont {Reddy}}, \bibinfo {author}
  {\bibfnamefont {J.}~\bibnamefont {Yang}}, \bibinfo {author} {\bibfnamefont
  {J.}~\bibnamefont {Seo}}, \bibinfo {author} {\bibfnamefont {K.}~\bibnamefont
  {Watanabe}}, \bibinfo {author} {\bibfnamefont {T.}~\bibnamefont {Taniguchi}},
  \bibinfo {author} {\bibfnamefont {L.}~\bibnamefont {Fu}},\ and\ \bibinfo
  {author} {\bibfnamefont {L.}~\bibnamefont {Ju}},\ }\href@noop {} {\bibinfo
  {title} {Fractional quantum anomalous hall effect in a graphene moire
  superlattice}} (\bibinfo {year} {2023}{\natexlab{c}}),\ \Eprint
  {https://arxiv.org/abs/2309.17436} {arXiv:2309.17436 [cond-mat.mes-hall]}
  \BibitemShut {NoStop}%
\bibitem [{\citenamefont {Jang}\ \emph
  {et~al.}(2023{\natexlab{b}})\citenamefont {Jang}, \citenamefont {Park},
  \citenamefont {Jung},\ and\ \citenamefont {Min}}]{jang-multiGr-prblett-2023}%
  \BibitemOpen
  \bibfield  {author} {\bibinfo {author} {\bibfnamefont {Y.}~\bibnamefont
  {Jang}}, \bibinfo {author} {\bibfnamefont {Y.}~\bibnamefont {Park}}, \bibinfo
  {author} {\bibfnamefont {J.}~\bibnamefont {Jung}},\ and\ \bibinfo {author}
  {\bibfnamefont {H.}~\bibnamefont {Min}},\ }\href
  {https://doi.org/10.1103/PhysRevB.108.L041101} {\bibfield  {journal}
  {\bibinfo  {journal} {Phys. Rev. B}\ }\textbf {\bibinfo {volume} {108}},\
  \bibinfo {pages} {L041101} (\bibinfo {year}
  {2023}{\natexlab{b}})}\BibitemShut {NoStop}%
\bibitem [{\citenamefont {McCann}(2006)}]{mccann-bGrEfield-prb-2006}%
  \BibitemOpen
  \bibfield  {author} {\bibinfo {author} {\bibfnamefont {E.}~\bibnamefont
  {McCann}},\ }\href {https://doi.org/10.1103/PhysRevB.74.161403} {\bibfield
  {journal} {\bibinfo  {journal} {Phys. Rev. B}\ }\textbf {\bibinfo {volume}
  {74}},\ \bibinfo {pages} {161403} (\bibinfo {year} {2006})}\BibitemShut
  {NoStop}%
\bibitem [{\citenamefont {Avetisyan}\ \emph
  {et~al.}(2009{\natexlab{a}})\citenamefont {Avetisyan}, \citenamefont
  {Partoens},\ and\ \citenamefont
  {Peeters}}]{avetisyan-multiGrEfield1-prb-2009}%
  \BibitemOpen
  \bibfield  {author} {\bibinfo {author} {\bibfnamefont {A.~A.}\ \bibnamefont
  {Avetisyan}}, \bibinfo {author} {\bibfnamefont {B.}~\bibnamefont
  {Partoens}},\ and\ \bibinfo {author} {\bibfnamefont {F.~M.}\ \bibnamefont
  {Peeters}},\ }\href {https://doi.org/10.1103/PhysRevB.79.035421} {\bibfield
  {journal} {\bibinfo  {journal} {Phys. Rev. B}\ }\textbf {\bibinfo {volume}
  {79}},\ \bibinfo {pages} {035421} (\bibinfo {year}
  {2009}{\natexlab{a}})}\BibitemShut {NoStop}%
\bibitem [{\citenamefont {Avetisyan}\ \emph
  {et~al.}(2009{\natexlab{b}})\citenamefont {Avetisyan}, \citenamefont
  {Partoens},\ and\ \citenamefont
  {Peeters}}]{avetisyan-multiGrEfield2-prb-2009}%
  \BibitemOpen
  \bibfield  {author} {\bibinfo {author} {\bibfnamefont {A.~A.}\ \bibnamefont
  {Avetisyan}}, \bibinfo {author} {\bibfnamefont {B.}~\bibnamefont
  {Partoens}},\ and\ \bibinfo {author} {\bibfnamefont {F.~M.}\ \bibnamefont
  {Peeters}},\ }\href {https://doi.org/10.1103/PhysRevB.80.195401} {\bibfield
  {journal} {\bibinfo  {journal} {Phys. Rev. B}\ }\textbf {\bibinfo {volume}
  {80}},\ \bibinfo {pages} {195401} (\bibinfo {year}
  {2009}{\natexlab{b}})}\BibitemShut {NoStop}%
\bibitem [{\citenamefont {Koshino}\ and\ \citenamefont
  {McCann}(2009)}]{koshino-triGrEfield-prb-2009}%
  \BibitemOpen
  \bibfield  {author} {\bibinfo {author} {\bibfnamefont {M.}~\bibnamefont
  {Koshino}}\ and\ \bibinfo {author} {\bibfnamefont {E.}~\bibnamefont
  {McCann}},\ }\href {https://doi.org/10.1103/PhysRevB.79.125443} {\bibfield
  {journal} {\bibinfo  {journal} {Phys. Rev. B}\ }\textbf {\bibinfo {volume}
  {79}},\ \bibinfo {pages} {125443} (\bibinfo {year} {2009})}\BibitemShut
  {NoStop}%
\bibitem [{\citenamefont {Zhang}\ \emph {et~al.}(2010)\citenamefont {Zhang},
  \citenamefont {Sahu}, \citenamefont {Min},\ and\ \citenamefont
  {MacDonald}}]{zhang-triGrEfield-prb-2010}%
  \BibitemOpen
  \bibfield  {author} {\bibinfo {author} {\bibfnamefont {F.}~\bibnamefont
  {Zhang}}, \bibinfo {author} {\bibfnamefont {B.}~\bibnamefont {Sahu}},
  \bibinfo {author} {\bibfnamefont {H.}~\bibnamefont {Min}},\ and\ \bibinfo
  {author} {\bibfnamefont {A.~H.}\ \bibnamefont {MacDonald}},\ }\href
  {https://doi.org/10.1103/PhysRevB.82.035409} {\bibfield  {journal} {\bibinfo
  {journal} {Phys. Rev. B}\ }\textbf {\bibinfo {volume} {82}},\ \bibinfo
  {pages} {035409} (\bibinfo {year} {2010})}\BibitemShut {NoStop}%
\bibitem [{\citenamefont {Elias}\ \emph {et~al.}(2011)\citenamefont {Elias},
  \citenamefont {Gorbachev}, \citenamefont {Mayorov}, \citenamefont {Morozov},
  \citenamefont {Zhukov}, \citenamefont {Blake}, \citenamefont {Ponomarenko},
  \citenamefont {Grigorieva}, \citenamefont {Novoselov}, \citenamefont {Guinea}
  \emph {et~al.}}]{elias_natphys2011}%
  \BibitemOpen
  \bibfield  {author} {\bibinfo {author} {\bibfnamefont {D.~C.}\ \bibnamefont
  {Elias}}, \bibinfo {author} {\bibfnamefont {R.}~\bibnamefont {Gorbachev}},
  \bibinfo {author} {\bibfnamefont {A.}~\bibnamefont {Mayorov}}, \bibinfo
  {author} {\bibfnamefont {S.}~\bibnamefont {Morozov}}, \bibinfo {author}
  {\bibfnamefont {A.}~\bibnamefont {Zhukov}}, \bibinfo {author} {\bibfnamefont
  {P.}~\bibnamefont {Blake}}, \bibinfo {author} {\bibfnamefont
  {L.}~\bibnamefont {Ponomarenko}}, \bibinfo {author} {\bibfnamefont
  {I.}~\bibnamefont {Grigorieva}}, \bibinfo {author} {\bibfnamefont
  {K.}~\bibnamefont {Novoselov}}, \bibinfo {author} {\bibfnamefont
  {F.}~\bibnamefont {Guinea}}, \emph {et~al.},\ }\href@noop {} {\bibfield
  {journal} {\bibinfo  {journal} {Nature Physics}\ }\textbf {\bibinfo {volume}
  {7}},\ \bibinfo {pages} {701} (\bibinfo {year} {2011})}\BibitemShut {NoStop}%
\bibitem [{\citenamefont {Vafek}\ and\ \citenamefont
  {Kang}(2020{\natexlab{b}})}]{vafek_prl2020}%
  \BibitemOpen
  \bibfield  {author} {\bibinfo {author} {\bibfnamefont {O.}~\bibnamefont
  {Vafek}}\ and\ \bibinfo {author} {\bibfnamefont {J.}~\bibnamefont {Kang}},\
  }\href {https://doi.org/10.1103/PhysRevLett.125.257602} {\bibfield  {journal}
  {\bibinfo  {journal} {Phys. Rev. Lett.}\ }\textbf {\bibinfo {volume} {125}},\
  \bibinfo {pages} {257602} (\bibinfo {year} {2020}{\natexlab{b}})}\BibitemShut
  {NoStop}%
\bibitem [{\citenamefont {Lu}\ \emph {et~al.}(2023{\natexlab{d}})\citenamefont
  {Lu}, \citenamefont {Zhang}, \citenamefont {Wang}, \citenamefont {Gao},
  \citenamefont {Yang}, \citenamefont {Guo}, \citenamefont {Gao}, \citenamefont
  {Ye}, \citenamefont {Han},\ and\ \citenamefont
  {Liu}}]{lu-GrCrOCl-natcom-2023}%
  \BibitemOpen
  \bibfield  {author} {\bibinfo {author} {\bibfnamefont {X.}~\bibnamefont
  {Lu}}, \bibinfo {author} {\bibfnamefont {S.}~\bibnamefont {Zhang}}, \bibinfo
  {author} {\bibfnamefont {Y.}~\bibnamefont {Wang}}, \bibinfo {author}
  {\bibfnamefont {X.}~\bibnamefont {Gao}}, \bibinfo {author} {\bibfnamefont
  {K.}~\bibnamefont {Yang}}, \bibinfo {author} {\bibfnamefont {Z.}~\bibnamefont
  {Guo}}, \bibinfo {author} {\bibfnamefont {Y.}~\bibnamefont {Gao}}, \bibinfo
  {author} {\bibfnamefont {Y.}~\bibnamefont {Ye}}, \bibinfo {author}
  {\bibfnamefont {Z.}~\bibnamefont {Han}},\ and\ \bibinfo {author}
  {\bibfnamefont {J.}~\bibnamefont {Liu}},\ }\href@noop {} {\bibfield
  {journal} {\bibinfo  {journal} {Nature Communications}\ }\textbf {\bibinfo
  {volume} {14}},\ \bibinfo {pages} {5550} (\bibinfo {year}
  {2023}{\natexlab{d}})}\BibitemShut {NoStop}%
\bibitem [{\citenamefont {Zhang}\ \emph {et~al.}(2022)\citenamefont {Zhang},
  \citenamefont {Dai},\ and\ \citenamefont {Liu}}]{zhang_prl2022}%
  \BibitemOpen
  \bibfield  {author} {\bibinfo {author} {\bibfnamefont {S.}~\bibnamefont
  {Zhang}}, \bibinfo {author} {\bibfnamefont {X.}~\bibnamefont {Dai}},\ and\
  \bibinfo {author} {\bibfnamefont {J.}~\bibnamefont {Liu}},\ }\href
  {https://doi.org/10.1103/PhysRevLett.128.026403} {\bibfield  {journal}
  {\bibinfo  {journal} {Phys. Rev. Lett.}\ }\textbf {\bibinfo {volume} {128}},\
  \bibinfo {pages} {026403} (\bibinfo {year} {2022})}\BibitemShut {NoStop}%
\bibitem [{\citenamefont {Reddy}\ \emph
  {et~al.}(2023{\natexlab{b}})\citenamefont {Reddy}, \citenamefont {Alsallom},
  \citenamefont {Zhang}, \citenamefont {Devakul},\ and\ \citenamefont
  {Fu}}]{Fu-27site-prb23}%
  \BibitemOpen
  \bibfield  {author} {\bibinfo {author} {\bibfnamefont {A.~P.}\ \bibnamefont
  {Reddy}}, \bibinfo {author} {\bibfnamefont {F.}~\bibnamefont {Alsallom}},
  \bibinfo {author} {\bibfnamefont {Y.}~\bibnamefont {Zhang}}, \bibinfo
  {author} {\bibfnamefont {T.}~\bibnamefont {Devakul}},\ and\ \bibinfo {author}
  {\bibfnamefont {L.}~\bibnamefont {Fu}},\ }\href
  {https://doi.org/10.1103/PhysRevB.108.085117} {\bibfield  {journal} {\bibinfo
   {journal} {Phys. Rev. B}\ }\textbf {\bibinfo {volume} {108}},\ \bibinfo
  {pages} {085117} (\bibinfo {year} {2023}{\natexlab{b}})}\BibitemShut
  {NoStop}%
\bibitem [{\citenamefont {Regnault}\ and\ \citenamefont
  {Bernevig}(2011{\natexlab{b}})}]{Bernevig-prx11}%
  \BibitemOpen
  \bibfield  {author} {\bibinfo {author} {\bibfnamefont {N.}~\bibnamefont
  {Regnault}}\ and\ \bibinfo {author} {\bibfnamefont {B.~A.}\ \bibnamefont
  {Bernevig}},\ }\href {https://doi.org/10.1103/PhysRevX.1.021014} {\bibfield
  {journal} {\bibinfo  {journal} {Phys. Rev. X}\ }\textbf {\bibinfo {volume}
  {1}},\ \bibinfo {pages} {021014} (\bibinfo {year}
  {2011}{\natexlab{b}})}\BibitemShut {NoStop}%
\bibitem [{\citenamefont {Thompson}\ \emph {et~al.}(2022)\citenamefont
  {Thompson}, \citenamefont {Aktulga}, \citenamefont {Berger}, \citenamefont
  {Bolintineanu}, \citenamefont {Brown}, \citenamefont {Crozier}, \citenamefont
  {in~'t Veld}, \citenamefont {Kohlmeyer}, \citenamefont {Moore}, \citenamefont
  {Nguyen}, \citenamefont {Shan}, \citenamefont {Stevens}, \citenamefont
  {Tranchida}, \citenamefont {Trott},\ and\ \citenamefont {Plimpton}}]{LAMMPS}%
  \BibitemOpen
  \bibfield  {author} {\bibinfo {author} {\bibfnamefont {A.~P.}\ \bibnamefont
  {Thompson}}, \bibinfo {author} {\bibfnamefont {H.~M.}\ \bibnamefont
  {Aktulga}}, \bibinfo {author} {\bibfnamefont {R.}~\bibnamefont {Berger}},
  \bibinfo {author} {\bibfnamefont {D.~S.}\ \bibnamefont {Bolintineanu}},
  \bibinfo {author} {\bibfnamefont {W.~M.}\ \bibnamefont {Brown}}, \bibinfo
  {author} {\bibfnamefont {P.~S.}\ \bibnamefont {Crozier}}, \bibinfo {author}
  {\bibfnamefont {P.~J.}\ \bibnamefont {in~'t Veld}}, \bibinfo {author}
  {\bibfnamefont {A.}~\bibnamefont {Kohlmeyer}}, \bibinfo {author}
  {\bibfnamefont {S.~G.}\ \bibnamefont {Moore}}, \bibinfo {author}
  {\bibfnamefont {T.~D.}\ \bibnamefont {Nguyen}}, \bibinfo {author}
  {\bibfnamefont {R.}~\bibnamefont {Shan}}, \bibinfo {author} {\bibfnamefont
  {M.~J.}\ \bibnamefont {Stevens}}, \bibinfo {author} {\bibfnamefont
  {J.}~\bibnamefont {Tranchida}}, \bibinfo {author} {\bibfnamefont
  {C.}~\bibnamefont {Trott}},\ and\ \bibinfo {author} {\bibfnamefont {S.~J.}\
  \bibnamefont {Plimpton}},\ }\href {https://doi.org/10.1016/j.cpc.2021.108171}
  {\bibfield  {journal} {\bibinfo  {journal} {Comp. Phys. Comm.}\ }\textbf
  {\bibinfo {volume} {271}},\ \bibinfo {pages} {108171} (\bibinfo {year}
  {2022})}\BibitemShut {NoStop}%
\bibitem [{\citenamefont {Leconte}\ \emph {et~al.}(2022)\citenamefont
  {Leconte}, \citenamefont {Javvaji}, \citenamefont {An}, \citenamefont
  {Samudrala},\ and\ \citenamefont {Jung}}]{G-HBN-MD}%
  \BibitemOpen
  \bibfield  {author} {\bibinfo {author} {\bibfnamefont {N.}~\bibnamefont
  {Leconte}}, \bibinfo {author} {\bibfnamefont {S.}~\bibnamefont {Javvaji}},
  \bibinfo {author} {\bibfnamefont {J.}~\bibnamefont {An}}, \bibinfo {author}
  {\bibfnamefont {A.}~\bibnamefont {Samudrala}},\ and\ \bibinfo {author}
  {\bibfnamefont {J.}~\bibnamefont {Jung}},\ }\href
  {https://doi.org/10.1103/PhysRevB.106.115410} {\bibfield  {journal} {\bibinfo
   {journal} {Phys. Rev. B}\ }\textbf {\bibinfo {volume} {106}},\ \bibinfo
  {pages} {115410} (\bibinfo {year} {2022})}\BibitemShut {NoStop}%
\bibitem [{\citenamefont {Nam}\ and\ \citenamefont
  {Koshino}(2017)}]{koshino-prb17}%
  \BibitemOpen
  \bibfield  {author} {\bibinfo {author} {\bibfnamefont {N.~N.~T.}\
  \bibnamefont {Nam}}\ and\ \bibinfo {author} {\bibfnamefont {M.}~\bibnamefont
  {Koshino}},\ }\href {https://doi.org/10.1103/PhysRevB.96.075311} {\bibfield
  {journal} {\bibinfo  {journal} {Phys. Rev. B}\ }\textbf {\bibinfo {volume}
  {96}},\ \bibinfo {pages} {075311} (\bibinfo {year} {2017})}\BibitemShut
  {NoStop}%
\bibitem [{\citenamefont {Long}\ \emph {et~al.}(2022)\citenamefont {Long},
  \citenamefont {Pantale{\'o}n}, \citenamefont {Zhan}, \citenamefont {Guinea},
  \citenamefont {Silva-Guill{\'e}n},\ and\ \citenamefont
  {Yuan}}]{shengjun-yuan-npj2022}%
  \BibitemOpen
  \bibfield  {author} {\bibinfo {author} {\bibfnamefont {M.}~\bibnamefont
  {Long}}, \bibinfo {author} {\bibfnamefont {P.~A.}\ \bibnamefont
  {Pantale{\'o}n}}, \bibinfo {author} {\bibfnamefont {Z.}~\bibnamefont {Zhan}},
  \bibinfo {author} {\bibfnamefont {F.}~\bibnamefont {Guinea}}, \bibinfo
  {author} {\bibfnamefont {J.~{\'A}.}\ \bibnamefont {Silva-Guill{\'e}n}},\ and\
  \bibinfo {author} {\bibfnamefont {S.}~\bibnamefont {Yuan}},\ }\href@noop {}
  {\bibfield  {journal} {\bibinfo  {journal} {npj Computational Materials}\
  }\textbf {\bibinfo {volume} {8}},\ \bibinfo {pages} {73} (\bibinfo {year}
  {2022})}\BibitemShut {NoStop}%
\bibitem [{\citenamefont {Moon}\ and\ \citenamefont
  {Koshino}(2014{\natexlab{c}})}]{koshino-hBN-prb14}%
  \BibitemOpen
  \bibfield  {author} {\bibinfo {author} {\bibfnamefont {P.}~\bibnamefont
  {Moon}}\ and\ \bibinfo {author} {\bibfnamefont {M.}~\bibnamefont {Koshino}},\
  }\href {https://doi.org/10.1103/PhysRevB.90.155406} {\bibfield  {journal}
  {\bibinfo  {journal} {Phys. Rev. B}\ }\textbf {\bibinfo {volume} {90}},\
  \bibinfo {pages} {155406} (\bibinfo {year} {2014}{\natexlab{c}})}\BibitemShut
  {NoStop}%
\bibitem [{\citenamefont {Koshino}\ and\ \citenamefont
  {Nam}(2020)}]{koshino-tbg-epc-prb20}%
  \BibitemOpen
  \bibfield  {author} {\bibinfo {author} {\bibfnamefont {M.}~\bibnamefont
  {Koshino}}\ and\ \bibinfo {author} {\bibfnamefont {N.~N.~T.}\ \bibnamefont
  {Nam}},\ }\href {https://doi.org/10.1103/PhysRevB.101.195425} {\bibfield
  {journal} {\bibinfo  {journal} {Phys. Rev. B}\ }\textbf {\bibinfo {volume}
  {101}},\ \bibinfo {pages} {195425} (\bibinfo {year} {2020})}\BibitemShut
  {NoStop}%
\bibitem [{\citenamefont {Xie}\ and\ \citenamefont
  {Liu}(2023)}]{xie-tbg-phonon-prb23}%
  \BibitemOpen
  \bibfield  {author} {\bibinfo {author} {\bibfnamefont {B.}~\bibnamefont
  {Xie}}\ and\ \bibinfo {author} {\bibfnamefont {J.}~\bibnamefont {Liu}},\
  }\href {https://doi.org/10.1103/PhysRevB.108.094115} {\bibfield  {journal}
  {\bibinfo  {journal} {Phys. Rev. B}\ }\textbf {\bibinfo {volume} {108}},\
  \bibinfo {pages} {094115} (\bibinfo {year} {2023})}\BibitemShut {NoStop}%
\end{thebibliography}%

\end{document}